\documentclass[utf8]{FrontiersinHarvard} 
\usepackage{url,hyperref,lineno,microtype,subcaption}
\usepackage[onehalfspacing]{setspace}
\usepackage{float}
\usepackage{amsmath}
\usepackage{makecell}

\nolinenumbers

\def\keyFont{\fontsize{8}{11}\helveticabold }
\def\firstAuthorLast{Vieten} 
\def\Authors{Manfred M. Vieten\,$^{1,*}$}
\def\Address{$^{1}$University of Konstanz, Konstanz, Germany }
\def\corrAuthor{Manfred M. Vieten}

\def\corrEmail{manfred.vieten@uni-konstanz.de}

\begin{document}
\onecolumn
\firstpage{1}

\title[Mass and forces in a Kaluza-Klein model]{Mass, forces, and the dark sector in a Kaluza-Klein model with 6 compact dimensions} 

\author[\firstAuthorLast ]{\Authors} 
\address{} 
\correspondance{} 

\extraAuth{}

\maketitle

\begin{abstract}
(1) Objective: This study seeks to calculate the meson and baryon masses and to investigate the dynamics within hadrons and lepton pairs by formulating a unified physical framework that, in addition to gravitation, simultaneously incorporates the strong, electromagnetic and weak interactions. (2) Method: A framework inspired by the Standard Model and based on Kaluza–Klein theory is formulated using an extended general-relativistic description in a spacetime of one temporal, three spatial, and six compactified dimensions. The model suggests that excitations, deformations of the 10-dimensional spacetime traveling at light-speed, contribute to mass and distinct spin characteristics in a 4-dimensional setting. Using the electron g-factor, charged-lepton masses, and masses of the proton, neutron, and $\pi^{0}, \pi^{+}, \rho^{0}, \rho^{+}, \phi, \psi, \Upsilon$ mesons, the parameters compactification radius $\rho$, weak interaction coupling $\alpha_{w}$, strong interaction coupling $\alpha_{s}$, flavor constants $A_i$ and the antineutrino/neutrino density ratio $\delta^3$ are determined and used to compute the masses of the hadrons. (3) Key results: Masses of 102 hadrons are obtained; comparison with data from 71 measurements shows relative errors below 0.05 in 70\%, below 0.1 in 89\% and below 0.2 in 100\%, with a correlation coefficient $r=0.997$ ($p<10^{-75}$) between the measured and calculated masses. Thirty-one additional hadron masses are predicted, and the model identifies asymmetries in electric and weak forces, as well as distance- and velocity-dependent interactions. (4) Implications: The framework offers testable predictions for hadron spectroscopy, predicts sterile, exclusive gravitationally interacting bosons as dark-matter candidates, states the equations for the elementary forces, and suggests possible connections between force asymmetries and dark energy.
\section{}

\tiny
 \keyFont{ \section{Keywords:} hadron masses, Kaluza-Klein theory, geodesic equation, fundamental forces, neutrino mass, dark matter, dark energy} 
\end{abstract}

\section{Introduction}
The Standard Model of particle physics~\cite{RN176, RN177, RN178, RN179, RN180, RN181, RN182} stands as one of the most triumphant theories in the realm of physics. It excels at elucidating the fundamental interactions among particles through bosons~\cite{RN183}: the photon governs electromagnetism, the eight gluons oversee the strong force, and the \texorpdfstring{$\mathrm{W}^{+}, \mathrm{W}^{-}$, and $\mathrm{Z}$}{W+, W-, and Z} bosons manage the weak force. A significant accomplishment of the Standard Model is its accurate determination of the electron's anomalous magnetic dipole moment via quantum corrections \cite{RN184}. Subsequently, with accuracy extending to 12 significant figures~\cite{RN185} between theoretical predictions and experimental data. Despite its remarkable successes, the Standard Model has important gaps, such as the mere partial success in calculating the masses of mesons and baryons, its inability to explain dark matter and dark energy~\cite{RN186}, and its exclusion of gravity. Although extensions like quantum gravity theories ~\cite{RN187} have been developed suggesting a spin-2 graviton as a potential mediator, no experimental evidence supports this. Another attempt to close Standard Model's gaps is found with supersymmetry (SUSY), but experiments came out empty-handed \cite{fischer2023promise, constantin2025lhc, ARBEY2012162}. String theory~\cite{RN197, RN198, RN199} is a further effort, a field rich in mathematical complexity, but still lacking testable predictions. In all these models quantization takes place and the forces between particles are accomplished via boson exchange. A totally different approach, temporarily absent from mainstream physics, is found in Kaluza-Klein models \cite{RN190,Rn191,RN192,RN193}. Here force is caused by curvature of the spacetime, reminiscent of Einstein's General Relativity, but being established in five dimensions instead of four. This paper presents a model that reinterprets the Standard Model through a *Kaluza-Klein-like theory* in ten adaptable dimensions: one temporal, three large-scale spatial, and six compact spatial dimensions. Within this ten-dimensional spacetime, an *excitation*—a transient deformation—consistently travels at the speed of light. In four-dimensional space, such an excitation appears as a gravitational wave, whereas in a compact dimension, a stable excitation has a clearly defined length based on the dimension’s radius. The six compact dimensions' radii dynamically change according to specific excitations, which relate to particle interactions. The Standard Model of Particle Physics (SM) is formulated in a flat, 4-dimensional spacetime. Moreover, the observed particle masses are too small to necessitate a curved spacetime description. This raises the question of how a Kaluza–Klein-type framework, which attributes all forces to spacetime curvature, can be reconciled with the empirically successful SM. For strong, electromagnetic and weak interactions, the corresponding forces appear to operate within this flat 4D background. However, one can still describe them geometrically by assigning each interaction its own metric space. This idea is crucial for the subsequent development of the model. Ordinary matter remains subject to the flat 4D spacetime, whereas the charges of the other forces are associated with strongly curved spaces that incorporate effects from the additional dimensions. Another essential fact is that, with the energy/mass described within a flat space, energy conservation is ensured. As the model is developed, an explicit spacetime structure is devised. Further steps involve elucidating the three fundamental interactions and devising formulas for computing particle masses, with particles identified in relation to the six compact dimensions. The essential parameters of the model are determined using empirical data, incorporating the known masses of charged leptons, proton, neutron, and mesons $\pi^{0}, \pi^{+}, \rho^{0}, \rho^{+}, \phi, \psi, \Upsilon$, as well as the magnetic moment of the electron. Key parameters include the compactification radius $\rho$, the weak coupling $\alpha_{w}$, the strong coupling $\alpha_{s}$, the mean $z$-values for baryons, scalar, and vector mesons $z_{Baryon}$, $z_{s\_Meson}$ and $z_{v\_Meson}$, the flavor constants $A_i$, and the antineutrino-to-neutrino density ratio $\delta^3$. Once established, no further free parameters exist. Subsequently, the model outlines rules for calculating the mass of scalar and vector mesons, along with spin-$\frac{1}{2}$ and spin-$\frac{3}{2}$ baryons without orbital angular momentum. It calculates masses for 102 hadrons, which are then compared to experimental data when available, with the relative error between observed and predicted masses being under 0.05 for most hadrons. The model also accommodates sterile bosons, which interact solely through gravitational forces, thus constituting dark matter. Finally, the evaluation of the model includes an analysis of the strong, electric, and weak forces, which are described with equations, numbers, and illustrations. The electric force, and to a lesser extent the weak force, show asymmetrical behavior, which may offer an explanation for dark energy. Ultimately, to evaluate the model, two experiments conducted in a laboratory setting and one observation from astronomical data are proposed. Analyzing lepton interactions within this framework yielded surprising findings. In $e^+e^+/e^-e^-$ experiments, particles are unable to approach distances less than approximately $9 \, fm$. During $e^+e^-$ collisions, particles come nearer but slow to nearly a halt at around $1 \, fm$ before attraction sets in. This indicates that particle transmutation could take place at this range, where particles exhibit substantial potential energy determined by the initial beam energy.

\section{Method} 
The proposed Kaluza-Klein-like model seeks to describe the particle masses and the forces between them through a 10-dimensional curved spacetime. Six compact cylindrical dimensions serve as sources for the three interactions, strong, electromagnetic, and weak. Each of the compactified dimensions, respectively, the three interactions form independent manifolds being felt by the respective charges only. This separation is necessary because a
formulation involving a mixture of interactions within a single manifold would imply that
all matter is simultaneously influenced by electric, weak, and strong forces in an
indistinguishable manner; for instance, an electric charge would influence neutral matter as well. Gravity couples to matter/energy as its source charge within 4D spacetime via the energy-momentum tensor. Together with the strong, electromagnetic, and weak interactions, gravity couples to their respective charges, while together determining the total force acting on a given particle. The proposed framework draws its conceptual foundations from the Standard Model (SM) of particle physics. It is intended as an alternative perspective on the domain rather than as a rival theoretical edifice. Accordingly, it introduces analogs to the strong, electromagnetic, and weak interactions. Moreover, it employs formulas that are corroborated by experimental data. Established physical principles, including the superposition principle, are also explicitly incorporated. Despite these parallels, the underlying mathematical formalisms of the two theoretical constructs differ substantially. The Standard Model is formulated within quantum mechanics, whose structure fundamentally relies on complex numbers and its forces are generated by virtual boson exchange. The quantum fields and wave functions of the SM representing particles are inherently complex, possessing both real and imaginary components. By contrast, the newly proposed model is cast as a classical field theory, closely resembling the formulation of general relativity. Its geometric content—metrics, distances, and curvature—is described entirely in terms of real-valued tensor fields. All forces are caused by the curvature of the 10-dimensional spacetime and are described by geodesics. The gravitational, strong, electric, and weak interactions act independently of each other. As a result, the description cannot rely on a single metric. Rather, each particle interaction is characterized by its own metric. The metric tensors associated with the strong, electromagnetic, and weak interactions \eqref{eq:metric_tensor_of_all_interactions} depend on the particle’s normalized separation $z$, the normalized velocity $\beta$, and the derivative of the normalized velocity with respect to $z$. This dependence applies to all components that are not constant:
\begin{align}
\label{eq:metric_tensor_of_all_interactions}
g_{\mu \nu}^{s,e,w} \! \left(z, \beta, \beta' \right) = 
\begin{bmatrix}
1 & 0 & 0 & 0 & 0 & 0 & 0 & 0 & 0 & 0 \\
0 & g_{11}^{s,e,w} \!  & 0 & 0 & g_{14}^{s} \!  & g_{15}^{s} \! & g_{16}^{s} \!  & g_{17}^{e} \!  & g_{18}^{w} & g_{19}^{w} \\
0 & 0 & -1 & 0 & 0 & 0 & 0 & 0 & 0 & 0 \\
0 & 0 & 0 & -1 & 0 & 0 & 0 & 0 & 0 & 0 \\
0 & g_{41}^{s} \!  & 0 & 0 & -1 & 0 & 0 & 0 & 0 & 0 \\
0 & g_{51}^{s} \!  & 0 & 0 & 0 & -1 & 0 & 0 & 0 & 0 \\
0 & g_{61}^{s} \!  & 0 & 0 & 0 & 0 & -1 & 0 & 0 & 0 \\
0 & g_{71}^{e} \!  & 0 & 0 & 0 & 0 & 0 & -1 & 0 & 0 \\
0 & g_{18}^{w} & 0 & 0 & 0 & 0 & 0 & 0 & -1 & 0 \\
0 & g_{91}^{w} & 0 & 0 & 0 & 0 & 0 & 0 & 0 & -1
\end{bmatrix}
\end{align}
as will be demonstrated in the following discussion. For \( z \gg 1 \), all off-diagonal elements approach zero, and the respective 11-components satisfy \( g_{11} \! \left(z \gg 1 \right) = -1 \). Gravitation, formulated within the theory of general relativity, is an integral part of the model. Nevertheless, for particles with masses up to tens of TeV, the associated Schwarzschild radius, $r_s=\frac{2 G m}{c^2}\approx 10^{-50}\,\text{m}$, lies far below the Planck length. Accordingly, gravitational contributions may be ignored in the following treatment. For matter, the non-compact dimensions of the model constitute a manifold that coincides with the 4D spacetime of special relativity, which, for instance, guaranties energy conservation. By contrast, charges established in the compactified dimensions create highly curved manifolds described by the dynamical metric tensors of the interactions. The force cannot be calculated from the potential energies using the Nabla operator, instead a generalized geodesic equation must be applied. The following will be shown; any of the three fundamental interactions - strong, electric, weak - can be expressed as a function of the distance between the interacting particles, their velocity, and the first and second derivatives of that velocity with respect to the interparticle separation. The velocity itself is determined by all three forces. Consequently, for relativistic velocity, the various interactions affect one another.

\subsection{Embedding and container space}
We begin with a 10-dimensional framework comprising a single time dimension in conjunction with nine spatial dimensions. The familiar concept of spacetime includes the time and three of these spatial dimensions, whereas the six additional dimensions are compactified. To facilitate computation, this 10-dimensional framework is embedded within a 20-dimensional container. The correlation between the 10D spacetime coordinates, $x^{i}$, and the container space coordinates, $y^{j}$, is expressed by: 
\begin{align}
\label{eq:compact_radii}
&y^{0}=c t \\
&y^{1}=x^{1} \nonumber\\
&y^{2}=x^{2} \nonumber\\
&y^{3}=x^{3} \nonumber\\
&y^{i}=r^{i} \cos \left(\varphi_{i}\right)=r^{i}(x^{1}) \cos \left(\frac{x^{i}}{r^{i}(x^{1})}\right) \quad \text{for } i=4,5,6,7,8,9 \nonumber\\
&y^{10}=0 \nonumber\\
&y^{11}=0 \nonumber\\
&y^{12}=0 \nonumber\\
&y^{13}=0 \nonumber\\
&y^{i+10}=r^{i} \sin \left(\varphi_{i}\right)=r^{i}(x^{1}) \sin \left(\frac{x^{i}}{r^{i}(x^{1})}\right) \quad \text{for } i=4,5,6,7,8,9 \nonumber
\end{align}
Here, $r^{i}(x^{1})$ represents the radius of the compactified dimension, which, as will be shown later, can depend on the distance $x^{1}$ to the partner particle. The container space is equipped with a simple flat metric of the form:  
\begin{align}
\eta_{a b}=\operatorname{diag}(1,9 \times (-1),1,9 \times (-1))
\end{align}
where the indices $a, b$ range from 0 to 19. The coordinate functions $y^a(x^i)$ form a smooth one-to-one map whose Jacobian has full rank everywhere, so the geometry of the x-space is faithfully represented as a submanifold of the 20-dimensional flat space. In this sense, the y-coordinates provide a smooth embedding of the x-coordinates. Although the minimum dimension of the y-space necessary is 16 ($y^{a}=0$ for $a=10,11,12,13$), twenty dimensions were chosen for symmetry reasons.   
The connection between the 10-dimensional physical space and the 20-dimensional container space is expressed as:  
\begin{align}
\label{eq:metric_tensor}
g_{\mu \nu}=\frac{\partial y^{a}}{\partial x^{\mu}} \frac{\partial y^{b}}{\partial x^{\nu}} \eta_{a b}
\end{align}
For the undeformed/static 10D spacetime, this results in a metric tensor of $g_{\mu \nu}=diag(1, -1, -1, -1, -1, -1, -1, -1, -1, -1)$. This is apparent since with $\eta_{a b}$ having non-vanishing elements on the main diagonal axis only and $y^i = x^i$ for $i=0,1,2,3$ the first four elements of $g_{\mu \nu}$ are identical with the first four elements of $\eta_{a b}$. For the other elements, combinations of $\left(\frac{\partial y^a}{\partial x^{\mu}}\right)^2 \eta_{aa}+ \left(\frac{\partial y^{a+10}}{\partial x^{\mu}}\right)^2 \eta_{(a+10)(a+10)} = -\sin{\left(\frac{x^{\mu}}{r^{\mu}} \right)}^2 - \cos{\left(\frac{x^{\mu}}{r^{\mu}} \right)}^2 = -1$ with $\mu = 4,5,...,9$ follow. If $r^{i}$ acquires a dependence on $x^{1}$, the metric becomes dynamical, leading to a substantially more intricate metric tensor, as symbolized by \eqref{eq:metric_tensor_of_all_interactions}, which governs the interactions in this model. The six compactified dimensions are classified according to their respective roles in the fundamental interactions, and, purely for bookkeeping purposes, are discussed in the following sequence: 
- \textbf{Strong interaction}: dimensions 4 to 6 
- \textbf{Electromagnetic interaction}: dimension 7 
- \textbf{Weak interaction}: dimensions 8 and 9.  

\subsection{Angular momentum of stable excitations in compactified dimensions}  
A particle is viewed as an aggregate of associated excitations that move as waves through one or more of the six compactified cylindrical dimensions. The signal speed around any cylinder consistently equals $c$, the speed of light. Completing one full circuit of a cylinder (encompassing double the typical phase space following that of the parameter interval of the mathematical group SU(2)) is expressed as:  
\begin{align}
\label{eq:double_phase_space}
4 \pi r_{i}=n_{i} \lambda_{i}
\end{align}
where $i=4,...,9$ is the index of the compactified dimension, $n_{i}$ an integer, and $\lambda_{i}$, the wavelength, is related to the frequency $f_{i}$ and the speed of light using the usual formula:  
\begin{align}
c=f_{i} \lambda_{i}
\end{align}
The excitation's momentum is:  
\begin{align}
p_{i}=\hbar k_{i}=\frac{2 \pi \hbar}{\lambda_{i}}=\frac{n_{i} \hbar}{2 r_{i}} 
\end{align}
where $k_{i}=2 \pi / \lambda_{i}$ represents the wave vector. Using these relations, the spin (angular momentum) of any excitation in a compactified dimension is calculated as:  
\begin{align}
\vec{s}_{i} =\vec{r}_{i}  \times \vec{p}_{i}  = n_{i}  \frac{\hbar}{2} \cdot \vec{e_s}
\end{align}
where $\vec{s}_{i} $, $\vec{r}_{i} $, $\vec{p}_{i} $,  are all perpendicular to each other, so that the spin of a particle becomes
\begin{align}
s=\sum_ {i=4} ^{9} r_{i} p_{i} = \frac{\hbar}{2} \sum_ {i=4} ^{9} n_{i}
\end{align}

\subsection{Electric interaction – dimension 7}  

The derivation begins with a point-particle current (as observed in four-dimensional special relativistic spacetime) with four-velocity $w^{\mu}$:  
\begin{align}
J_{e}^{\mu}=\frac{e}{3} \cdot o_{7} \cdot \delta^{3}(\vec{x}-\vec{y}) \cdot w^{\mu}
\end{align}
A charge of $e/3$ explains the quark charge values. The associated covariant Liénard–Wiechert potential~\cite{RN205}, produced by a point particle with four-velocity $v^{\mu}$, is expressed as:
\begin{align}
A_{e}^{\mu}=\frac{\frac{e}{3} \cdot n_{7}}{4 \pi \varepsilon_{0} c^{2}} \cdot \frac{v^{\mu}}{\left|\vec{y}-\vec{x}_{0}\right|}
\end{align}
The potential energy between these two particles is:  
\begin{align}
\label{eq:electric_potential_energy}
&U_{no}^{e}(\left|\vec{x}-\vec{x}_{0}\right|) = \iiint_{-\infty}^{\infty} J_{e}^{\mu} A_{\mu}^{e} d y_{1} d y_{2} d y_{3} \\
&= \frac{e^{2} n_{7} o_{7}}{4 \pi \varepsilon_{0} c^{2} 9} \cdot \frac{w^{\mu} v_{\mu}}{\left|\vec{x}-\vec{x}_{0}\right|}
= \frac{\alpha_{e} c \hbar n_{7} o_{7} f_{no}(z)}{9 \rho \cdot z} \nonumber
\end{align}
where $\left|\vec{x}-\vec{x}_{0}\right|$ is abbreviated as $\rho \cdot z$, with $\rho$ being a positive value constant with units of length and $z$ a dimensionless non-negative parameter. The invariant expression can be written as
\begin{align}
\label{eq:invariant_fno}
f_{no} \!\left(z \right) = \frac{w^{\alpha} v_{\alpha}}{c^2} = \frac{g_{\alpha\beta} w^{\alpha} v^{\beta}}{c^2},
\end{align}
where the velocities are augmented by components in the higher dimensions:
\begin{align}
\label{eq:velocity}
w_{o}^{\mu} = \left[\gamma_{o},\, \gamma_{o} \beta_{o},\, 0,\, 0,\, | \text{signum}(o_{i}) |\right] \cdot c
\end{align}
and
\begin{align}
v_{n}^{\mu} = \left[\gamma_{n},\, \gamma_{n} \beta_{n},\, 0,\, 0,\, | \text{signum}(n_{i}) |\right] \cdot c,
\end{align}
with $i = 4,5,6,7,8,9$. The function $signum$ is defined so that $signum(0) = 0$. The function $f_{no}(z)$ encapsulates $z$-dependent terms, including the metric tensor $g_{\mu \nu}(z)$, the normalized velocity $\beta(z) = v(z)/c$, and the Lorentz factor $\gamma(z) = 1/\sqrt{1 - \beta(z)^2}$. The electric energy of an excitation $o_{7}$ comprises both rest and kinetic energy, using the Planck relation:
\begin{align}
\label{eq:electric_energy}
E_{o}^{e}(z) = \hbar \omega\left(r_{o} \right) \cdot \gamma_{o}(z)   
= \frac{o_{7} c \hbar \cdot\gamma_{o}(z) }{r_{o}(z)} 
\end{align}
Since total energy is conserved, we obtain:
\begin{align}
0 &= \frac{\partial}{\partial z} \left(E_{o}^{e}(z) + U_{no}^{e}(z) \right) \\
&= 
\hslash  c \left(\frac{o_{7} \left(\frac{d}{d z}\gamma_{o} \! \left(z \right)\right)}{r \! \left(z \right)}-\frac{o_{7} \gamma_{o} \! \left(z \right) \left(\frac{d}{d z}r \! \left(z \right)\right)}{r \! \left(z \right)^{2}}\right)+\frac{\alpha_{e} c \hslash  n_{7} o_{7} \left(\frac{d}{d z}f_{\mathit{no}} \! \left(z \right)\right)}{9 \rho  z}-\frac{\alpha_{e} c \hslash  n_{7} o_{7} f_{\mathit{no}} \! \left(z \right)}{9 \rho  \,z^{2}}
\nonumber
\end{align}
Solving this differential equation allows us to determine the cylinder radius of dimension 7, where the integers $o_{7}$ and $n_{7}$ represent excitation numbers:
\begin{align}
\label{eq:electric_r_with_c}
r \! \left(z \right) = 
-\frac{9 \gamma_{o} \! \left(z \right) \rho  z}{f_{\mathit{no}} \! \left(z \right) n_{7} \alpha_{e}-9 C \rho  z}
\end{align}
In the limit as $z \to \infty$, the undisturbed radius, see Eq. \eqref{eq:undisturbed_es_radii}, allows us to calculate the constant $C$
\begin{align}
\lim_{z \rightarrow \infty} r_{o7}(z) = \frac{\gamma_{o} \! \left(\infty \right)}{C}= \frac{9 \rho}{\alpha_{e} \operatorname{signum}(o_{7})}
\end{align}
with the result:
\begin{align}
C = \frac{\gamma_{o} \! \left(\infty \right) \alpha_{e} \mathrm{signum}\! \left(o_{7} \right)}{9 \rho}
\end{align}
where $\rho$ is identified as the compactification radius, which will be determined later. Substituting this into Eq. \eqref{eq:electric_r_with_c}, we obtain:
\begin{align}
\label{eq:electric_radius}
r_{o7}(z) = \frac{9 \gamma_{o} \! \left(z \right) \rho  \,\mathrm{signum}\! \left(o_{7} \right)}{\alpha_{e} \left(\gamma_{o} \! \left(\infty \right)-\frac{f_{\mathit{no}} \left(z \right) n_{7} \mathrm{signum}\left(o_{7} \right)}{z}\right)}
\end{align}
By substituting Eq. \eqref{eq:electric_radius} into Eq. \eqref{eq:electric_energy}, the electric energy takes the form
\begin{align}
\label{eq:electric_mass_no-field}
E_{o}^{e}(z) = \frac{\alpha_{e} c \hslash \left({| o_{7} |} \gamma_{o} \! \left(\infty \right)-\frac{o_{7} f_{\mathit{no}} \left(z \right) n_{7}}{z} \right) }{9 \rho} \, .
\end{align}
This term represents the energy of an excitation with excitation number $o_{7}$, corresponding to a sinusoidal mode of wavelength $\lambda = \frac{4 \pi r_{7}}{\left| o_{7} \right|}$, along with the associated potential energy. There is, however, an additional mechanism for extracting energy from the electric dimension. When both right- and left-moving waves are present with equal wavelengths $\lambda_A = \frac{4 \pi r_{7}}{\left| o_{R} \right|} = \frac{4 \pi r_{7}}{\left| o_{L} \right|}$ and excitation numbers related by $o_{R} = -o_{L}$, they generate an extra energy consisting of two terms, each analogous to \eqref{eq:electric_mass_no-field}. Together, these terms yield the electric energy
\begin{align}
\label{eq:electric_mass_flavour_part}
E_{oA}^{e} = \frac{\alpha_{e} c \hslash \left({| o_{R}| + | o_{L}|} \right)\gamma_{o} \! \left(\infty \right) }{9 \rho} = \frac{\alpha_{e} c \hslash A\gamma_{o} \! \left(\infty \right) }{9 \rho} \, ,
\end{align}
while their potentials cancel each other. The quantity $A$ is interpreted as an excitation corresponding to a standing wave that carries equal numbers of positive and negative excitations. As a result, it generates neither an additional force nor an additional spin. Therefore, $A$ must be zero or a positive even integer. Since the mass is obtained by adding the potential energy \eqref{eq:electric_potential_energy} and dividing the total energy by $c^2$, the electric mass becomes
\begin{align}
\label{eq:electric_mass}
m_{o7} = \frac{E_{o}^{e}(z) + E_{oA}^{e} + U_{no}^{e}(z)}{c^2} = \frac{\alpha_{e} \hslash  \left({| o_{7} |} +A \right)\gamma_{o} \! \left(\infty \right)}{9 c \rho} \, ,
\end{align}
which is independent of $z$.

\subsection{Strong interaction – dimensions 4, 5, and 6}  

At this point, the color representation is introduced as  
\begin{align}
c_{\text{red}}=\left(\begin{array}{c} 1 \\ -1 \\ 0 \end{array}\right), \quad
c_{\text{green}}=\left(\begin{array}{c} 0 \\ 1 \\ -1 \end{array}\right), \quad
c_{\text{blue}}=\left(\begin{array}{c} -1 \\ 0 \\ 1 \end{array}\right) ,
\end{align}
which contrasts with the Standard Model description, where complex Gell-Mann matrices are employed to encode the three-quark content of baryons and the quark–antiquark composition of mesons. In the present framework, an analogous internal structure is obtained using these real three-component vectors, which couple the three compactified spatial directions labeled four through six. Anti-colors are represented by multiplying each component of the corresponding color vector by $-1$.
The calculation of the cylinder radii follows the same approach as in the electric case. Again, starting with a point-particle current of four-velocity $w^{\mu}$, the current in each relevant dimension is:  
\begin{align}
J_{s}^{\mu}=e_{s} \cdot \delta^{3}(\vec{x}-\vec{y}) \cdot w^{\mu} n_{i}, \quad i=4,5,6
\end{align}
The vector potential, assuming a simplified ansatz~\cite{RN206,RN207,RN208,RN209}, depends linearly on the distance between the two color charges:  
\begin{align}
A_{s}^{\mu}=-\frac{e_{s}}{4 \pi \varepsilon_{0} c^{2}} \cdot \frac{\left|\vec{y}-\vec{x}_{0}\right|}{\rho^{2}} v^{\mu} o_{i}=-\frac{e_{s}}{4 \pi \varepsilon_{0} c^{2}} \cdot \frac{v^{\mu} o_{i} z}{\rho} 
\end{align}
The potential energy between two particles is then:  
\begin{align}
U_{no}^{s}(z) &= \iiint_{-\infty}^{\infty} J_{s}^{\mu} A_{\mu}^{s} d y_{1} d y_{2} d y_{3} \\
&= -\frac{e_{s}^{2} n o}{4 \pi \varepsilon_{0} c^{2}} \cdot w^{\mu} v_{\mu} \cdot n_{i} o_{i} \cdot \frac{z}{\rho} \nonumber \\
&= -\frac{\alpha_{s} c \hbar n_{i} o_{i} f_{no}(z) \cdot z }{\rho} \nonumber
\end{align}
The rest and kinetic color energy of a cylindrical excitation have the same form as in the electric case. To account for the extra energy produced by standing waves, we included this contribution already at this stage of the derivation.
\begin{align}
\label{eq:strong_rest_kinetic_energy}
E_{o}^{s}(z)=\left(\frac{o_{i} c \hbar\cdot \gamma_{o}(z)}{r_{o}(z)}+\frac{\alpha_{s} c \hbar A \gamma_{o} \! \left(0\right) }{\rho}\right) 
\end{align}
As before, energy conservation gives:
\begin{align}
0 &= \frac{\partial}{\partial z} \left(E_{o}^{s}(z)+U_{no}^{s}(z)\right) \\
&= \frac{o_{i} \hslash  c \left(\frac{d}{d z}\gamma_{o} \! \left(z \right)\right)}{r \! \left(z \right)}-\frac{o_{i} \hslash  c \gamma_{o} \! \left(z \right) \left(\frac{d}{d z}r \! \left(z \right)\right)}{r \! \left(z \right)^{2}}-\frac{\alpha_{s} \hslash  c n_{i} o_{i} f_{\mathit{no}} \! \left(z \right)}{\rho}-\frac{\alpha_{s} \hslash  c n_{i} o_{i} z \left(\frac{d}{d z}f_{\mathit{no}} \! \left(z \right)\right)}{\rho}
 \nonumber
\end{align}
Solving for the radii of the color cylinders ($i=4,5,6$), we obtain:
\begin{align}
r_{o_{i}}(z) = -\frac{\rho  o_{i} \gamma_{o} \! \left(z \right)}{-f_{\mathit{no}} \! \left(z \right) n_{i} z \alpha_{s} o_{i} -C \rho  o_{i}}
\end{align}
The constant is determined by employing the undisturbed radius \eqref{eq:undisturbed_es_radii}:
\begin{align}
\lim _{z \rightarrow 0} r_{o_{i}}(z)=\frac{\gamma_{o} \! \left(0 \right)}C = \frac{\rho}{\alpha_{s} \operatorname{signum}(o_{i})}
\end{align}
which leads to:
\begin{align}
C = \frac{\gamma_{o} \! \left(0\right) \alpha_{s} \mathrm{signum}\! \left(o_{i} \right)}{\rho}
\end{align}
Thus, the cylinder radius is:
\begin{align}
\label{eq:strong_radius}
r_{o_{i}}(z) = \frac{\gamma_{o} \! \left(z \right) \rho  \,\mathrm{signum}\! \left(o_{i} \right)}{\alpha_{s} \left(\gamma_{o} \left(0\right) +f_{\mathit{no}} \! \left(z \right) n_{i} z \,\mathrm{signum}\! \left(o_{i} \right)\!\right)}
\end{align}
Without including potential energy, the color energy simplifies to:
\begin{align}
\label{eq:Strong_mass_no-field}
E_{o}^{s}(z) = \frac{\alpha_{s}  \hslash  c \left(\gamma_{o} \! \left(0\right) \left( {| o_{i} |}+A \right)+f_{\mathit{no}} \! \left(z \right) n_{i} z o_{i} \right)}{\rho}
\end{align}
Finally, the color mass is:
\begin{align}
m_{o_{i}} = \frac{E_{o}^{s}(z) + U_{no}^{s}(z)}{c^{2}} = \frac{\alpha_{s} \hslash   \left({| o_{i} |}+A \right) \gamma_{o} \! \left(0\right) }{c \rho}
\end{align}

\subsection{Weak interaction – dimensions 8 and 9}  

The point-particle currents associated with the weak interaction are given by:
\begin{align}
J_{w}^{\mu} = e_{w} \cdot \delta^{3}(\vec{x}-\vec{y}) \cdot w^{\mu} n_{i}, \quad i=8,9
\end{align}
Within the Standard Model, the weak interaction is mediated by exchange of the gauge bosons \(W^+\), \(W^-\), and \(Z^0\), and is commonly described by a Yukawa potential that explicitly depends on the masses of these bosons. In the present framework, however, no exchange of virtual bosons is invoked. Rather, spacetime curvature is responsible for generating the interaction potential, which is deduced by dimensional analysis. For the electromagnetic interaction, the vector potential scales as \(1/z\), whereas for the strong interaction it increases linearly with \(z\). The electromagnetic force is treated as effectively one-dimensional, while the strong force extends over three spatial dimensions. By analogy, the weak interaction is assigned two effective dimensions, which naturally leads to a logarithmic dependence of the corresponding potential as the appropriate functional form for this dimensionality:
\begin{align}
A_{w}^{\mu} = \frac{e_{w}}{4 \pi \varepsilon_{0} c^{2}} \cdot \frac{v^{\mu} o_{i} n_{i} \cdot \ln (z)}{\rho} 
\end{align}
The associated potential energy, following the same derivation as in previous cases, is given by:
\begin{align}
\label{eq:weak_potential_energy}
U_{no}^{w}(z) &= \iiint_{-\infty}^{\infty} J_{w}^{\mu} A_{\mu}^{w} d y_{1} d y_{2} d y_{3} \\
&= \frac{e_{w}^{2}}{4 \pi \varepsilon_{0} c^{2}} \cdot \frac{w^{\mu} v_{\mu}}{\rho} o_{i} n_{i} \cdot \ln (z) \nonumber\\
&= \frac{\alpha_{w} c \hbar o_{i}  n_{i}  f_{no}(z)\cdot \ln (z)}{\rho} \nonumber
\end{align}
Following the same considerations as in previous cases, the rest and kinetic energy can be expressed as:
\begin{align}
\label{eq:weak_kinetic_energy}
E_{o}^{w}(z) = \left(\frac{o_{i} c \hbar \cdot \gamma_{o}(z)}{r_{o_{i}}(z)} + \frac{\alpha_{w} c \hbar A  \gamma_{o} \! \left(1 \right)}{\rho} \right) 
\end{align}
As before, total energy conservation requires:
\begin{align}
0 &= \frac{\partial}{\partial z} (E_{o}^{s}(z) + U_{no}^{s}(z)) \\
&= \frac{o_{i} \hslash  c \left(\frac{d}{d z}\gamma_{o} \! \left(z \right)\right)}{r \! \left(z \right)}-\frac{o_{i} \hslash  c \gamma_{o} \! \left(z \right) \left(\frac{d}{d z}r \! \left(z \right)\right)}{r \! \left(z \right)^{2}}+\frac{\alpha_{w} \hslash  c o_{i} \left(\frac{d}{d z}f_{\mathit{no}} \! \left(z \right)\right) n_{i} \ln \! \left(z \right)}{\rho}+\frac{\alpha_{w} \hslash  c o_{i} f_{\mathit{no}} \! \left(z \right) n_{i}}{\rho  z}
 \nonumber
\end{align}
From this, the radius of the compactified weak interaction dimensions ($i=8,9$) is found to be:
\begin{align}
r_{o_{i}}(z) = -\frac{\gamma_{o} \! \left(z \right) \rho}{\ln \! \left(z \right) f_{\mathit{no}} \! \left(z \right) n_{i} \alpha_{w}-C \rho}
\end{align}
\textbf{Neutrino screening effect}: An important point that has not yet been addressed is the screening effect that arises from nearby neutrinos and antineutrinos. Particles that interact through the weak force are affected by the ambient distributions of neutrinos and antineutrinos. Consequently, beyond a possible direct interaction with a specific partner particle, there is an additional contribution originating from the surrounding neutrino and antineutrino background. In accordance with Eq.~\eqref{eq:weak_potential_energy}, this requires the inclusion of an additional term of the form
\begin{align}
\label{eq:delta_definition}
U_{no}^{w}(z_\nu,z_{\bar{\nu}}) &=  \frac{\alpha_{w}  c \hbar \, o_{i} \, f_{no}(z)}{\rho} \left( n_{\nu}\ln z_{\nu}  + n_{\bar{\nu}}\ln z_{\bar{\nu}} \right) \\
&= -\frac{\alpha_{w}  c \hbar \, f_{no}(z)}{\rho} \, o_{i} \,\ln(\delta)\, \nonumber
\end{align}
, where $n_{\nu}=-1$ and $n_{\bar{\nu}}=1$ (see Table \ref{tab:fermions_first_generation}). $\delta=\frac{z_{\nu}}{z_{\bar{\nu}}}$ denotes the ratio involving particle separation to the mean neutrino distance $z_{\nu}$ and the mean antineutrino distance $z_{\bar{\nu}}$. This is equivalent to the cubic root of the ratio of antineutrino density $\rho_{\bar{\nu}}$ to neutrino density $\rho_{\nu}$. In a homogeneous density, this relationship holds for any given volume $V$, where:
\begin{align}
V = n_{\nu} z_{\nu}^{3} \rho^{3} = n_{\bar{\nu}} z_{\bar{\nu}}^{3} \rho^{3}
\end{align}
which gives:
\begin{align}
\frac{\rho_{\bar{\nu}}}{\rho_{\nu}} = \frac{n_{\bar{\nu}}}{V} \cdot \frac{V}{n_{\nu}} = \frac{n_{\bar{\nu}} n_{\nu} z_{\nu}^{3} \rho^{3}}{n_{\nu} n_{\bar{\nu}} z_{\bar{\nu}}^{3} \rho^{3}} = \frac{z_{\nu}^{3}}{z_{\bar{\nu}}^{3}} = \delta^{3}
\end{align}
This establishes the neutrino-to-antineutrino distance ratio:
\begin{align}
\label{eq:delta_neutrino_density}
\delta = \frac{z_{\nu}}{z_{\bar{\nu}}} = \left(\frac{\rho_{\bar{\nu}}}{\rho_{\nu}}\right)^{\frac{1}{3}}
\end{align}
The presence of $\delta$ modifies the undisturbed cylinder radii of the weak interaction.
Consequently, fixing the free constant requires the undisturbed radius \eqref{eq:undisturbed_weak_radius}:
\begin{align}
\lim _{z \rightarrow 1} r_{o_{i}}(z) = \frac{\gamma_{o} \! \left(1 \right)}{C} = \frac{\rho}{\alpha_{w} \left(\operatorname{signum}(o_{i}) + \ln (\delta) \cdot f_{no}(1)\right)}
\end{align}
which leads to the final expression:
\begin{align}
\label{eq:weak_radius}
r_{o_{i}}(z) &= \frac{\gamma_{o} \! \left(z \right) \rho  \,\mathrm{signum}\! \left(o_{i} \right)}{\alpha_{w} \left(\left(1+f_{\mathit{no}} \! \left(1\right) \ln \! \left(\delta \right) \mathrm{signum}\! \left(o_{i} \right)\right) \gamma_{o} \! \left(1\right)-\ln \! \left(z \right) f_{\mathit{no}} \! \left(z \right) n_{i} \mathrm{signum}\! \left(o_{i} \right)\right)}
\end{align}
The general neutrino-antineutrino distribution, characterized by $\delta$, contributes only a very small correction to the energy of a particle:
\begin{align}
\label{eq:weak_mass_no-field}
E_{o}^w(z) =  \frac{\alpha_{w} \hslash  c \left(\left(f_{\mathit{no}} \! \left(1\right) \ln \! \left(\delta \right) o_{i} +{| o_{i} |}+A\right) \gamma_{o} \! \left(1\right)-o_{i} \ln \! \left(z \right) n_{i} f_{\mathit{no}} \! \left(z \right) \right) }{\rho}
\end{align}
Finally, the weak interaction mass is:
\begin{align}
\label{eq:weak_particle_mass}
m_{o_{i}} = \frac{E_{o}^w(z) + U_{no}^{w}(z)}{c^{2}} = \frac{\hslash  \alpha_{w} \left(\left(f_{\mathit{no}} \! \left(1\right) \ln \! \left(\delta \right) o_{i} +{| o_{i} |+A}\right) \gamma_{o} \! \left(1\right) \right)}{c \rho}
\end{align}

\subsection{Masses solely dependent on the weak sector}  
The situation differs for particles affected exclusively by the weak interaction. In this case, only nearby neutrinos and antineutrinos influence the radii of the weak interaction cylinders. As a result, \eqref{eq:weak_potential_energy} simplifies to:
\begin{align}
U_{o}^{w}(z) = 0
\end{align}
In the rest frame, the structure of the excitation energy mirrors that found in the general weak interaction scenario \eqref{eq:weak_kinetic_energy}. However, $r_{o_{i}}(z)$ is given by Eq. \eqref{eq:weak_inf_radius}, which results in the energy of a purely weakly interacting particle as
\begin{align}
E_{o}^{w} = \frac{o_{i} c \hbar}{r_{o_{i}}(z)} + \frac{\alpha_{w} c \hbar A \ ln(\delta) }{\rho} = \frac{\alpha_{w} o_{i} \hslash  c \ln \! \left(\delta \right) f_{\mathit{no}} \! \left(z \right) }{\rho} + \frac{\alpha_{w} c \hbar A \ ln(\delta) }{\rho}
\end{align}
Here, the additional term $\ln(\delta)$ in addition to $A$ accommodates the varied cylinder radii characteristic of exclusive weakly interacting particles. Rewritten, the excitation energy is
\begin{align}
E_{o}^w = \frac{\alpha_{w} \hslash  \ln \! \left(\delta \right) c \left(f_{\mathit{no}} \! \left(z \right) o_{i} +A \right)}{\rho}
\end{align}
Finally, the corresponding mass for $i=8,9$ is:
\begin{align}
\label{eq:weak_mass}
m_{o_{i}} = \frac{E_{o}^w + U_{o}^{w}(z)}{c^{2}} = \frac{\alpha_{w} \hslash  \ln \! \left(\delta \right) \left(f_{\mathit{no}} \! \left(z \right) o_{i} +A \right)}{c \rho}
\end{align}

\subsection{Lorentz factors at the interaction fix points}  
The Lorentz factors for all three interactions at their respective fix points are considered individually. These fix points represent the initial conditions where a particle begins to be affected by the respective forces when its velocity is zero, thus:
   \begin{align}
   \gamma_{s}(0) = 1 && \gamma_{e}(\infty) = 1 && \gamma_{w}(1) = 1
   \end{align}

\subsection{The undisturbed cylinder radii}  

The undisturbed radius of a cylinder is its natural radius, unaffected by any potential energy dependent on $z$. The energy of a particle is supposed to be a positive constant and consists, without contributions from "magnetism", of the excitation energy and the potential energy. An expression with the term $F(z)$ symbolizing the z-dependent part of the potential energy can be written as 
\begin{align}
\label{eq:general_particle_energy_without_magnetism}
\frac{\alpha  {| o |} \hslash  c \cdot \gamma_{o}(z)}{\rho} = 
\frac{o \hslash  c \cdot \gamma_{o}(z)}{r(z)}+\frac{\alpha  c \hslash  n o f_{\mathit{no}} \! \left(z \right) F \! \left(z \right)}{\rho}
\end{align}
which solving for $r(z)$ results in
\begin{align}
r \! \left(z \right) = 
\frac{\rho \cdot \gamma_{o}(z)}{\alpha  \left(\mathrm{signum}\! \left(o \right) \cdot \gamma_{o}(z) - n F \! \left(z \right) f_{\mathit{no}} \! \left(z \right)\right)}
\end{align}
The undisturbed radius is received for $F(z) = 0$
\begin{align}
r_{\mathit{undisturbed}} = 
\frac{\rho}{\alpha  \,\mathrm{signum}\! \left(o \right)}
\end{align}
The expressions for electric - $F(z)=\frac{1}{z}$ - and strong - $F(z)=z$ - interactions are as follows.
\begin{align}
\label{eq:undisturbed_es_radii}
r \! \left(\infty \right) = 
\frac{9 \rho}{\alpha_{e} \mathrm{signum}\! \left(o_{7} \right)} &&
r \! \left(0\right) = 
\frac{\rho}{\alpha_{s} \mathrm{signum}\! \left(o_{i} \right)} 
\end{align}
Eq. \eqref{eq:general_particle_energy_without_magnetism} for the weak interaction - $F(z)=ln(z)$ - gets an additional term caused by the neutrino screening effect.
\begin{align}
\frac{\alpha_{w} {| o_{i} |} \hslash  c \cdot \gamma_{o}(z)}{\rho} 
=
\frac{o_{i} \hslash  c \cdot \gamma_{o}(z)}{r(z)}+\frac{\alpha_{w} \hslash  c o_{i} n_{i} \ln \! \left(z \right) f_{\mathit{no}} \! \left(z \right)}{\rho}
-\frac{\alpha_{w} \hslash  c o_{i} \ln \! \left(\delta \right) f_{\mathit{no}} \! \left(z \right)}{\rho}
\end{align}
Here, the radius is
\begin{align}
r \! \left(z \right) = 
\frac{\rho \cdot \gamma_{o}(z)}{\alpha_{w} \left( \mathrm{signum}\! \left(o_{i} \right)  \cdot \gamma_{o}(z)- n_{i} f_{\mathit{no}} \! \left(z \right) \ln \! \left(z \right) + f_{\mathit{no}} \! \left(z \right) \ln \! \left(\delta \right)\right)}
\end{align}
and the undisturbed radius;
\begin{align}
\label{eq:undisturbed_weak_radius}
r \! \left(1\right) = 
\frac{\rho}{\alpha_{w} \left(\mathrm{signum}\! \left(o_{i} \right)+f_{\mathit{no}} \! \left(1\right) \ln \! \left(\delta \right)\right)}
\end{align}
Again, a different situation occurs when an exclusively weak-interacting particle interacts with the surrounding neutrinos and antineutrinos only. The energy equation here is
\begin{align}
\label{eq:undisturbed_inf_radius}
0 = \frac{o_{i} \hslash  c}{r(z)}-\frac{\alpha_{w} \hslash  c o_{i} \ln \! \left(\delta \right) f_{\mathit{no}} \! \left(z \right)}{\rho}
\end{align}
, which yields a computed radius of
\begin{align}
\label{eq:weak_inf_radius}
r(z)=\frac{\rho}{\alpha_{w} f_{\mathit{no}} \! \left(z \right) \ln \! \left(\delta \right)}
\end{align}

\subsection{Particle mass from strong, electric, and weak interactions}    
For quarks and charged leptons, the total particle mass $m_{o}$ is given by
\begin{align}
\label{eq:quark_masses_no_mag}
m_{o} &= \frac{\hslash}{c \rho} \cdot \left[ \alpha_{s} \left({| o_{4} |}+{| o_{5} |}+{| o_{6} |}\right)+\frac{\alpha_{e} {| o_{7} |}}{9} \right.\\ &\left. +\alpha_{w} \sum_{i=8}^{9} \left(f_{\mathit{no}} \! \left(1\right) \ln \! \left(\delta \right) o_{i}+{| o_{i} |}\right)+ \left(\alpha_{s}+\frac{\alpha_{e}}{9}+\alpha_{w}\right) \cdot A
\right]
\end{align}
, while the masses of purely weak interacting elementary particles are 
\begin{align}
\label{eq:weak_masses_no_mag}
m_{o} &= \frac{\alpha_{w} \hbar \cdot ln(\delta)}{c \rho} \left[  \sum_{i=8}^{9}  \   o_{i}  \ f_{no}(z) +A \right]
\end{align}
The constants $A$ are referred to as flavor constants and, as stated above, must be zero or even positive integers.

\subsection{Particle identification}  
\subsubsection{Excitation numbers of fermions} 
The fermion spectrum is dictated by the topology of the model, together with constraints imposed by experimental data. In this framework, charges of the fermions are fixed uniquely and without ambiguity:  
\begin{itemize}
    \item  A quark carries one of the three colors: red, green, or blue, along with its respective excitation numbers.  
    \item  The electric excitation number is:
    \begin{itemize}
      \item $-1$ for a down quark  
      \item $+2$ for an up quark  
      \item $-3$ for an electron  
      \end{itemize}
    \item Each excitation number contributes a spin of $\hbar / 2$ times the excitation number. To ensure an overall spin of $\hbar / 2$, the weak excitations are adjusted accordingly.  
    \item A neutrino possesses spin $\hbar / 2$, is composed purely of weak-sector excitations, and is represented as a superposition of the two weak compactified dimensions. Consequently, it must carry a weak charge of either $+1$ or $-1$. Considering the process
\begin{align}
d \rightarrow u + e^- + \bar{\nu}_e \nonumber
\end{align}
arising in neutron $\beta$ decay, together with the corresponding transformation of excitation numbers,
\begin{align}
\begin{array}{c}
\left(\begin{array}{c}
1 \\
-1 \\
0 \\
-1 \\
(+2) \\
(0)
\end{array}\right) 
\rightarrow
\left(\begin{array}{c}
1 \\
-1 \\
0 \\
+2 \\
(-1) \\
(0)
\end{array}\right)
+
\left(\begin{array}{c}
0 \\
0 \\
0 \\
-3 \\
(+2) \\
(0)
\end{array}\right) +
\left(\begin{array}{c}
0 \\
0 \\
0 \\
0 \\
(+1) \\
(0)
\end{array}\right) \nonumber
\end{array}
\end{align}
one obtains a uniquely determined outcome. The neutrino's excitation number $(-1)$, is the same as the weak excitation contribution of a u-quark (see Table \ref{tab:fermions_first_generation}). 
\begin{table}[H]
    \caption{Excitation numbers of first generation fermions}
    \centering
    \begin{tabular}{|l|c|c|c|c|c|c|c|c|c|}
        \hline
         Particle & $e^-$ & $\nu_{e}$ & $u$ red & $u$ green & $u$ blue & $d$ red & $d$ green & $d$ blue\\
         \hline
         strong 4   & 0 & 0 & 1 & 0 & -1 & 1 & 0 & -1\\
         \hline
         strong 5   & 0 & 0 & -1 & 1 & 0 & -1 & 1 & 0\\
         \hline
         strong 6   & 0 & 0 & 0 & -1 & 1 & 0 & -1 & 1\\
         \hline
         electric 7 & -3 & 0 & +2 & +2 & +2 & -1 & -1 & -1\\
         \hline
         weak 8 & (2) & (-1) & (-1) & (-1) & (-1) & (+2) & (+2) & (+2)\\
         \hline
         weak 9 & (0) & (0) & (0) & (0) & (0) & (0) & (0) & (0)\\
         \hline
    \end{tabular}
    \label{tab:fermions_first_generation}
\end{table}
But, unlike quarks, neutrinos lack an associated partner particle. Therefore, neutrino mass determination is performed using Equation \eqref{eq:weak_mass}, while other particles with weak mass contribution adhere to Equation \eqref{eq:weak_particle_mass}.      
\end{itemize}
The eighth and ninth dimensions, responsible for the weak force, consistently remain in a super-positional state. As a result, the excitation numbers are represented in brackets, indicating the interchangeability of the excitation numbers associated with dimensions 8 and 9.

\subsubsection{Excitation numbers of photons and gluons}  

The excitation numbers can be inferred from known decay processes. For example, in electron-positron annihilation \cite{RN214}, two photons are produced:  
\begin{align}
e^{+} + e^{-} \rightarrow 2 \gamma \nonumber
\end{align}
This corresponds to the following excitation number transformation:
\begin{align}
\begin{array}{c}
\left(\begin{array}{c}
0 \\
0 \\
0 \\
+3 \\
(-2) \\
(0)
\end{array}\right) +
\left(\begin{array}{c}
0 \\
0 \\
0 \\
-3 \\
(+2) \\
(0)
\end{array}\right)
\rightarrow
\left(\begin{array}{c}
0 \\
0 \\
0 \\
0 \\
(-2) \\
(0)
\end{array}\right) +
\left(\begin{array}{c}
0 \\
0 \\
0 \\
0 \\
(2) \\
(0)
\end{array}\right) \nonumber
\end{array}
\end{align}
Two real, measurable photons are created. Photons neither possess strong nor electric charge and they are spin one bosons. Therefore, the weak excitation number must be plus or minus $2$. As follows, using Eq. \eqref{eq:weak_masses_no_mag}, the photon mass is:
\begin{align}
\label{eq:Photon_mass}
m_{\gamma} = 
\frac{ \alpha_{w} \hslash  }{c \rho} \cdot ln(\delta) \cdot ( A  \pm 2 f_{no}(1)  )
\end{align}
As will be demonstrated later, it has zero rest mass. The decay of the $J/\psi$ meson offers a powerful probe of the internal dynamics of the gluon. Within the Standard Model, one of the primary decay channels of $J/\psi$ involves the emission of three gluons \cite{BESIII2022Multiplicity}. Although individual gluons cannot be directly observed as they are interacting strongly, the picture of three-gluon decay is supported by experimental data: measured branching ratios, hadronic final-state multiplicities, and comparisons of partial decay widths \cite{ParticleDataGroup:2024cfk} all show striking consistency with QCD-based theoretical calculations, which themselves are compatible with the structure of this model as well.
\begin{align}
J/\psi \left(  c \bar{c} \right) \rightarrow 3 g \nonumber
\end{align}
emphasizes the internal structure of the gluon and illustrates how it corresponds to the excitation-number mapping associated with one among several possible color–anticolor configurations.
\begin{align}
\uparrow\begin{array}{c}
\left(\begin{array}{c}
1 \\
-1 \\
0 \\
+2 \\
(-1) \\
(0)
\end{array}\right) 
\uparrow\left(\begin{array}{c}
-1 \\
1 \\
0 \\
-2 \\
(+1) \\
(0)
\end{array}\right) 
\rightarrow
\left(\begin{array}{c}
1 \\
-2 \\
1 \\
0 \\
(+2) \\
(0)
\end{array}\right) \big\uparrow +
\left(\begin{array}{c}
1 \\
1 \\
-2 \\
0 \\
(-2) \\
(0)
\end{array}\right) \big\downarrow +
\left(\begin{array}{c}
-2\\
1 \\
1 \\
0 \\
(+2) \\
(0)
\end{array}\right) \nonumber
\end{array} \big\uparrow
\end{align}
Another experimentally confirmed decay mode is
\begin{align}
J/\psi \left(  c \bar{c} \right) \rightarrow \gamma + 2 g \nonumber
\end{align}
with the excitation number transform
\begin{align}
\uparrow\begin{array}{c}
\left(\begin{array}{c}
1 \\
-1 \\
0 \\
+2 \\
(-1) \\
(0)
\end{array}\right) 
\uparrow\left(\begin{array}{c}
-1 \\
1 \\
0 \\
-2 \\
(+1) \\
(0)
\end{array}\right) 
\rightarrow
\left(\begin{array}{c}
0 \\
0 \\
0 \\
0 \\
(+2) \\
(0)
\end{array}\right) \big\uparrow +
\left(\begin{array}{c}
1 \\
1 \\
-2\\
0 \\
(-2) \\
(0)
\end{array}\right) \big\downarrow +
\left(\begin{array}{c}
-1 \\
-1 \\
2 \\
0 \\
(+2) \\
(0)
\end{array}\right) \nonumber
\end{array} \big\uparrow
\end{align}
The arrows directed upward and downward represent a specific configuration of spin orientations (one among multiple allowed arrangements), illustrating that spin is conserved. Specific aspects related to spin statistics and chirality are not addressed in this work and will need to be investigated in subsequent studies. Consistent with the Standard Model (SM), both the strong and electric charges remain conserved, whereas the weak charge does not. Under these assumptions, and employing the equation \eqref{eq:quark_masses_no_mag}, the gluon mass can be expressed as
\begin{align}
\label{eq:gluon_mass}
m_{\text{gluon}} = \frac{\hslash}{c \rho} \cdot \left( 4 \alpha_{s}  +  2 \alpha_{w} \right)
\end{align}
where the screening effect due to neutrinos has been neglected. The gluon thus emerges as a massive spin-1 boson, consistent with the spin of the Standard Model gluon but different in mass. In the Standard Model there are 8 gluons. In contrast, this framework yields only 6 gluons that mediate color charges, because it represents them with a real-valued vector rather than with the full set of complex Gell‑Mann matrices. Configurations in which a gluon consists of red–antired, blue–antiblue, or green–antigreen are ruled out, since these combinations would not give rise to color force. Table \ref{tab:intermediate_bosons} 
\begin{table}[H]
    \caption{Excitation numbers of the photon and the color changing gluons}
    \centering
    \begin{tabular}{|l|c|c|c|c|c|c|c|c|}
        \hline
         Particle & $\gamma$ & $g_{r\overline{g}}$ & $g_{r\overline{b}}$ & $g_{g\overline{b}}$ & $g_{g\overline{r}}$ & $g_{b\overline{g}}$ & $g_{b\overline{r}}$\\
         \hline
         strong 4 & 0 & 1 & 2 & 1 & -1 & -1 & -2\\
         \hline
         strong 5 & 0 & -2 & -1 & 1 & 2 & -1 & 1\\
         \hline
         strong 6 & 0 & 1 & -1 & -2 & -1 & 2 & 1\\
         \hline
         electric 7 & 0 & 0 & 0 & 0 & 0 & 0 & 0\\
         \hline
         weak 8 & $(\pm 2)$ & $(\pm 2)$ & $(\pm 2)$ & $(\pm 2)$ & $(\pm 2)$ & $(\pm 2)$ & $(\pm 2)$\\
         \hline
         weak 9 & (0) & (0) & (0) & (0) & (0) & (0) & (0)\\
         \hline
    \end{tabular}
    \label{tab:intermediate_bosons}
\end{table}
summarizes the photon and color-changing gluons, which are distinct from the eight gluons appearing in the Standard Model.

\subsubsection{Higher-generation particles and antiparticles}  

For higher-generation particles, the excitation settings remain identical to those of the first generation. Antiparticles are obtained by multiplying all excitation numbers by $-1$. 

\section{Results}
\subsection{The compactification radius \texorpdfstring{\(\rho\)}{rho}}
\label{subsec:Compactification_raduis}
The time required for a particle excitation in the compactified electric dimension, traveling at speed $c$, to traverse the entire circumference of the cylinder once is:
\begin{align}
\Delta T = \frac{4 \pi |r|}{c}
\end{align}
From this, the electric current is calculated as:
\begin{align}
I = \frac{n_e e}{3 \Delta T} = \frac{n_e e c}{12 \pi |r|}
\end{align}
The classical magnetic dipole moment, defined with winding number $n=2$ and current loop area $A$, is:
\begin{align}
\mu = n I A = 2 \cdot \frac{n_e e c}{12 \pi |r|} \cdot \pi |r|^2 = \frac{n_e e c |r|}{6}
\end{align}
For a single electron, with $n_e=-3$, $|r| = \frac{9 \rho}{\alpha_{e}}$, the Bohr magneton $\mu_{B}=\frac{n_e e \hbar}{6 m_{e}}$, and the z-component of spin $S_{e} = n_e \hbar / 2$ (corresponding to the spin of the electric dimension), the magnetic dipole moment is:
\begin{align}
\mu_{S} = -\frac{g_{e} \mu_{B} S_{e}}{\hbar} = -\frac{n_e^2 g_{e} e \hbar}{12 m_{e}}
\end{align}
By examining these two equations and employing the electron's measured g-factor of $g_{e} = 2.00231930436182(52)$ in conjunction with $\hbar$, the reduced Planck constant, $c$, the velocity of light, $\alpha_{e}$, the electromagnetic coupling constant, and $m_{e}$, the mass of the electron (refer to Table \ref{tab:Measured_masses}), the value for the compactification radius $\rho$ is calculated as:
\begin{align}
\label{eq:rho_value}
\rho = \frac{g_{e} \hbar \alpha_{e}}{6 m_{e} c} = 9.4040252(14) \times 10^{-16} \text{ m}
\end{align}
Once the numerical value of the compactification radius $\rho$ is known, the g-factor of any quark can be determined. Using the mass $m_{\mathit{n0}}$  from \eqref{eq:quark_masses_no_mag} with $A=0$, the g-factor is given by
\begin{align}
\label{eq:general_g_factor}
g_n = \frac{18 c \rho  m_{\mathit{n0}}}{n_{e} \hslash  \alpha_{e} }
\end{align}
where we identify the g-factor as mainly dependent on the first-generation masses. The following magnetic dipole moment of a quark is
\begin{align}
\label{eq:particle dipole momentum}
\mu_{n} = \frac{g_n e_{e} n_{e} \hslash}{12 m_{n}} = \frac{3 c \rho  e_{e}m_{\mathit{n0}} }{2 \alpha_{e} m_{n}}
\end{align}

\subsection{Weak coupling \texorpdfstring{\(\alpha_{w}\)}{alpha\_w}, neutrino distribution ratio \texorpdfstring{\(\delta\)}{delta}, and lepton flavor constants \texorpdfstring{\(A_i\)}{A_i}}  
\label{subsec:Weak_coupling}
The masses of the three charged leptons $(e, \mu, \tau)$ are calculated using Eq. \eqref{eq:quark_masses_no_mag} but without an active color charge:
\begin{align}
\label{eq:lepton_masses}
m_{e^\pm, \mu^\pm, \tau^\pm} = \frac{\hbar}{c \rho} \left[ \frac{\alpha_{e}}{3} + 2 \alpha_{w} (1\mp \ln (\delta) \cdot f_{no}(1)) + \left( \alpha_{s}  + \frac{\alpha_{e}}{9} + \alpha_{w}\right) \cdot A_{e,\mu ,\tau} \right]
\end{align}
At this stage, the weak coupling $\alpha_{w}$ and the neutrino distribution ratio $\delta$ are unknown. However, by writing the masses of the electron \eqref{eq:electron_mass_no_mag} and the positron \eqref{eq:positron_mass_no_mag}, with $f_{no}(1)=-1$ (see Eq. \eqref{eq:fno_lepton}) and assuming $A_{e}=0$
\begin{align}
\label{eq:electron_mass_no_mag}
m_{e^{-}} &= \frac{\hbar}{c r_{e}} + \frac{\hbar}{c r_{w}} = \frac{\hbar}{c \rho} \left[ \frac{\alpha_{e}}{3} + 2 \alpha_{w} (1 - \ln (\delta)) \right] \end{align}
\begin{align}
\label{eq:positron_mass_no_mag}
m_{e^{+}} &= \frac{\hbar}{c r_{e}} + \frac{\hbar}{c r_{w}} = \frac{\hbar}{c \rho} \left[ \frac{\alpha_{e}}{3} + 2 \alpha_{w} (1 + \ln (\delta)) \right]
\end{align}
, the system allows us to determine $\alpha_{w}$ and estimate $\delta$. 
\begin{align}
\label{eq:alpha_w_value}
\alpha_{w} &= \frac{3 c \rho m_{\text{average}} - \hbar \alpha_{e}}{6 \hbar} = 1.41040(26) \times 10^{-6} 
\end{align}
Here is $m_{\text{average}}=\frac{1}{2} \left( m_{e^{-}} + m_{e^{+}} \right)$. It is known that $\alpha_{w} \approx 10^{-6}$ \cite{RN212}, which agrees well with the result in \eqref{eq:alpha_w_value}. This supports the assumption that $A_{e} = 0$. The second argument supporting this result comes from the higher-dimensional structure of the photon. In order for the photon to be massless, the condition $A_{\text{first generation}} = 0$ must be met. From this point onward, calculations are performed under the assumption:
\begin{align}
\label{eq:First_generation_flavor_constants}
A_{e} = A_{\nu_{e}} = A_{d} = A_{u} = 0
\end{align}
The experimental upper bound on the positron-electron mass difference,  
\begin{align}
\frac{\left( m_{e^+} - m_{e^-} \right)}{m_{\text{average}}} < 8 \times 10^{-9} \quad \text{\cite{RN204, RN213}}
\end{align}
allows for the calculation of an interval with $\delta$ within:
\begin{align}
1  \ \le \delta < \exp \left(  \frac{m_{\text{average}} c \rho}{5 \cdot \,10^{8} \alpha_{w} \hslash} \right) = 1.0000034530(64) 
\label{eq:delta_value} 
\end{align}
For $1  \ < \delta < 1.0000034530(64)$, the density of antineutrinos is marginally greater than that of neutrinos. To calculate the flavor constants of the charged leptons, the measured masses of muon and tau, along with Eq. \eqref{eq:lepton_masses}, are used to determine their values. The results are $A_{\mu} =  0.93$ and $A_{\tau} =  15.66$. We use $\alpha_{s}=0.53520582(60)$, a result that is independently received later (see Table \ref{tab:field_multipliers}). Flavor constants must be zero or even integers. Taking into account the statistical, observational, and numerical errors of the calculation results, the flavor constants are close to integers. However, $A_{\mu}$ cannot take a value other than 1. Therefore, the muon is regarded as a superposition of the electron and a respective second-generation lepton, leading to the results which are used in all further calculations. 
According to Eq. \eqref{eq:weak_masses_no_mag}, first-generation neutrinos are considered to be massless. Neutrino oscillation data \cite{esteban2020fate} provide squared mass differences among the three neutrino generations. Based on this empirical evidence (normal ordering, without using atmospheric data), the mass for the second-generation neutrino is determined as $m_{\nu _{\mu}}=0.0086(1) \frac{eV}{c^2}$, and for the third-generation neutrino it is $m_{\nu _{\tau}}=0.0501(3) \frac{eV}{c^2}$. Consequently, it is possible to derive the flavor constants for both second- and third-generation neutrinos. Given the hypothesis that flavor constants are even positive integers, these values must simultaneously satisfy this condition while ensuring $1 \le \delta < 1.0000034530(64)$, with preference for combinations near the upper value of the interval \eqref{eq:delta_value}. The suitable arrangement based on Eq. \eqref{eq:weak_masses_no_mag} is found with $\delta = 1.0000029100(64)$ as 
\begin{align}
\label{eq:neutrino_flavor_values} 
A_{\nu _{\mu}} = 
\frac{m_{\nu _{\mu}} c \rho}{\alpha_{w} \hslash  \ln \! \left(\delta \right)} = 9.99 &&
A_{\nu _{\tau}} = 
\frac{m_{\nu _{\tau}} c \rho}{\alpha_{w} \hslash  \ln \! \left(\delta \right)} = 58.17
\end{align}
Rounded to $10$ and $58$, these values have been recorded in Table \ref{tab:exact_flavor_constants}, which lists all flavor constants. The neutrino masses are described by Eq. \eqref{eq:weak_masses_no_mag} with $f_{no}^{weak} = 0$ (see Eq. \eqref{eq:fno_weak}) and the obove stated neutrino masses. Those numbers are compatible with the experimental results
\begin{itemize}
    \item  \textbf{KATRIN collaboration}: $m_{\nu} < 0.45 \frac{eV}{c^{2}}$ \cite{Katrin2025}  
    \item   \textbf{Chandra astrophysical observations}: $\sum m_{\nu} < 0.33 \frac{eV}{c^{2}}$ \cite{RN218}  
\end{itemize}

\subsection{Hadron mass formulas}  
Using Eq. \eqref{eq:quark_masses_no_mag}, the mass contributions from the strong, electric, and weak interactions for quarks can be expressed as:  
\begin{align}
m_{d, s, b} &= \frac{\hbar}{c \rho} \left[ 2 \alpha_{s} + \frac{1}{9} \alpha_{e} + 2 \alpha_{w} +\left(\alpha_{s} +\frac{\alpha_{e}}{9}+\alpha_{w} \right) A_{d, s, b} \right] + \frac{2 \alpha_{w} \hbar}{c \rho} \cdot \ln (\delta) \cdot f_{no}(1)
\label{eq:reduced_dsb-quark_masses} 
\end{align}
\begin{align}
m_{\overline{d}, \overline{s}, \overline{b}} &= \frac{\hbar}{c \rho} \left[ 2 \alpha_{s} + \frac{1}{9} \alpha_{e} + 2 \alpha_{w} +\left(\alpha_{s} +\frac{\alpha_{e}}{9}+\alpha_{w} \right) A_{d, s, b} \right] - \frac{2 \alpha_{w} \hbar}{c \rho} \cdot \ln (\delta) \cdot f_{no}(1)
\label{eq:reduced_dsb-anti-quark_masses} 
\end{align}
\begin{align}
m_{u, c, t} &= \frac{\hbar}{c \rho} \left[ 2 \alpha_{s} + \frac{2}{9} \alpha_{e} + \alpha_{w} +\left(\alpha_{s} +\frac{\alpha_{e}}{9}+\alpha_{w} \right) A_{u, c, t} \right] - \frac{ \alpha_{w} \hbar}{c \rho} \cdot \ln (\delta) \cdot f_{no}(1)
\label{eq:reduced_uct-quark_masses} 
\end{align}
\begin{align}
m_{\overline{u}, \overline{c}, \overline{t}} &= \frac{\hbar}{c \rho} \left[ 2 \alpha_{s} + \frac{2}{9} \alpha_{e} + \alpha_{w} +\left(\alpha_{s} +\frac{\alpha_{e}}{9}+\alpha_{w} \right) A_{u, c, t} \right] + \frac{ \alpha_{w} \hbar}{c \rho} \cdot \ln (\delta) \cdot f_{no}(1)
\label{eq:reduced_uct-anti-quark_masses} 
\end{align}
For baryons, finding $f_{no}(1)=-2$ (see Eq. \eqref{eq:fno_baryon}), and for mesons, $f_{no}(1)=-3$ (see Eq. \eqref{eq:fno_meson}), the mass disparity between quarks and antiquarks is negligible, approximately $0.01 \frac{eV}{c^2}$; hence, it can often be disregarded in the computation of hadron masses. Variations in quark masses are predominantly governed not just by interaction dynamics but are largely impacted by flavor constants $A_{d,s,b,u,c,t}$. As noted above, all first-generation constants assume a value of zero \eqref{eq:First_generation_flavor_constants}. When ignoring magnetic field effects, the masses of first-generation mesons, denoted as $\widetilde{m}$, can be expressed as:  
\begin{align}
\widetilde{m}_{\pi^{0}} &= \frac{\widetilde{m}_{d \bar{d}}+\widetilde{m}_{u \bar{u}}}{2} = \frac{m_{d}+m_{\bar{d}}}{2} + \frac{m_{u}+m_{\bar{u}}}{2} = \widetilde{m}_{\rho^{0}} \\
\widetilde{m}_{\pi^{+}} &= \widetilde{m}_{u \bar{d}} = m_{u} + m_{\bar{d}} = \widetilde{m}_{\rho^{+}}. \nonumber
\end{align}
\subsubsection{Magnetism}
Each quark and antiquark behave as a dipole and possesses a magnetic field corresponding to the excited compactified dimension 7. The Biot-Savart law in three-dimensional space is given by:
\begin{align}
\vec{B}(\vec{R}) = \frac{\mu_{0}}{4 \pi} \int_{C} \frac{I d \vec{l} \times \hat{R}}{R^{2}}.
\end{align}
Here, the denominator $R^{2}$ captures the usual weakening of the magnetic field with increasing separation $R$ from the dipole. However, in a compactified dimension, the field is confined to the curled-up dimension and does not exhibit the same spatial falloff. We therefore replace the distance scale $R$ by the compactification radius $\rho$, which leads to the numerical value of the strong coupling constant $\alpha_s$ reported in subsection~\ref{subsec:strong_coupling_and_zValues}. This value of $\alpha_s$ is then used in Eq.~\eqref{eq:gluon_mass} to compute the gluon mass; the resulting value is used to show the consistency between the calculated $\alpha_s$ and the measurements of the strong coupling (see subsection~\ref{subsec:Gluon_mass}), which supports the above substitution. The expression therefore becomes:
\begin{align}
\label{B_field_quark}
\vec{B}_{o}(\vec{z}) = \frac{\mu_{0}}{4 \pi }\int_{C} \frac{I_{o} d \vec{l} \times \hat{z}}{\rho^2}
= \frac{\mu_{0}}{4 \pi  \rho^2} \int_{0}^{4 \pi} I_{o} r_{e} d \theta 
\begin{bmatrix} 
\sin\left(\frac{1}{2} \alpha \right) \\ 
-\cos\left(\frac{1}{2} \alpha \right) \\ 
0
\end{bmatrix}
\end{align}
which simplifies to:
\begin{align}
\label{eq:B-field quark}
\vec{B}_{o}(\vec{z}) = -\frac{1}{4 \pi \rho^2} \mu_{0} |o_{7}| \cdot \frac{e_{e}}{3} \cdot c 
\begin{bmatrix} 
\sin\left(\frac{1}{2} \alpha \right) \\ 
-\cos\left(\frac{1}{2} \alpha \right) \\ 
0
\end{bmatrix}
\end{align}
where:
\begin{align}
I_{o} = \frac{o_{7} e_{e} c}{3 \cdot  |r_{e}| \cdot 4 \pi}
\end{align}
\begin{align}
d \vec{l} \times \hat{z} = r_{e} d \theta 
\begin{bmatrix} 
\cos \left(\frac{1}{2} \alpha \right) \\ 
\sin \left(\frac{1}{2} \alpha \right) \\ 
0
\end{bmatrix} 
\times 
\begin{bmatrix} 0 \\ 0 \\ 1 \end{bmatrix},
\end{align}
and $\alpha$ denotes the angle in the compactified dimension. A similar equation holds for the antiquark, with substitution \( o \longrightarrow n \). Since quarks and antiquarks share a common compactified dimension but with a phase shift of \( \frac{\pi}{2} \), the corresponding magnetic field is:
\begin{align}
\label{eq:B-field antiquark}
\vec{B}_{\bar{n}}(\vec{z}) = -\frac{1}{4\pi \rho^2} \mu_{0} |n_{7}| \cdot \frac{e_{e}}{3} \cdot c 
\begin{bmatrix} 
\sin\left(\frac{1}{2} (\alpha + \frac{\pi}{2} )\right) \\ 
-\cos\left(\frac{1}{2} (\alpha+\frac{\pi}{2})\right) \\ 
0
\end{bmatrix}.
\end{align}
Consequently, whenever an expression involving the charge of a quark is combined with one involving the charge of the corresponding antiquark, a normalization factor of $\sqrt{2}$ must be included. This requirement arises from the condition that the product of the quark charge and the antiquark charge should remain a straightforward, unmodified product of the two charges.
\begin{align}
\sqrt{2} \cdot \frac{n_{q} e_{e}}{3} \cdot \begin{bmatrix} 
\sin\left(\frac{1}{2} \alpha \right) \\ 
-\cos\left(\frac{1}{2} \alpha \right) \\ 
0
\end{bmatrix} \cdot \frac{n_{\bar{q}} e_{e}}{3} \cdot \begin{bmatrix} 
\sin\left(\frac{1}{2} (\alpha + \frac{\pi}{2} )\right) \\ 
-\cos\left(\frac{1}{2} (\alpha+\frac{\pi}{2})\right) \\ 
0
\end{bmatrix}  =  \left( \frac{n_{q} e_{e}}{3} \right ) \cdot \left(\frac{n_{\bar{q}} e_{e}}{3} \right)
\end{align}

\subsubsection{Magnetic field's contribution on hadron masses}

The energy contribution due to the magnetic field leaking into three-dimensional space for mesons is given by
\begin{align}
\label{eq:N_in_meson_mass}
m_{B}(o \bar{n}) =  \int_{0}^{2 \pi} \int_{0}^{\pi} \int_{\frac{z \rho}{2}}^{\infty}
\frac{\left(\vec{B}_{o}^2 + 2 \sqrt{2}  \vec{B}_{o} \cdot \vec{B}_{\bar{n}} +\vec{B}_{\bar{n}}^2\right)\cdot \rho^4}{2 \mu_{0} c^{2} R^{2}} \cdot \sin \vartheta d R d \vartheta d \varphi
\end{align}
, where the strength of  $\vec{B}$ weakens with the distance $R^2$ from the dipole. Evaluating the integral, we obtain:
\begin{align}
m_{B}(o \bar{n}) = 
\begin{cases} 
\frac{\hslash  \alpha_{e} \left({| n_{7} |}-{| o_{7} |}\right)^2}{9 c z \rho} , & \text{(Scalar mesons)} \\[10pt]
\frac{\hslash  \alpha_{e} \left({| n_{7} |}+{| o_{7} |}\right)^2}{9 c z \rho} , & \text{(Vector mesons)}
\end{cases}
\end{align}
For baryons, the energy contribution associated with magnetic fields spreading into three-dimensional space reduces the baryon’s mass. Hence, a negative sign appears in the equation below.
\begin{align}
\label{eq:N_in_baryon_mass}
m_{B}(nop) = -\int_{0}^{2 \pi} \int_{0}^{\pi} \int_{\frac{z \rho}{2}}^{\infty} 
\frac{\left(\vec{B}_{n} + \vec{B}_{o} + \vec{B}_{p} \right)^{2}  \cdot \rho^4}{2 \mu_{0} c^{2}  R^{2}} \cdot \sin \vartheta d R d \vartheta d \varphi.
\end{align}
This mass reduction can be checked by comparing the proton mass with the neutron mass. Computing the above integral gives:
\begin{align}
\label{eq:magnetic_mass_or_baryons}
m_{B}(nop) = 
\begin{cases} 
-\frac{\hslash  \alpha_{e} \left(| n_{7} |^{2} + 2 | n_{7} | | o_{7} | + | o_{7} |^{2} + | p_{7} |^{2} \right)}{9 c z \rho} , & \text{(Spin-1/2; 2 or 3 quark flavors)} \\[10pt]
-\frac{\hslash  \alpha_{e} \left(| n_{7} | + | o_{7} | + | p_{7} | \right)^{2}}{9 c z \rho} , & \text{(Spin-3/2; 2 or 3 quark flavors)} \\[10pt]
-\frac{\hslash  \alpha_{e} \left({| p_{7} |}^{2}+2 {| n_{7} |} {| p_{7} |}+{| n_{7} |}^{2}+|o_{7}|^{2}\right)}{9 c z \rho}, & \text{(Spin-3/2; one quark flavor)}
\end{cases}
\end{align}

\subsubsection{Magnetic dipole's contribution to hadron masses}
Another contribution to the hadron mass arises from the dipole moment \eqref{eq:particle dipole momentum} of a quark/antiquark within the magnetic field of its partner quark/antiquark:
\begin{align}
\vec{\mu}_{o} &= \frac{g \! \left(o \right) o_{7} e_{e} \hslash}{12 m_{o}} 
\begin{bmatrix} 
\sin \! \left(\frac{\alpha}{2}\right) \\ 
-\cos \! \left(\frac{\alpha}{2}\right) \\ 
0 
\end{bmatrix}, \quad
\vec{\mu}_{\bar{n}} = \frac{g \! \left(n \right) n_{7} e_{e} \hslash}{12 m_{n}} 
\begin{bmatrix} 
\sin \! \left(\frac{1}{2}(\alpha+\frac{\pi}{2})\right) \\ 
-\cos \! \left(\frac{1}{2}(\alpha+\frac{\pi}{2})\right) \\ 
0 
\end{bmatrix}.
\end{align}

The additional mass resulting from the dipole moment of $\vec{B}(z)$ within the curled-up dimension is given by:
\begin{align}
m_{M}(o \bar{n}) &= \frac{\vec{\mu}_{o} \cdot \vec{B}_{\bar{n}}(\vec{z}) 
+ \vec{\mu}_{\bar{n}} \cdot \vec{B}_{o}(\vec{z})}{c^2} \cdot \sqrt{2} \\ 
&= 
\begin{cases} 
-\frac{\hslash}{2 c \rho} \cdot \left[\frac{\left(2 \alpha_{s}+\frac{{| o_{7} |} \alpha_{e}}{9}+{| o_{8} |} \alpha_{w} \right) {| n_{7} |}}{2 \alpha_{s}+\frac{{| o_{7} |} \alpha_{e}}{9}+{| o_{8} |} \alpha_{w} +\mathit{Bo}}+\frac{\left(2 \alpha_{s}+\frac{{| n_{7} |} \alpha_{e}}{9}+{| n_{8} |} \alpha_{w} \right) {| o_{7} |}}{2 \alpha_{s}+\frac{{| n_{7} |} \alpha_{e}}{9}+{| n_{8} |} \alpha_{w} +\mathit{Bn}} \right], 
& \text{(Scalar mesons)} \\[10pt]
+\frac{\hslash}{2 c \rho} \cdot \left[\frac{\left(2 \alpha_{s}+\frac{{| o_{7} |} \alpha_{e}}{9}+{| o_{8} |} \alpha_{w} \right) {| n_{7} |}}{2 \alpha_{s}+\frac{{| o_{7} |} \alpha_{e}}{9}+{| o_{8} |} \alpha_{w} +\mathit{Bo}}+\frac{\left(2 \alpha_{s}+\frac{{| n_{7} |} \alpha_{e}}{9}+{| n_{8} |} \alpha_{w} \right) {| o_{7} |}}{2 \alpha_{s}+\frac{{| n_{7} |} \alpha_{e}}{9}+{| n_{8} |} \alpha_{w} +\mathit{Bn}} \right]
 , 
& \text{(Vector mesons)}
\end{cases} \nonumber
\end{align}
$\sqrt{2}$ is the normalizing factor and $\mathit{Bi} = 
\left(\alpha_{s}+\frac{\alpha_{e}}{9}+\alpha_{w} \right) \mathit{Ai}
$.

For baryons, the additional mass contribution due to dipole interactions is:
\begin{align}\label{eq:dipole_mass_baryons}
m_{M}(nop)  &= \left(
  (\vec{\mu}_{o} +  \vec{\mu}_{p})\cdot \vec{B}_{n}(\vec{z}) 
+ (\vec{\mu}_{p} +  \vec{\mu}_{n})\cdot \vec{B}_{o}(\vec{z})
+ (\vec{\mu}_{n} +  \vec{\mu}_{o})\cdot \vec{B}_{p}(\vec{z}) \right)
 \cdot \frac{1}{c^2 } \\ 
&= \frac{\hslash}{2 c \rho} \cdot
\begin{cases} 
\frac{\left(2 \alpha_{s}+\frac{{| o_{7} |} \alpha_{e}}{9}+{| o_{8} |} \alpha_{w} \right) {| n_{7} |}}{2 \alpha_{s}+\frac{{| o_{7} |} \alpha_{e}}{9}+{| o_{8} |} \alpha_{w} +\mathit{Bo}}+\frac{\left(2 \alpha_{s}+\frac{{| n_{7} |} \alpha_{e}}{9}+{| n_{8} |} \alpha_{w} \right) {| o_{7} |}}{2 \alpha_{s}+\frac{{| n_{7} |} \alpha_{e}}{9}+{| n_{8} |} \alpha_{w} +\mathit{Bn}}
 \\[10pt] 
\left(\frac{2 \alpha_{s}+\frac{{| n_{7} |} \alpha_{e}}{9}+{| n_{8} |} \alpha_{w}}{2 \alpha_{s}+\frac{{| n_{7} |} \alpha_{e}}{9}+{| n_{8} |} \alpha_{w} +\mathit{Bn}}+\frac{2 \alpha_{s}+\frac{{| p_{7} |} \alpha_{e}}{9}+{| p_{8} |} \alpha_{w}}{2 \alpha_{s}+\frac{{| p_{7} |} \alpha_{e}}{9}+{| p_{8} |} \alpha_{w} +\mathit{Bp}}\right) {| o_{7} |} \\
+\left(\frac{2 \alpha_{s}+\frac{{| o_{7} |} \alpha_{e}}{9}+{| o_{8} |} \alpha_{w}}{2 \alpha_{s}+\frac{{| o_{7} |} \alpha_{e}}{9}+{| o_{8} |} \alpha_{w} +\mathit{Bo}}+\frac{2 \alpha_{s}+\frac{{| p_{7} |} \alpha_{e}}{9}+{| p_{8} |} \alpha_{w}}{2 \alpha_{s}+\frac{{| p_{7} |} \alpha_{e}}{9}+{| p_{8} |} \alpha_{w} +\mathit{Bp}}\right) {| n_{7} |} \\
+\left(\frac{2 \alpha_{s}+\frac{{| n_{7} |} \alpha_{e}}{9}+{| n_{8} |} \alpha_{w}}{2 \alpha_{s}+\frac{{| n_{7} |} \alpha_{e}}{9}+{| n_{8} |} \alpha_{w} +\mathit{Bn}}+\frac{2 \alpha_{s}+\frac{{| o_{7} |} \alpha_{e}}{9}+{| o_{8} |} \alpha_{w}}{2 \alpha_{s}+\frac{{| o_{7} |} \alpha_{e}}{9}+{| o_{8} |} \alpha_{w} +\mathit{Bo}}\right) {| p_{7} |}
 \\[10pt]
\frac{\left(2 \alpha_{s}+\frac{{| p_{7} |} \alpha_{e}}{9}+{| p_{8} |} \alpha_{w} \right) {| n_{7} |}}{  \left(2 \alpha_{s}+\frac{{| p_{7} |} \alpha_{e}}{9}+{| p_{8} |} \alpha_{w} +\mathit{Bp} \right)}+\frac{ {| p_{7} |} \left(2 \alpha_{s}+\frac{{| n_{7} |} \alpha_{e}}{9}+{| n_{8} |} \alpha_{w} \right)}{ \left(2 \alpha_{s}+\frac{{| n_{7} |} \alpha_{e}}{9}+{| n_{8} |} \alpha_{w} +\mathit{Bn} \right)}
\end{cases} \nonumber
\end{align}
The first equation applies to spin-1/2 baryons containing two or three quark flavors, the second equation pertains to spin-3/2 baryons composed of at least two distinct quark flavors, and the third equation corresponds to spin-3/2 baryons whose constituent quarks all share the same flavor.

\subsubsection{Weak and strong magnetism}

Since all compactified dimensions display similar structures, it is natural to expect that magnetism might also originate from the weak and strong sectors. Yet, when one computes the corresponding strong and weak magnetic contributions, the results deviate from experimental values substantially. Because strong and weak interactions always involve two active dimensions, their respective mass contributions are expected to cancel each other out.

\subsubsection{Meson mass formula}  
\begin{align}
\label{meson_full_mass_appendix}
m_{o \bar{n}} = m_{o} + m_{\bar{n}} + m_{B}(o \bar{n}) + m_{M}(o \bar{n}).
\end{align}
Using this framework, the mass formulas for the first-generation mesons are given by:
\begin{align}
\label{eq:first_gereration_meson_masses}
m_{\pi^{0}} &= \frac{(m_{d} + m_{u} + m_{\bar{d}} + m_{\bar{u}})}{2} - \frac{3  \hslash }{2 c \rho}, \\
m_{\pi^{+}} &= m_{u} + m_{\bar{d}} + \frac{\hslash  \alpha_{e}}{9 c z \rho}-\frac{3  \hslash }{2 c \rho}, \nonumber \\
m_{\rho^{0}} &= \frac{(m_{d} + m_{u} + m_{\bar{d}} + m_{\bar{u}})}{2} + \frac{10 \hslash  \alpha_{e}}{9 c z \rho}+\frac{3  \hslash }{2 c \rho}, \nonumber \\
m_{\rho^{+}} &= m_{u} + m_{\bar{d}} + \frac{\hslash  \alpha_{e}}{c z \rho}+\frac{3  \hslash }{2 c \rho}. \nonumber
\end{align}

\subsubsection{Baryon mass formula}  
\begin{align}
\label{eq:baryon_full_mass}
m_{nop} = m_{n} + m_{o} + m_{p} + m_{B}(nop) + m_{M}(nop)
\end{align}
Baryons consist of three quarks and exist, when without orbital rotation, in two spin states: $\hbar / 2$ and $3 \hbar / 2$. For spin $\hbar / 2$, one of the quarks (the $p$-quark) undergoes a spin flip, breaking the symmetry. As a result, different quark configurations—despite having the same quark content—can lead to different baryon masses. A clear example of this is the mass difference between the $\Lambda^{0}$ (1116 $MeV/c^2$) and the $\Sigma^{0}$ (1193 $MeV/c^2$) baryons, both of which share the quark content $(d, u, s)$. This pattern of symmetry breaking, together with the selection rules stated below, modifies how the additional mass terms are computed. Quarks are characterized by their strong, electric, and weak charges. The strong charge is flavor-independent and thus identical for all quark types, while the electric and weak charges occur in flavor-specific pairs: for an up quark they are $\frac{2 e_e}{3}$ and $- e_w$, and for a down quark they are $\frac{-e_e}{3}$ and $2 e_w$. An analogous charge structure applies to all higher generations. The only distinction is the presence of a nonzero flavor constant, which affects the mass but leaves the charge pattern unchanged. Since quarks are fermions, configurations must combine states with different charges, leading to the first selection rule. For three-flavor quark states with differing charges, one has the freedom to construct two distinct combinations, which motivates the second selection rule. The third selection rule is introduced to exclude redundant configurations in which quark states are identical and not flipped.

\subsubsection{Selection rules for spin $\hbar / 2$-baryons} 
\begin{enumerate}
\item  For baryons consisting of a mixture of u-like (u, c, t) and d-like (d, s, b) quarks, the non-flipped quarks always represent a u-d-like combination.
\item  In case of a baryon consisting of a u-d-mixture built of three different quark species, there always exist two combinations with a different quark flipped. These are baryons with identical excitation numbers but dissimilar mass. 
\item  For those baryons consisting of purely u- or d-like quarks the non-flipped quark pair consists of different flavors.
\end{enumerate}

\subsubsection{Spin $3\hbar / 2$-baryons} 

For spin $3 \hbar / 2$ baryons, each quark interacts with its two neighboring quarks, making the order of quarks irrelevant. Consequently, each unique quark combination corresponds to a single mass value, and all quarks are treated equally. There is, however, a particular exception in the case where three identical quarks form a spin-$3\hbar/2$ baryon. Under these circumstances, an exclusion principle—stating that no more than two identical quarks may occupy the same compactified tube—forces a configuration in which two quarks reside within one cylindrical tube, while the third quark is confined to a separate tube. This exclusion principle is a consequence of the compactified dimensions, which have a circumference of $4\pi r$ and are capable of hosting two identical quarks.


\subsection{Determination of the strong coupling constant \texorpdfstring{$\alpha_{s}$}{alpha	extunderscore s} and the mean distances between quarks in scalar mesons \texorpdfstring{$z_{s\_Mesons}$}{z	extunderscore scalar mesons}, vector mesons \texorpdfstring{$z_{v\_Mesons}$}{z	extunderscore vector mesons}, and baryons \texorpdfstring{$z_{Baryons}$}{z	extunderscore baryons}}
\label{subsec:strong_coupling_and_zValues}
To extract the strong coupling constant $\alpha_{s}$, which governs the hadronic binding energy, we make use of the precisely determined mass of the neutral pion \eqref{eq:first_gereration_meson_masses} and the data in Table \ref{tab:Measured_masses}. 
\begin{table}[tbp]
    \centering
    \caption{Measured lepton and hadron masses used for calibration}
    \includegraphics[width=\linewidth,keepaspectratio]{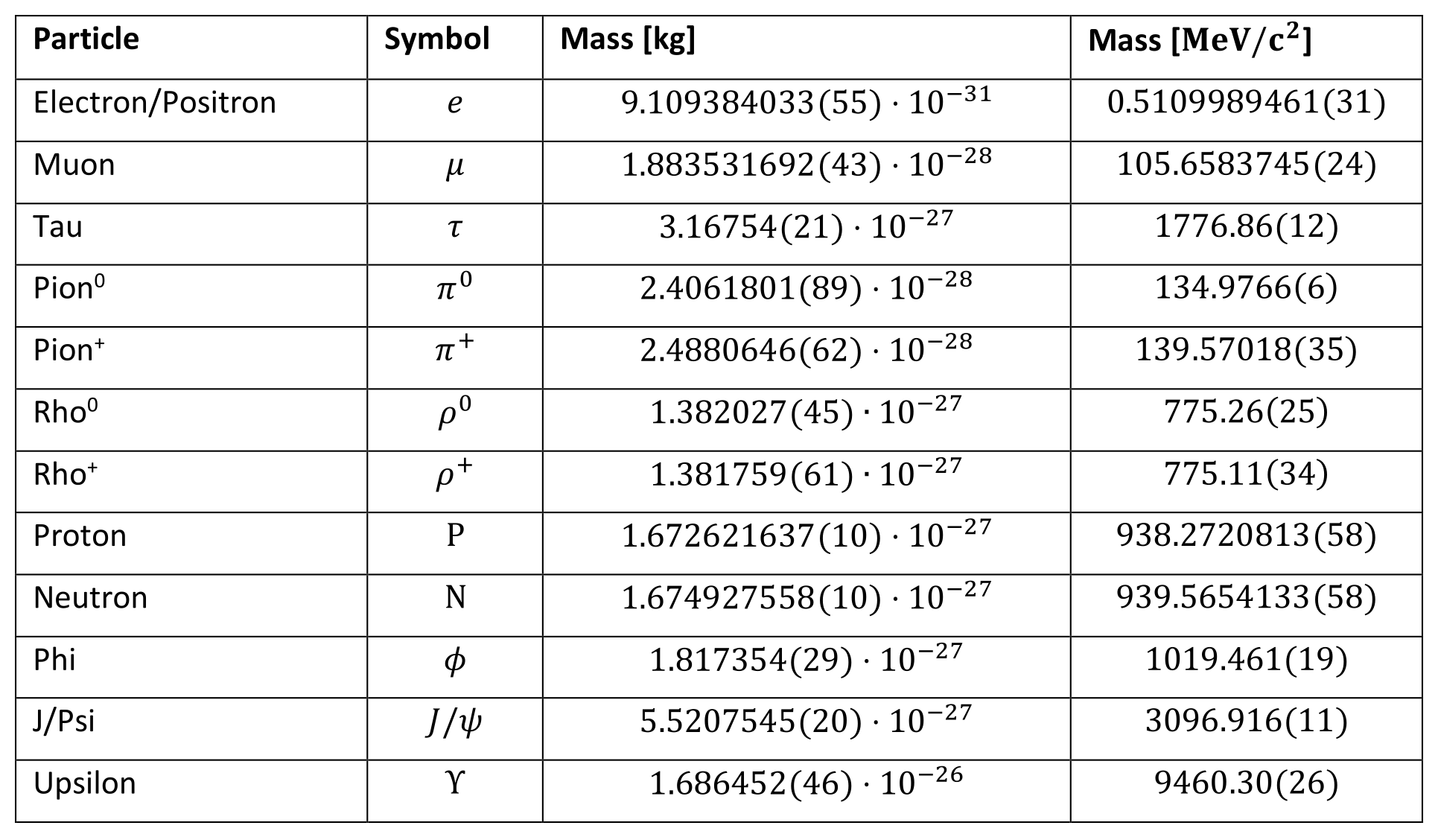}
    \label{tab:Measured_masses}
\end{table}
The pion mass is among the most accurately measured particle masses and, in this model, is independent of $z$. While the average value of $z$ may differ between various quark configurations, there are clear differences between the $z$-values associated with scalar mesons, vector mesons, and baryons. As a practical compromise for a more detailed calculation, the quantities $z_{s\_Mesons}$, $z_{v\_Mesons}$, and $z_{Baryons}$ are determined from the experimentally observed masses and the corresponding relations for $\pi^+$, $\rho^0+\rho^+$ \eqref{eq:first_gereration_meson_masses}, the proton
\begin{align}
\label{eq:proton_mass}
m_{\mathit{Proton}} = m_{d} + 2 m_{u} - \frac{13 \hslash  \alpha_{e}}{9 z c \rho}+\frac{3  \hslash }{2 c \rho}
\end{align}
and the neutron
\begin{align}
\label{eq:neutron_mass}
m_{\mathit{Neutron}} = 2 m_{d} + m_{u} - \frac{10 \hslash  \alpha_{e}}{9 z c \rho}+\frac{3  \hslash }{2 c \rho}
\end{align}
where the proton has quark composition $udu$, with one $u$-quark flipped, and the neutron has quark composition $dud$, with one $d$-quark flipped. The results are found in Table \ref{tab:field_multipliers}.
\begin{table}[H]
    \centering
    \caption{Tabulated values are given for the strong coupling constant $\alpha_s$, as well as for $z_{Baryon}$,  $z_{s\_Mesons}$, and $z_{v\_Mesons}$, which denote the mean z-values corresponding to baryons, scalar mesons, and vector, mesons, respectively. The listed uncertainties are purely statistical; systematic uncertainties are not taken into account.}
    \begin{tabular}{|c|c|}
        \hline
        $\alpha_s = 0.53520582(60)$ &
        $z_{Baryons} = 0.03880668(59)$ \\
        \hline
        $z_{s\_Mesons} = 0.0370376(43)$ &
        $z_{v\_Mesons} = 0.1509(24)$
        \\ \hline
    \end{tabular}
    \label{tab:field_multipliers}
\end{table}

\subsection{Gluon mass}
\label{subsec:Gluon_mass}
Within the Standard Model of Particle Physics, the dominant interaction among quarks inside hadrons is mediated by the exchange of virtual gluons. In contrast, in the Kaluza–Klein framework, the strong interaction is interpreted as a manifestation of spacetime curvature. If these two descriptions are regarded as complementary, one may associate the gluon energy in the Standard Model with the gluon mass in the Kaluza–Klein picture \eqref{eq:gluon_mass}, leading to
\begin{align}
\label{eq:gluon_mass_value}
m_{\text{gluon}} = 449.21473(50)\,\frac{\text{MeV}}{c^2}.
\end{align}
Using this relation and applying the renormalization group evolution equation from Ref.~\cite{RN178} to run $\alpha_s$ up to the $Z^{0}$-boson mass scale ($m_{Z^0}=91.1880(20)\,\text{GeV}/c^2$), we arrive at the result
\begin{align}
\label{eq:renormalization_group_evolution}
\alpha_s(m_{Z^0}) = \frac{\alpha_s}{1 + \frac{(11 N_c - 2 N_f)}{12 \pi}\,\alpha_s \,\ln \!\left( \frac{m_{Z^0}^2}{m_{\text{gluon}}^2} \right)} = 0.11974,
\end{align}
for three colors ($N_c = 3$) and five active quark flavors ($N_f = 5$). This value is close to the latest value from ATLAS \cite{atlascollaboration2023precisedeterminationstrongcouplingconstant} for the strong coupling at the $Z^0$ mass scale, $\alpha_s^{ATLAS}(m_{Z^0}) = 0.1183(9)$. The presence of real, as opposed to purely virtual, gluons was experimentally confirmed for the first time in 1979, when planar three-jet configurations in $e^+e^-$ annihilation were detected in measurements performed by four independent collaborations at DESY in Hamburg, Germany \cite{Brandelik1979,Barber1979,Berger1979,Bartel1980}. Their observations were consistent with the QCD expectation of a spin one massless gluon. Nevertheless, the data did not allow for a strict null statement for the gluon mass; instead, only an upper bound could be inferred. The following calculation, which was not presented in the original publications, provides the range of gluon masses consistent with experimental data. The characteristic gluon energy in these measurements was $E_{\text{gluon}} \approx 10\,\text{GeV}$, and the observed half-opening angle of the jet was $\theta_{\text{jet}} = 0.1\,\text{to} \,0.2\,\text{rad}$. In the gluon rest frame, the fragmentation into hadrons is approximately isotropic, implying a typical emission angle of $\theta = \pi/2$. Under a Lorentz boost, the transformation of angles is given by
\begin{align}
\label{eq:lorentz_boost}
\tan(\theta_{\text{jet}}) \approx \frac{\sin(\theta)}{\gamma \left(\cos(\theta) + \beta\right)} = \frac{1}{\gamma \beta} \, ,
\end{align}
from which one obtains $\beta \approx 1 / \sqrt{\tan^{2}(\theta_{\text{jet}}) + 1} \approx 0.995$. Using $\gamma = E_{\text{gluon}} / (m_{\text{gluon}} c^{2}) \approx 10$, this leads to
\begin{align}
\label{eq:gluonMassUpperLimit}
m_{\text{gluon}} = \frac{E_{\text{gluon}}}{\gamma c^{2}}  \lesssim  1\, \frac{\text{GeV}}{c^2} \, .
\end{align}
Thus, the experimental findings are compatible with the gluon mass of the present model \eqref{eq:gluon_mass_value}.


\subsection{Mass generation via compactified dimensions}  

In higher-dimensional spaces, excitations consistently propagate at the speed of light and are perceived as point particles within the framework of four-dimensional spacetime. As a consequence, particles obtain mass via the energy associated with the Planck relation located in higher-dimensional space.


\subsection{Hadron flavor constants}  

The average z-values, $z_{Baryon}$, $z_{s\_Meson}$, $z_{v\_Meson}$ and the coupling constants are assumed to be independent of particle generation. The quark flavor constants $A_{s}, A_{c}, A_{b}$ are determined using the known formulas \eqref{meson_full_mass_appendix} and the measured masses Table \ref{tab:Measured_masses}) of the $\phi, \psi$ and $\Upsilon$ mesons.     
\begin{align}
\label{eq:vector_mesons_for_flavor_constants}
m_{\phi} \! \left(s \overline{s}\right) = &
2 m_{s} + \frac{\hslash  }{c \rho} \cdot \left( \frac{4  \alpha_{e}}{9 z_{v} }+\frac{ m_{d}}{m_{s}} \right)\\
m_{\psi} \! \left(c \overline{c}\right) = &
2 m_{c} + \frac{\hslash  }{c \rho} \cdot \left( \frac{16  \alpha_{e}}{9 z_{v} }+\frac{ 2 m_{u}}{m_{c}} \right)\nonumber \\
m_{\Upsilon} \! \left(b \overline{b}\right) = &
2 m_{b} + \frac{\hslash  }{c \rho} \cdot \left( \frac{4  \alpha_{e}}{9 z_{v} }+\frac{ m_{d}}{m_{b}} \right)\nonumber
\end{align}
The flavor constants are obtained by solving the system of three Eqs. \eqref{eq:vector_mesons_for_flavor_constants} where the masses are dependent on the flavor constants, producing $A_{s}=2.05$, $A_{c}=11.41$, and $A_{b}=42.50$. We interpret the flavor constants as arising from standing waves in the compactified cylinders, implying that the corresponding excitation numbers occur in positive–negative pairs and therefore must be even integers. Incorporating statistical, experimental, and numerical uncertainties in the obtained values, the hadrons' flavor constants are as collected in Table \ref{tab:exact_flavor_constants}. 
\begin{table}[H]
    \centering
    \caption{Flavor constants}
    \begin{tabular}{|c|c|c|c|c|c|}
        \hline
        $A_{e} = 0$ & $A_{\mu} = 1$ & $A_{\tau} = 16$ & 
        $A_{\nu_{e}} = 0$ & $A_{\nu_{\mu}} =$ 10 & $A_{\nu_{\tau}} =$ 58
        \\ \hline        
        $A_{d}=0$ &
        $A_{u}=0$ &
        $A_{s}=2$ &
        $A_{c}=12$ &
        $A_{b}=42$ &
        $A_{t}=?$
        \\ \hline
    \end{tabular}
    \label{tab:exact_flavor_constants}
\end{table}
In the SM, the couplings $\alpha_s$ and $\alpha_e$ are treated as scale-dependent (running) quantities, while in the present model the corresponding couplings are constant. In this framework, the hadronic flavor constants vary from one flavor to another, incorporating the effects of the magnetic field and the magnetic dipole moment, and thereby effectively taking over the role of the SM running couplings. A direct correspondence between the SM running coupling and the introduced hadronic flavor constants can be readily established. The scale dependence of the coupling is governed by the renormalization group evolution in \eqref{eq:renormalization_group_evolution}. Rather than using the Z-boson mass, one instead employs the masses of mesons built from quark–antiquark pairs of identical flavor. By using the experimentally measured hadron masses to extract the SM running $\alpha_s$ and then comparing these extracted values with the model mass formula \eqref{meson_full_mass_appendix}, one obtains a relation connecting the SM running $\alpha_s$ with the hadronic flavor constants $A$ as well as the magnetic and dipole contributions.

\subsection{Quark masses}
Using Eqs.~\eqref{eq:reduced_dsb-quark_masses}–\eqref{eq:reduced_uct-anti-quark_masses} together with the hadron flavor constants listed in Table~\ref{tab:exact_flavor_constants}, the masses formulas between quarks and antiquarks are different. However, the mass differences are so small that they do not show up in the quoted numerical values (Table \ref{tab:Quark_masses_all}).
\begin{table}[tbp]
    \centering
    \caption{Quark masses}
    \includegraphics[width=\linewidth,keepaspectratio]{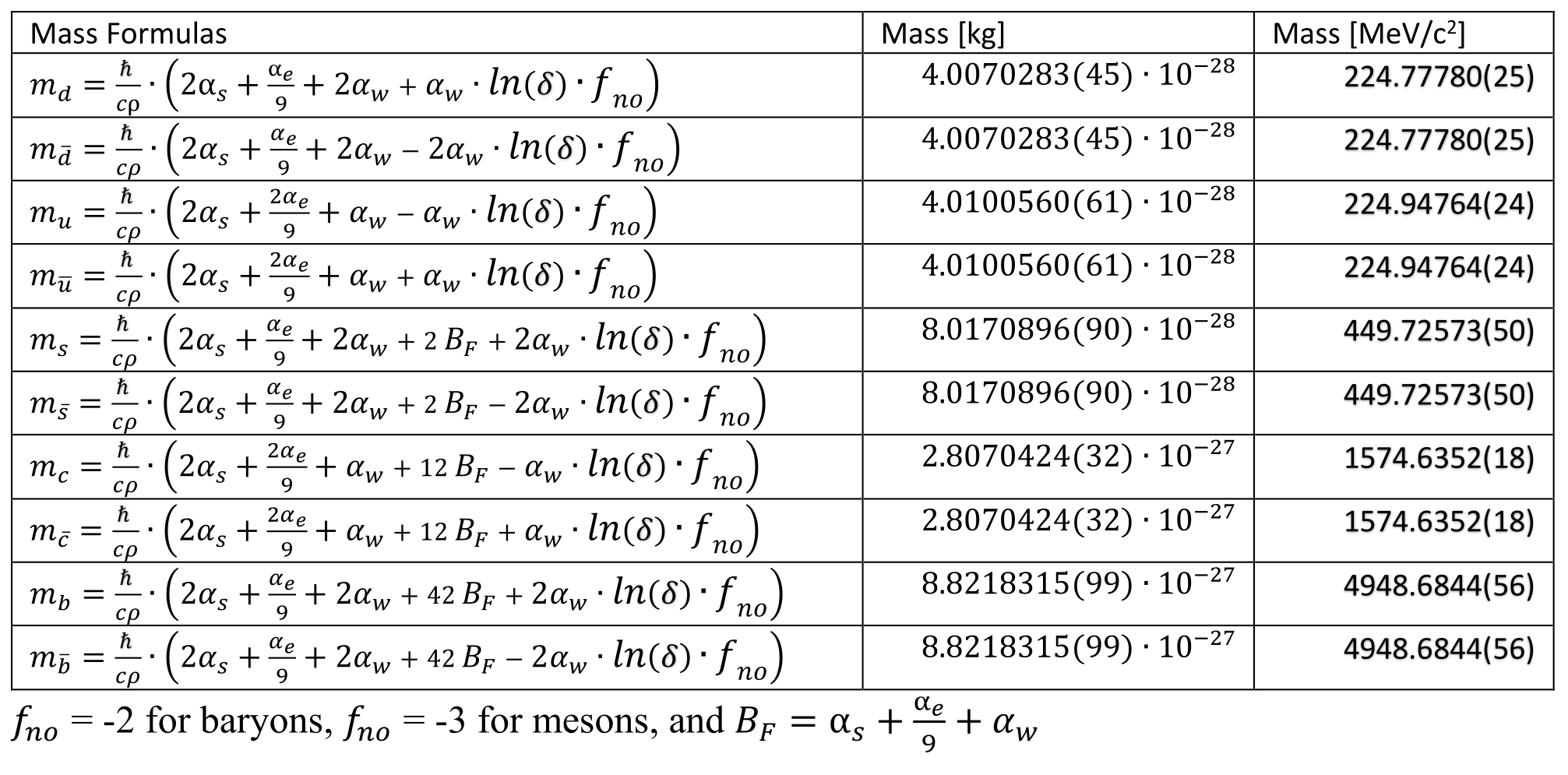}
    \label{tab:Quark_masses_all}
\end{table}

\subsection{Hadron masses}
\label{subsec:Hadron_mass}
The derived values aid in the estimation of hadron masses, leading to 102 masses that have been computed and juxtaposed with the 71 masses that have been experimentally determined. The particle reference list was obtained from Wikipedia \cite{RN215, RN216}, which specifies the lowest-mass states for each hadron, limited to those without orbital angular momentum. The information is structured into four Tables, which include the particle name and symbol, quark configuration, observed and computed masses, standard deviations, the relative error between observed and computed values, the mass formula, and the magnetic ratio. The magnetic mass values listed in Tables \ref{tab:Vector_mesons}, \ref{tab:hbar2_baryons}, and \ref{tab:3hbar2_baryons} are positive and are therefore added to the quark masses, while the magnetic mass in Table \ref{tab:Scalar_mesons} is negative and is consequently subtracted from the quark masses in that table.
\begin{table}[tbp]
    \centering
    \caption{Scalar mesons}
    \includegraphics[width=\linewidth,keepaspectratio]{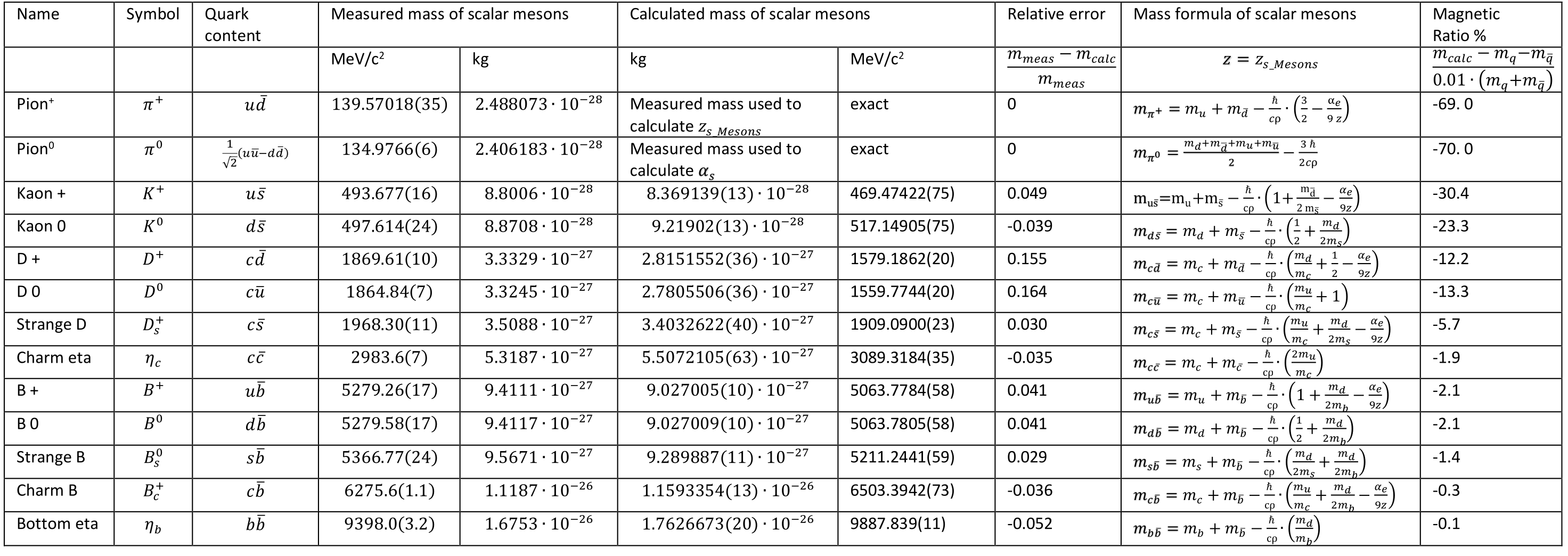}
    \label{tab:Scalar_mesons}
\end{table}

\begin{table}[tbp]
    \centering
    \caption{Vector mesons}
    \includegraphics[width=\linewidth,keepaspectratio]{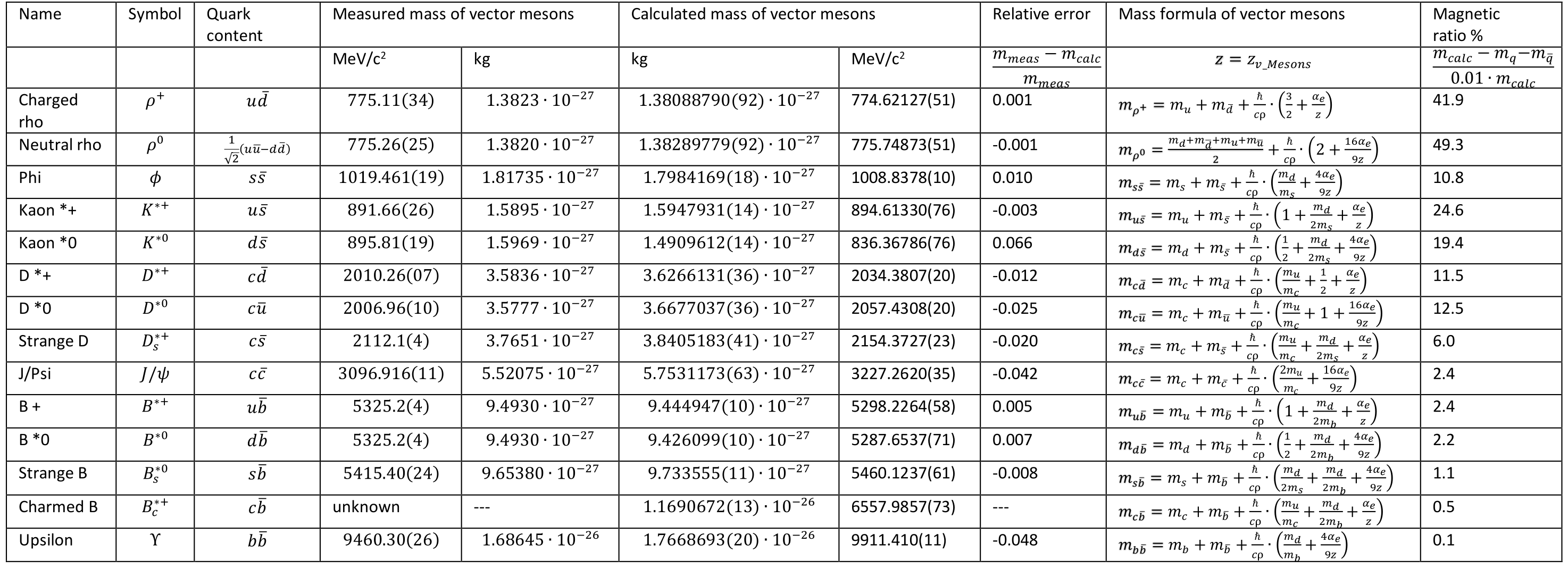}
    \label{tab:Vector_mesons}
\end{table}

\begin{table}[tbp]
    \centering
    \caption{Baryons with spin $\hbar/2$}
    \includegraphics[width=\linewidth,keepaspectratio]{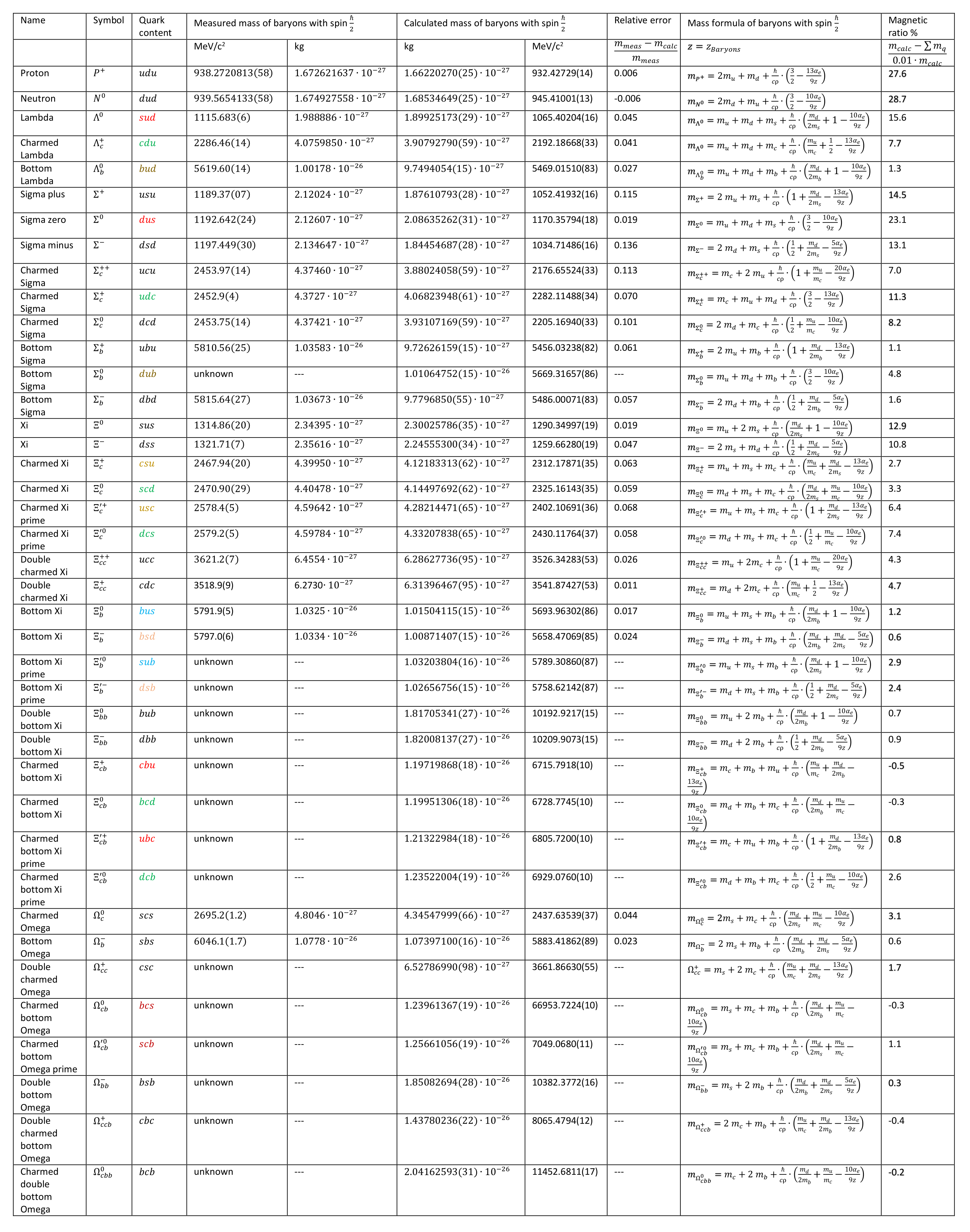}
    \label{tab:hbar2_baryons}
\end{table}

\begin{figure}
    \begin{center}
    \includegraphics[width=1\textwidth]{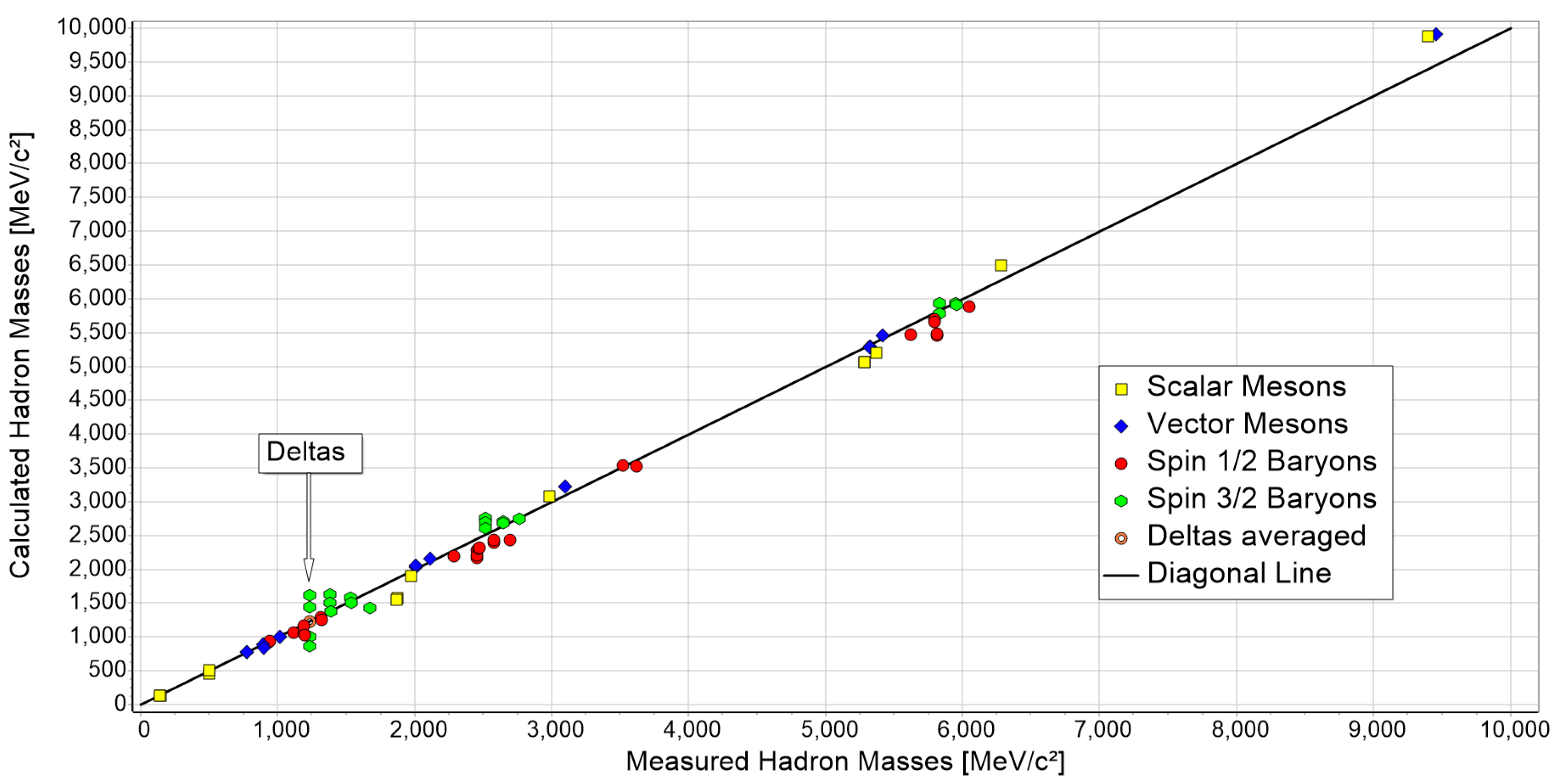}
    \end{center}
    \caption{Correlation between measured and calculated masses of hadrons without orbital angular momentum. The diagonal line represents the ideal case where measured and calculated masses are identical.}
    \label{fig:Masses_correlated_figure}
\end{figure}

\begin{table}[tbp]
    \centering
    \caption{Baryons with spin $3\hbar/2$}
    \includegraphics[width=\linewidth,keepaspectratio]{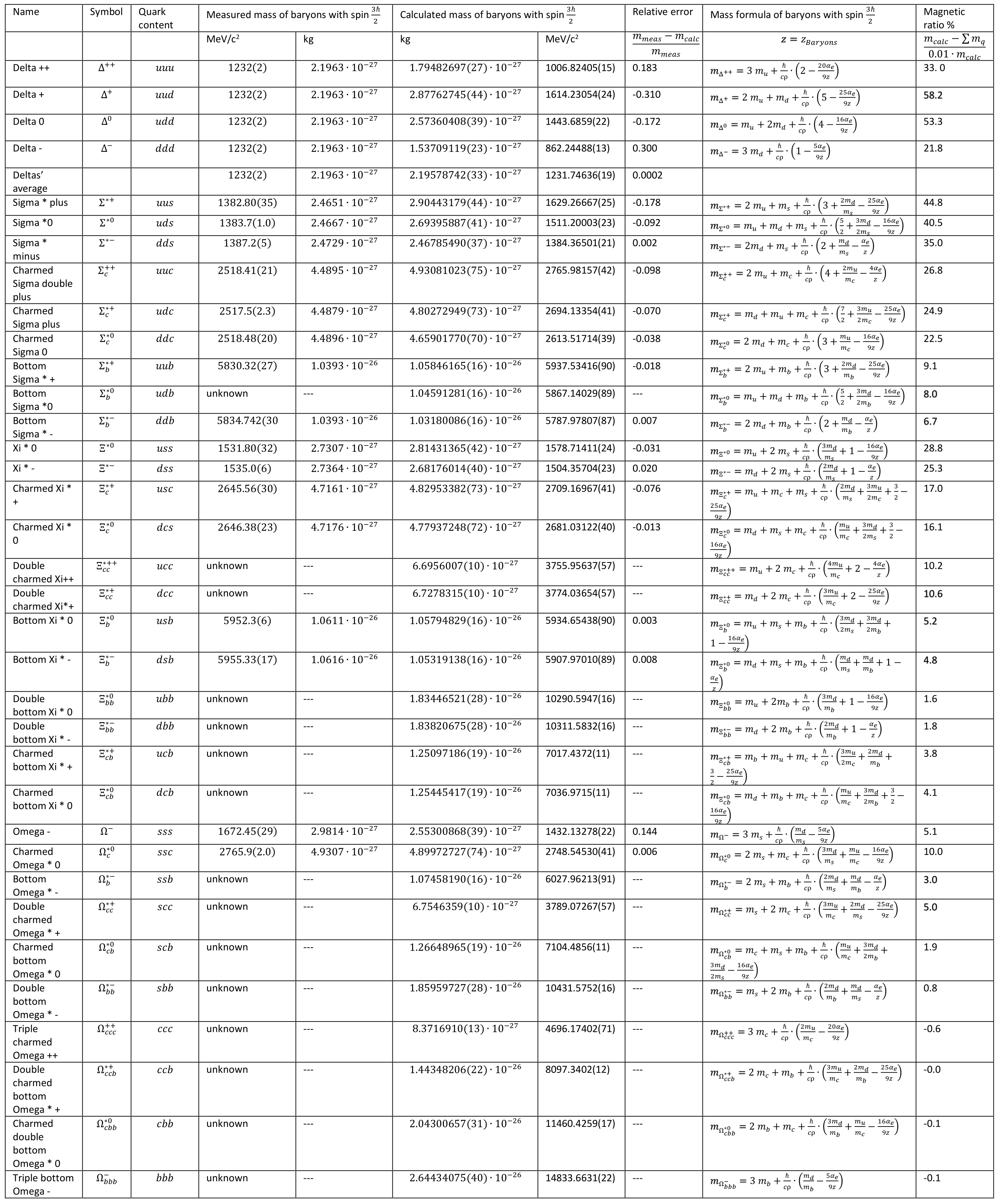}
    \label{tab:3hbar2_baryons}
\end{table}
Figure \ref{fig:Masses_correlated_figure} shows the relationship between the measured and predicted hadron masses, yielding a Pearson correlation coefficient of $r = 0.997$ with a significance level of $p < 10^{-75}$.

\subsection{Sterile bosons}
\label{Sterile elementary Particles}
Eq. (\ref{eq:quark_masses_no_mag}) suggests the possible presence of sterile bosons with spin zero, represented by coefficients $A_{i}$ and without charges. Under suitable conditions, such a sterile bosons could be generated in a collision, with a mass of the (most general) form
\begin{align}
\label{eq:sterile_particle}
m_{sterile} &= \frac{\hbar}{c \rho} \left[ \alpha_{s} (A_{4}+A_{5}+A_{6})
+ \frac{\alpha_{e}}{9} A_{7} + \alpha_{w} ( A_{8} + A_{9} ) \right] 
\end{align}
where $A_{i}=|n_{pi}|+|n_{ni}|$ is either zero or a positive even integer, given by the sum of the positive and negative excitation numbers, obeying $|n_{pi}|=|n_{ni}|$ for $i=4,\dots,9$. Sterile bosons lack strong, electromagnetic, and weak charges; they are spin-zero bosons that possess mass and consequently interact exclusively through gravitational force.


\subsection{Particle acceleration and force} \label{subsec:calculationForce} 

In a coordinate system with the two interacting particles aligned along the $z=z^{1}$-axis with $z^{2}=z^{3}=0$, the velocity is expressed as a function depending on the separation between the particles. Using the metric tensor given in Eq. \eqref{eq:metric_tensor}, the Christoffel symbols are determined independently for each interaction in the usual way, by performing a summation over repeated indices ranging from $0$ to $9$:
\begin{align}
\Gamma_{\nu \xi}^{\mu} = \frac{1}{2} g^{\mu \kappa}_\texttt{s,e,w} \left(\partial_{\nu} g^\texttt{s,e,w}_{\kappa \xi} + \partial_{\xi} g^\texttt{s,e,w}_{\kappa \nu} - \partial_{\kappa} g^\texttt{s,e,w}_{\nu \xi} \right).
\end{align}
The acceleration is then determined using the geodesic equation:
\begin{align}
\label{eq:geodesic_equation}
a_{\texttt{s,e,w}}^{\mu} = -\Gamma_{\nu \xi}^{\mu} u^{\nu} u^{\xi}.
\end{align}
The forces exerted on the particle can be described by:
\begin{align}
\label{eq:all_forces}
F_{\texttt{s,e,w}}^{\mu}(z) = \frac{E_{o}^{\texttt{s,e,w}}}{c^2} \cdot a_{\texttt{s,e,w}}^{\mu}.
\end{align}
The energies involved are of Eqs. \eqref{eq:electric_mass_no-field}, \eqref{eq:Strong_mass_no-field}, and \eqref{eq:weak_mass_no-field}, where in all cases $A$ is zero. 

\subsection{The constant scalar \texorpdfstring{\( f_{no} \! \left(z \right)  \)}{f\_no(z) }} \label{sec:constant_scalar_fno}

In conventional four-dimensional spacetime, the velocity four-vector $u^{\alpha}$ is defined as:
\begin{align}
u^{\alpha} = \gamma \beta^{\alpha} c
\end{align}
where it adheres to the conservation relation:
\begin{align}
u^{\alpha} u_{\alpha} = c^{2}, \quad \text{where } \alpha = 0,1,2,3.
\end{align}
Within a 10-dimensional spacetime, the conservation principle extends based on the quantity of compact active dimensions $n_{cd}$:
\begin{align}
\label{eq:invariant_velocity_squared_compact}
u^{\alpha} u_{\alpha} = (1 - n_{cd}) c^{2}.
\end{align}
The additional dimensions are indexed as $\alpha = 4, 5, 6, 7, 8, 9$.

The computation of force is feasible solely when the function \( f_{\text{no}}(z) \) is available. Thankfully, it stays the same since its derivative is zero, indicating that it is constant.
\begin{align}
\label{eq:derivative_of_fno}
\frac{d}{d \tau}f_{\mathit{no}} \! \left(z \right) = 
\frac{\frac{d}{d \tau}\left(g_{\alpha\beta} w^{\alpha} v^{\beta}\right)}{c^{2}}=\frac{g_{\alpha\beta}}{c^2} \! \left(a_{w}^{\alpha} v^{\beta}+w^{\alpha} a_{v}^{\beta}\right)
 = 
-\frac{g_{\alpha\beta}}{c^2} \! \left(\frac{m_{v} a_{v}^{\alpha} v^{\beta}}{m_{w}}+\frac{m_{w} w^{\alpha} a_{w}^{\beta}}{m_{v}}\right)=0
\end{align}
This result aligns with Newton's third law, \( m_{w} \cdot a_{w}^\alpha = -m_{v} \cdot a_{v}^\alpha \), and is evidently zero since the derivative of Eq. \eqref{eq:invariant_velocity_squared_compact} is zero. The value of \( f_{\text{no}}(z) \) is determined when both particles are stationary, ensuring their validity at any speed. 
Particularly for particles limited to weak interactions, it is observed
\begin{align}
\label{eq:fno_weak}
    f_{\text{no}}^{weak}(z,\beta_{o} = 0,\beta_{n} = 0) &= 0 \\
    &= g_{00} + g_{kk}  \nonumber
\end{align}
, where k= 8,9 denotes contributions of the two weak dimensions in superposition. The charged leptons show the following.
\begin{align}
\label{eq:fno_lepton}
    f_{\text{no}}^{\text{leptons}}(z,\beta_{o} = 0,\beta_{n} = 0) &= -1 \\
    &= g_{00} + g_{77}  +  g_{kk} \nonumber
\end{align}
For quarks located within a baryon relative to one of the other two accompanying quarks, we find:
\begin{align}
\label{eq:fno_baryon}
    f_{\text{no}}^{\text{baryons}}(z,\beta_{o} = 0,\beta_{n} = 0) &= -2 \\
    &= g_{00} + g_{ii}  + g_{77}  + g_{kk} \nonumber
\end{align}
, and in the context of quarks inside a meson:
\begin{align}
\label{eq:fno_meson}
    f_{\text{no}}^{\text{mesons}}(z,\beta_{o} = 0,\beta_{n} = 0) &= -3 \\
    &= g_{00} + g_{ii}  + g_{jj}  + g_{77}  + g_{kk} \nonumber
\end{align}
, where \( i, j = 4,5,6 \) (with \( i \neq j \)) pertain to active color contributions. The invariant's result remains constant but varies for distinct particle types:
\begin{align}
\label{eq:invariant_fno_summary}
    f_{\mathit{no}}^{\text{weak}}(z) = 0 &&
    f_{\mathit{no}}^{\text{leptons}}(z) = -1 &&
    f_{\mathit{no}}^{\text{baryons}}(z) = -2 &&
    f_{\mathit{no}}^{\text{mesons}}(z) = -3.
\end{align}
This constancy holds provided that the interparticle forces are symmetric. This condition is always satisfied for the strong interaction and for particles whose electric excitation numbers have equal magnitude, $|n_7| = |o_7|$, which in turn implies that the weak excitation numbers also match in magnitude. The only exception is the interaction between a u-quark and a d-quark or their heavy cousins. Even in the ud case, the strong force remains dominant, and the electric and weak forces become relevant only for $z \leq 0.01$ weakening the constancy of $f_{no}$.

\subsection{Coulomb force between two charges moving at low speeds}  

Calculating the electric force between two charges, $q_{1}=\frac{e}{3} \cdot o_{7}$ and $q_{2}=\frac{e}{3} \cdot n_{7}$ is performed for slow movement, where \textbf{$\left| \beta \right| \ll 1$} and \textbf{$f_{no}=-1$} are used.  
The radius expression from Eq. \eqref{eq:electric_radius} simplifies to:
\begin{align}
r_{o_{7}}(z) = \frac{9 \rho \operatorname{signum}(o_{7})}{\alpha_{e} \left(1+\frac{n_{7} \operatorname{signum}(o_{7})}{z} \right)}
\end{align}
$E_{o}^e$ is the energy of Eq. \eqref{eq:electric_mass_no-field}, and it interacts with the Christoffel symbols. Although solving Eq. \eqref{eq:all_forces} completely is intricate, the initial terms of the series expansion in $\rho$ are still tractable:
\begin{align}
\label{eq:develop_Coulomb}
F^{1}(R) &= - \frac{E_{o}^e}{c^2} \cdot \Gamma^1 _{\mu\nu} u^{\mu} u^{\nu} \\
&= - \frac{E_{o}^e}{c^2} \cdot \Gamma_{77}^{1} \cdot c^{2} + \dots \nonumber \\
&= \frac{\alpha_{e} \hbar c n_{7} o_{7}}{9 R^{2}} + \frac{18 \hslash  {| n_{7}|} c o_{7}^{2} \beta_{n} \rho^{3}}{\alpha_{e} R^{5}} + O(\rho^4) \nonumber
\end{align}
where $R = \rho \cdot z$ is the physical distance in meters and $\alpha_{e} = \frac{e^2}{4 \pi \epsilon_0 \hbar c}$. Since $\rho$ is of  the order of a femtometer, for distances greater than $10^{-13} \mathrm{~m}$, the first term in Eq. \eqref{eq:develop_Coulomb} dominates.  This term is equivalent to the well-known Coulomb force formula from classical electrostatics:
\begin{align}
F = \frac{q_{o} q_{n}}{4 \pi  \epsilon_{0} R^{2}}
\end{align}
where $q_{o}$ and $q_{n}$ represent the two charges, and $\epsilon_{0}$ is the vacuum permittivity.  

\subsection{Forces and their asymptotic behavior}  
\label{sec:Force_calculation_and_asymptotic}
\begin{figure}
    \begin{center}
    \includegraphics[width=1\textwidth]{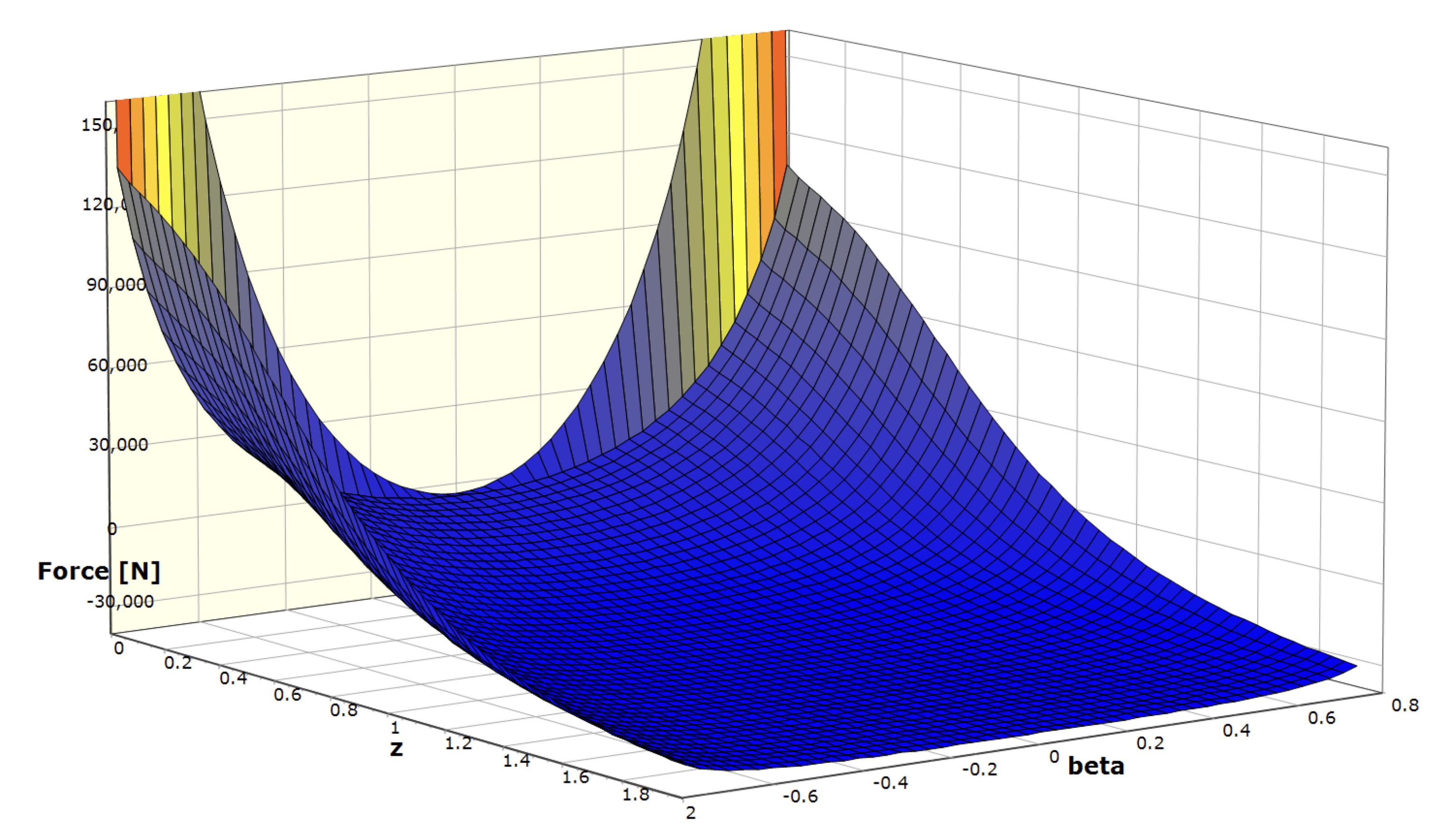}
    \end{center}
    \caption{Interactions between $ud$ quarks inside baryons: electromagnetic and weak forces start to play an important role for $z < 0.01$.}
    \label{fig:Figure_StrongForce}
\end{figure}

\begin{figure}
    \begin{center}
    \includegraphics[width=1\textwidth]{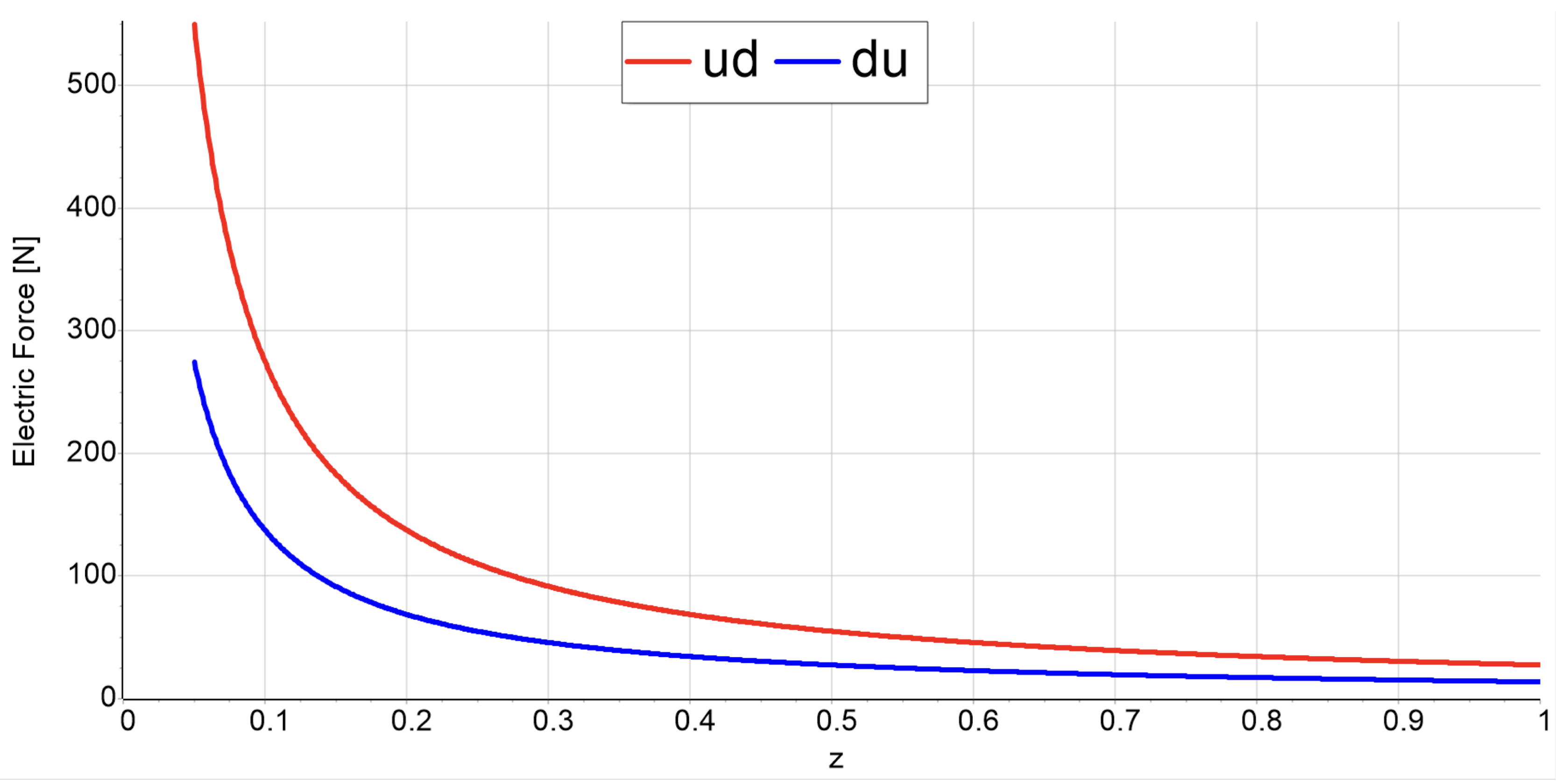}
    \end{center}
    \caption{Electric force exerted on a u-quark when it interacts with a d-quark, and vice versa, within a baryon. Shown here for $\beta=0.44$.}
    \label{fig:Figure_ElectricForceAsymmetrical}
\end{figure}

By establishing the invariant expression $f_{no}(z)$ (refer to Eqs. \eqref{eq:invariant_fno} and \eqref{eq:invariant_fno_summary}) and acknowledging its uniformity across particle types, the model provides the equations of the strong, electric, and weak forces. The full analytical forms of the forces are rather lengthy. The strong force is given by
\begin{align}
\label{eq:Strong_force_full}
\includegraphics[width=1\textwidth]{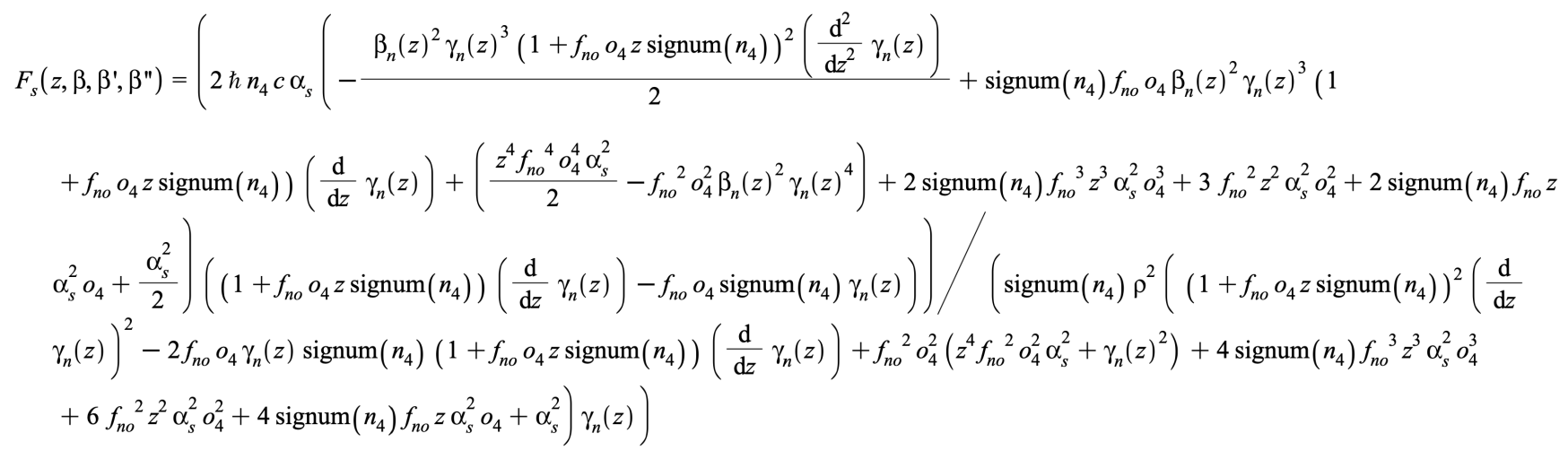}
\end{align}
, the electric force by
\begin{align}
\label{eq:Electric_force_full}
\includegraphics[width=1\textwidth]{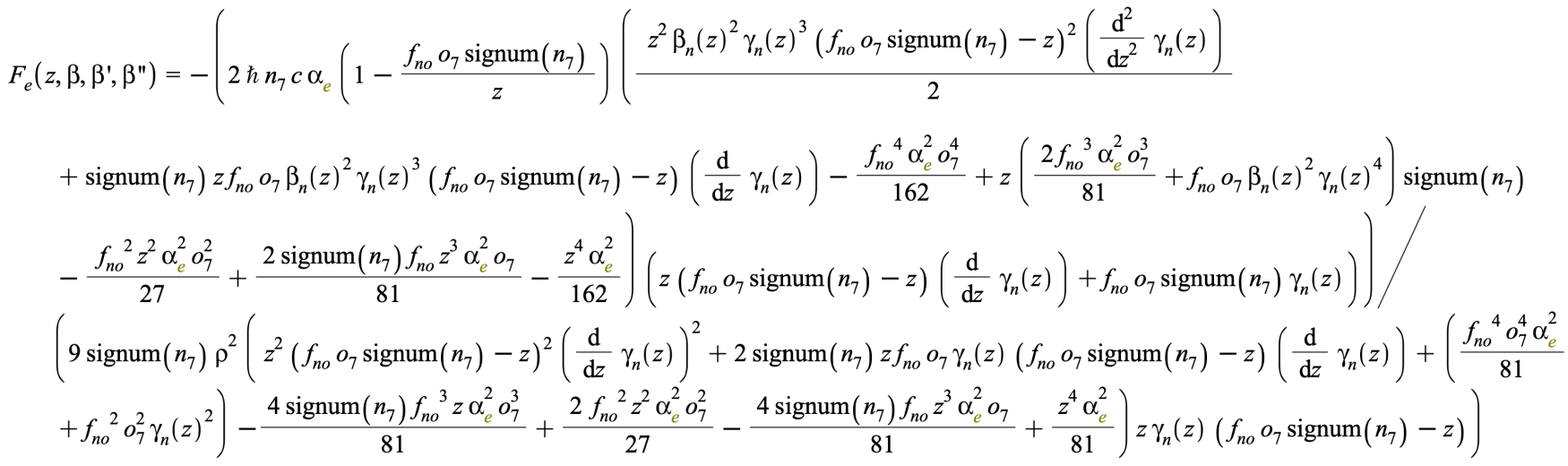}
\end{align}
, and the weak force by
\begin{align}
\label{eq:Weak_force_full}
\includegraphics[width=1\textwidth]{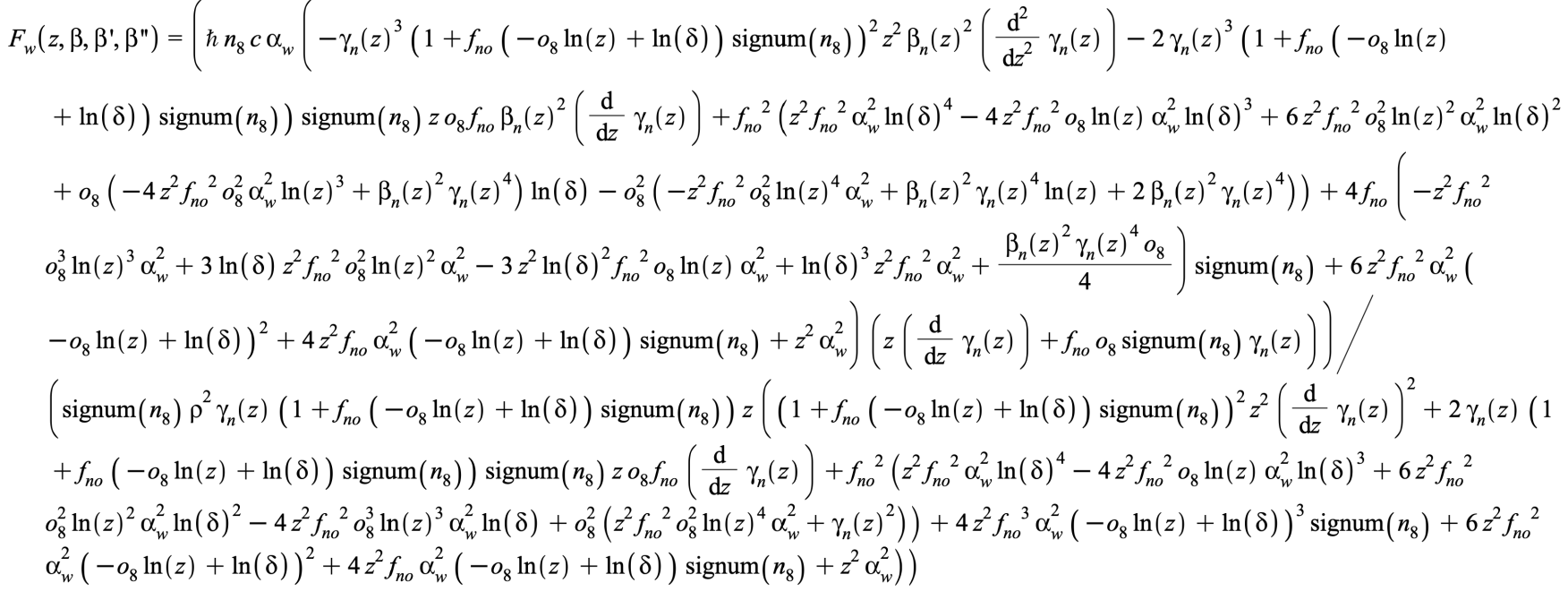}
\end{align}
Their asymptotic behavior shows that the electric and weak forces go to zero as $z \to \infty$, while the strong force tends to a finite, non-vanishing value in that limit. 
\begin{itemize}
    \item $\lim \limits_{z \to \infty}F_{s} \! \left(z , \beta , \frac{d}{d z}\beta \! \left(z \right), \frac{d^{2}}{d z^{2}}\beta \! \left(z \right)\right)
 = -\frac{\alpha_{s} \hslash c f_{\mathit{no}} n_{4} o_{4}}{\rho^{2}}
$
    \item $\lim \limits_{z \to \infty}F_{e} \! \left(z , \beta , \frac{d}{d z}\beta \! \left(z \right), \frac{d^{2}}{d z^{2}}\beta \! \left(z \right)\right)
 = 0$
    \item $\lim \limits_{z \to \infty}F_{w} \! \left(z , \beta , \frac{d}{d z}\beta \! \left(z \right), \frac{d^{2}}{d z^{2}}\beta \! \left(z \right)\right)
 = 0$
\end{itemize}
However, residual strong forces have a very limited range, where the force is $F_s^{residual}(z \ge 5) \approx 0$.  Each of the three forces depends on $z$, the normalized separation between the two constituent particles, as well as on the particle’s velocity and on the first and second derivatives of this velocity with respect to $z$. Owing to these interdependencies, which cause the various interactions to influence one another, it is difficult to construct individual force–distance curves for each interaction in isolation. Consequently, Fig. \ref{fig:Figure_StrongForce} 
presents the total combined strong, electric, and weak forces for a $ud$ interaction inside a baryon as functions of $z$ and $\beta$. An intriguing outcome is the asymmetric impact of electromagnetic and weak interactions on d- and u-quarks inside nucleons (see Fig. \ref{fig:Figure_ElectricForceAsymmetrical}). 
In a representation of the electric force for quarks at $\beta = 0.44$ relative to the center of gravity, the difference in force reaches approximately $137\,\mathrm{N}$ at $z = 0.1$, whereas the corresponding disparity in the weak force at the same position is only about $0.12\,\mathrm{N}$. These values should be interpreted with caution, as they are derived from a specific configuration. Nevertheless, they indicate a pronounced asymmetry in the electromagnetic forces acting on ud-quark configurations. This asymmetry becomes evident when one inspects the exact expression for the force, which involves different excitation numbers for the $u$-quark and the $d$-quarks, thereby producing different force values for the two quark types. In contrast to the explicitly specified distances in this model, quantum field theory describes particles in terms of wavefunctions, which implies that a particle cannot be assigned an exact position in space. Currently, it is still an open question how the predictions of this classical description and the formulation of quantum field theory can be integrated into a single, coherent interpretational framework.

\subsection{Deceleration in lepton collisions}
\label{ep-deceleration}
\begin{figure}
    \begin{center}
    \includegraphics[width=1\textwidth]{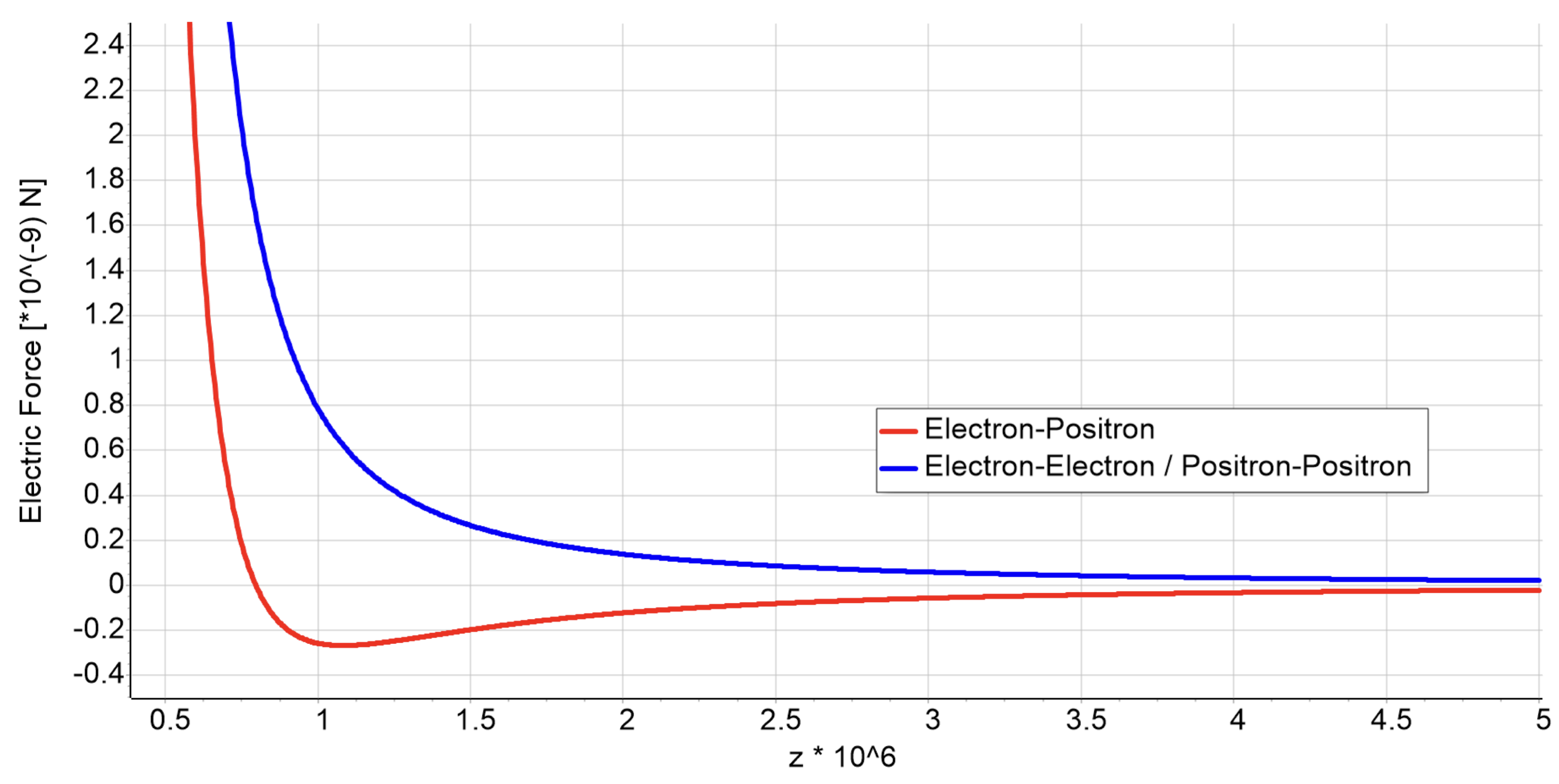}
    \end{center}
    \caption{Electromagnetic interactions involving $e^+e^-$ pairs and $e^+e^+/e^-e^-$ pairs at an invariant mass of $434 \, \text{MeV}/c^2$ are depicted for the range $4.5 \times 10^{5} \le z \le 5 \times 10^{6}$, where deceleration sets in.}
    \label{fig:FElectronPositron_at10^6}
\end{figure}

\begin{figure}
    \begin{center}
    \includegraphics[width=1\textwidth]{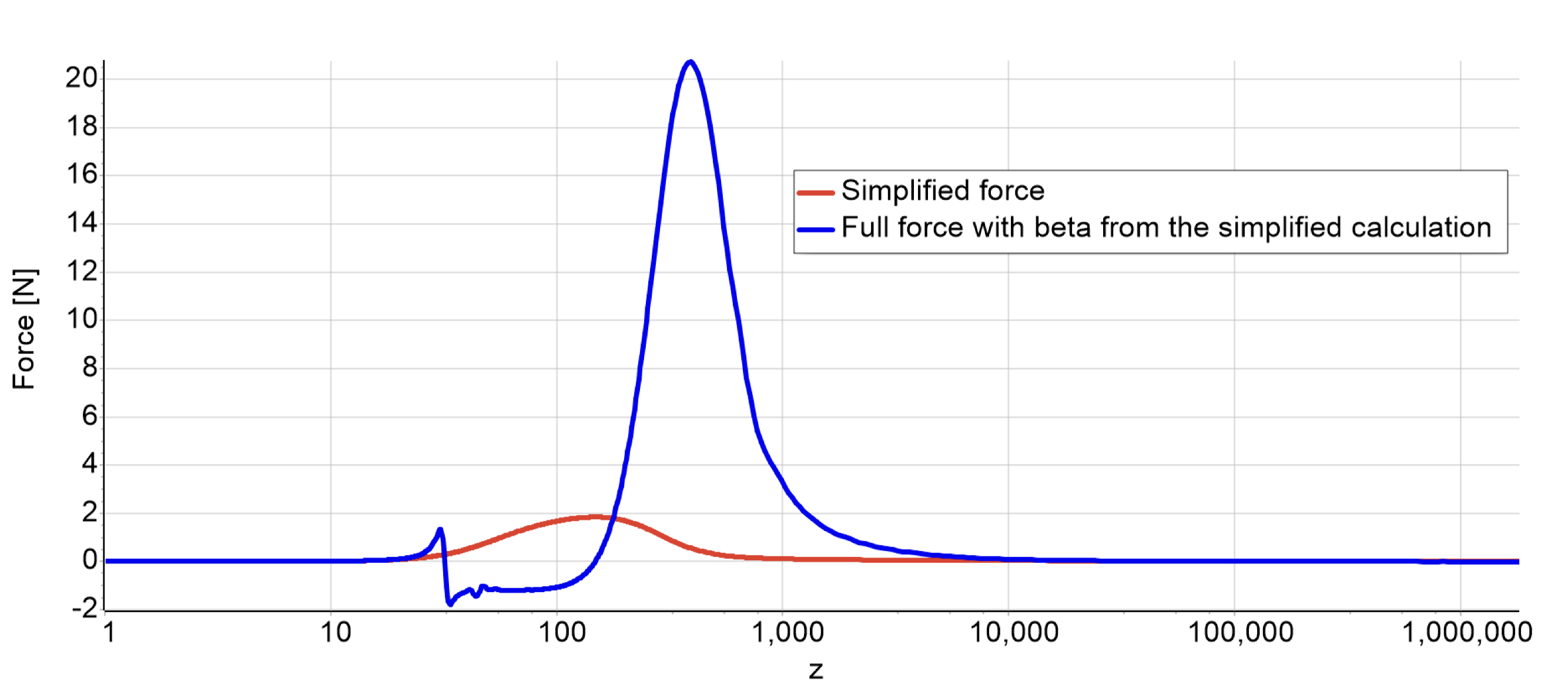}
    \end{center}
    \caption{Simplified versus approximated full force for $z \ge 1$. The blue curve is obtained by incorporating both the first and second derivatives of the velocity with respect to $z$ from the simplified calculation, so that the full force expression is evaluated. This results in a strongly overestimated force profile that also shows artifacts caused by the numerical evaluation of the simplified solution. However, it can clearly be seen that the onset of the force occurs at larger $z$ values and is more pronounced than predicted by the simplified model.}
    \label{fig:Figure_FullVersusSimplifiedForces}
\end{figure}

\begin{figure}
    \begin{center}
    \includegraphics[width=1\textwidth]{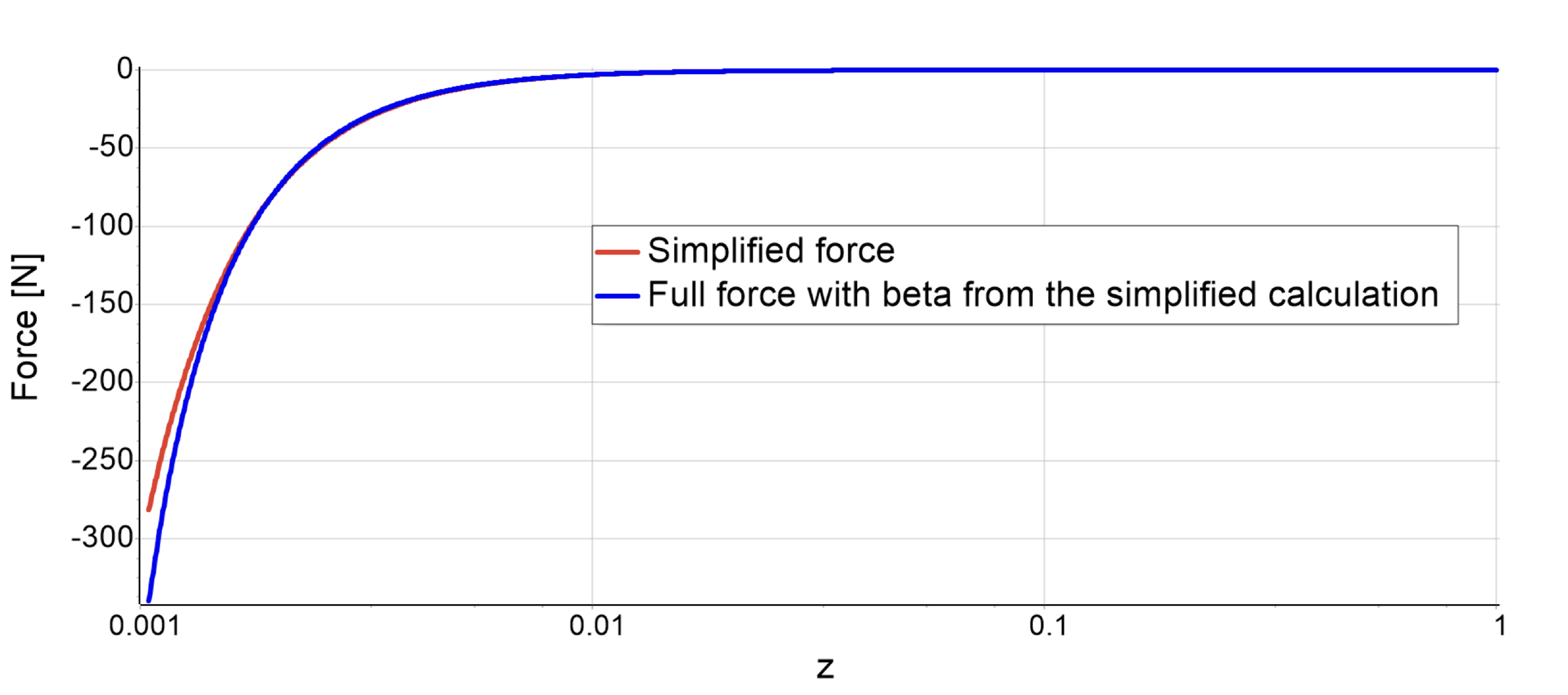}
    \end{center}
    \caption{Comparison of the simplified force with the approximated full force in the interval $10^{-3} < z \le 1$. }
    \label{fig:Figure_FullVersusSimplifiedForces_z<1}
\end{figure}

\begin{figure}
    \begin{center}
    \includegraphics[width=1\textwidth]{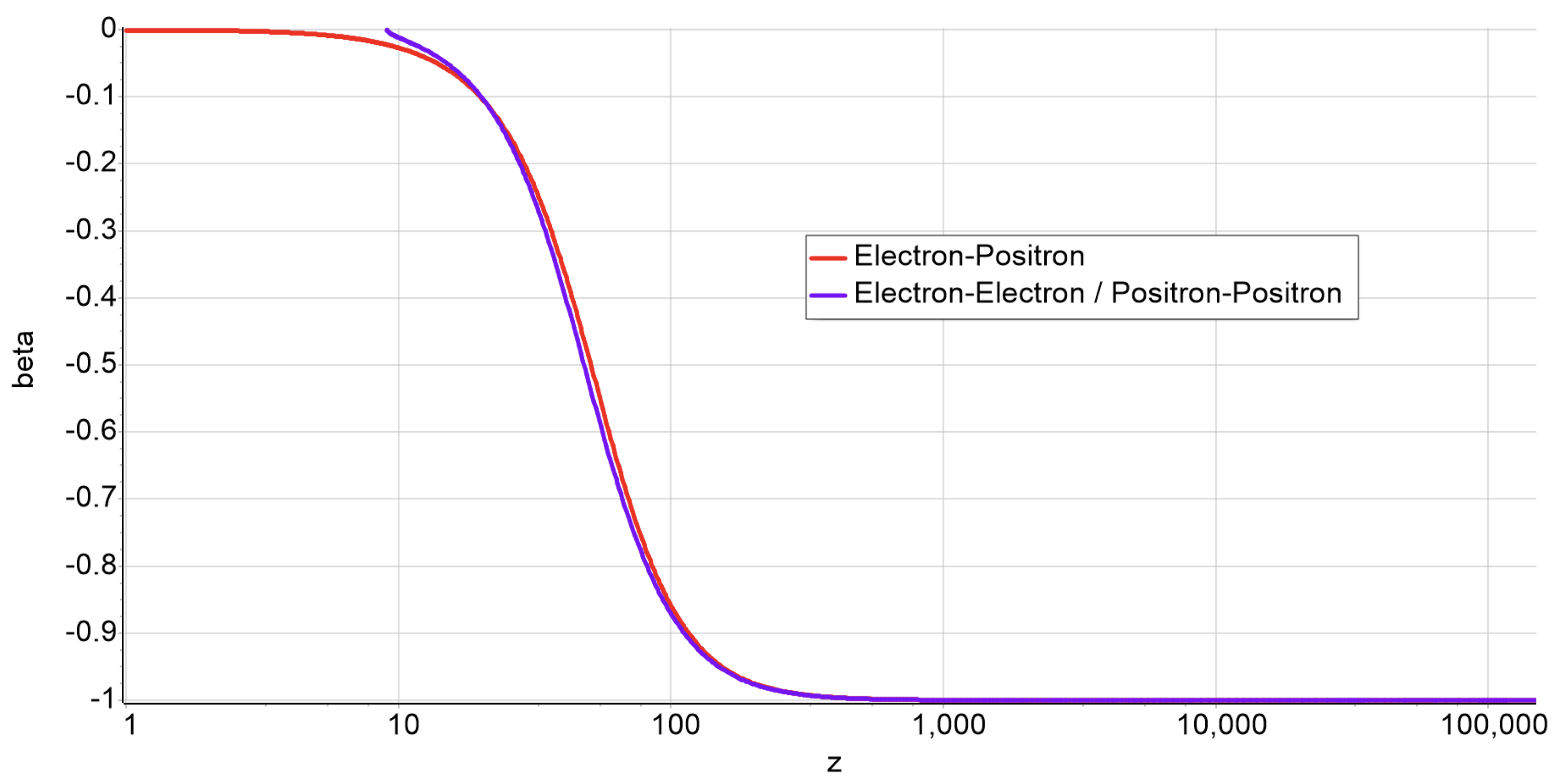}
    \end{center}
    \caption{Normalized velocity $\beta$ as a function of $z$ of the simplified equation with relativistic velocity $\beta_{0}$ at $z=10^{16}$ for $e^- e^- / e^+ e^+$ and $e^- e^+$ combinations.}
    \label{fig:Figure_SimpleForce_ee_pp_ep}
\end{figure}

\begin{figure}
    \begin{center}
    \includegraphics[width=1\textwidth]{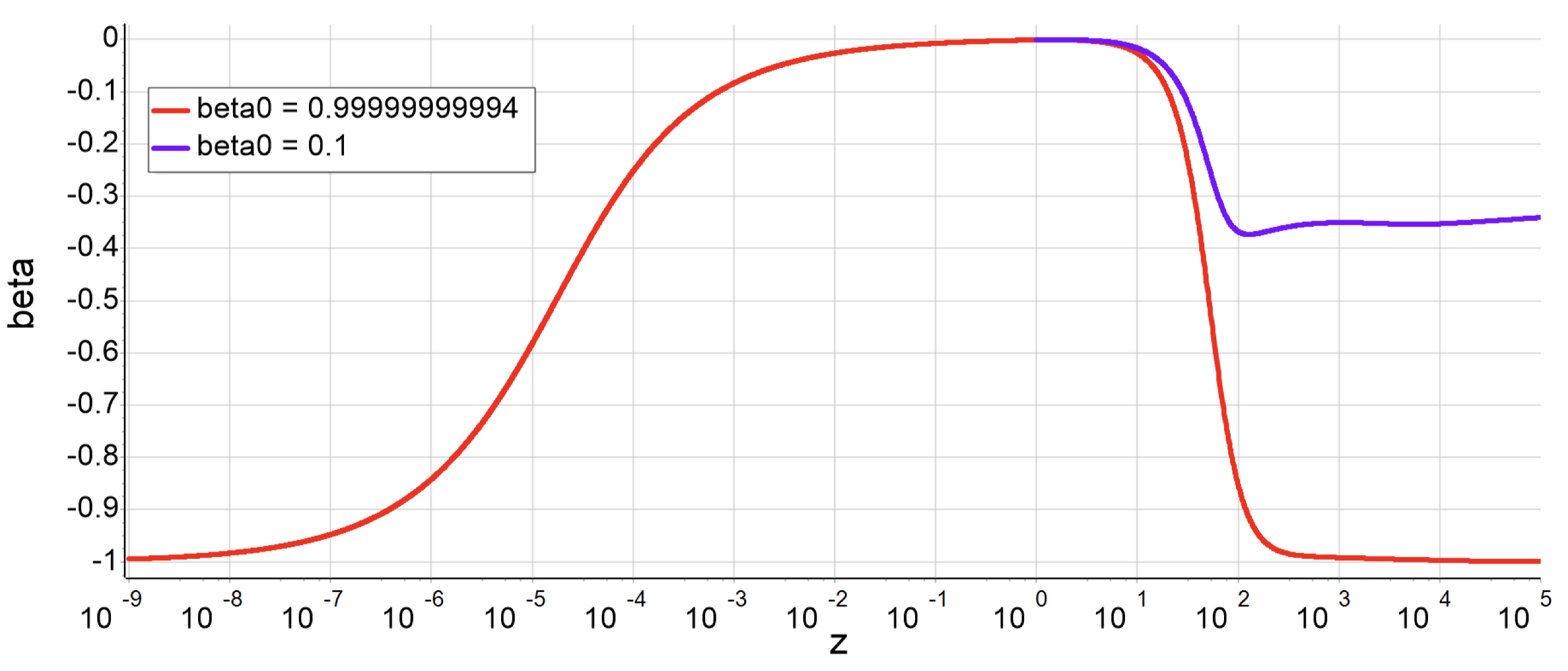}
    \end{center}
    \caption{Normalized velocity $\beta$ as a function of $z$ of the simplified equation with two different $\beta_{0}$-values at $z=10^{16}$ for $e^+e^-$ collisions.}
    \label{fig:Figure_SimpleBeta}
\end{figure}

\begin{figure}
    \begin{center}
    \includegraphics[width=1\textwidth]{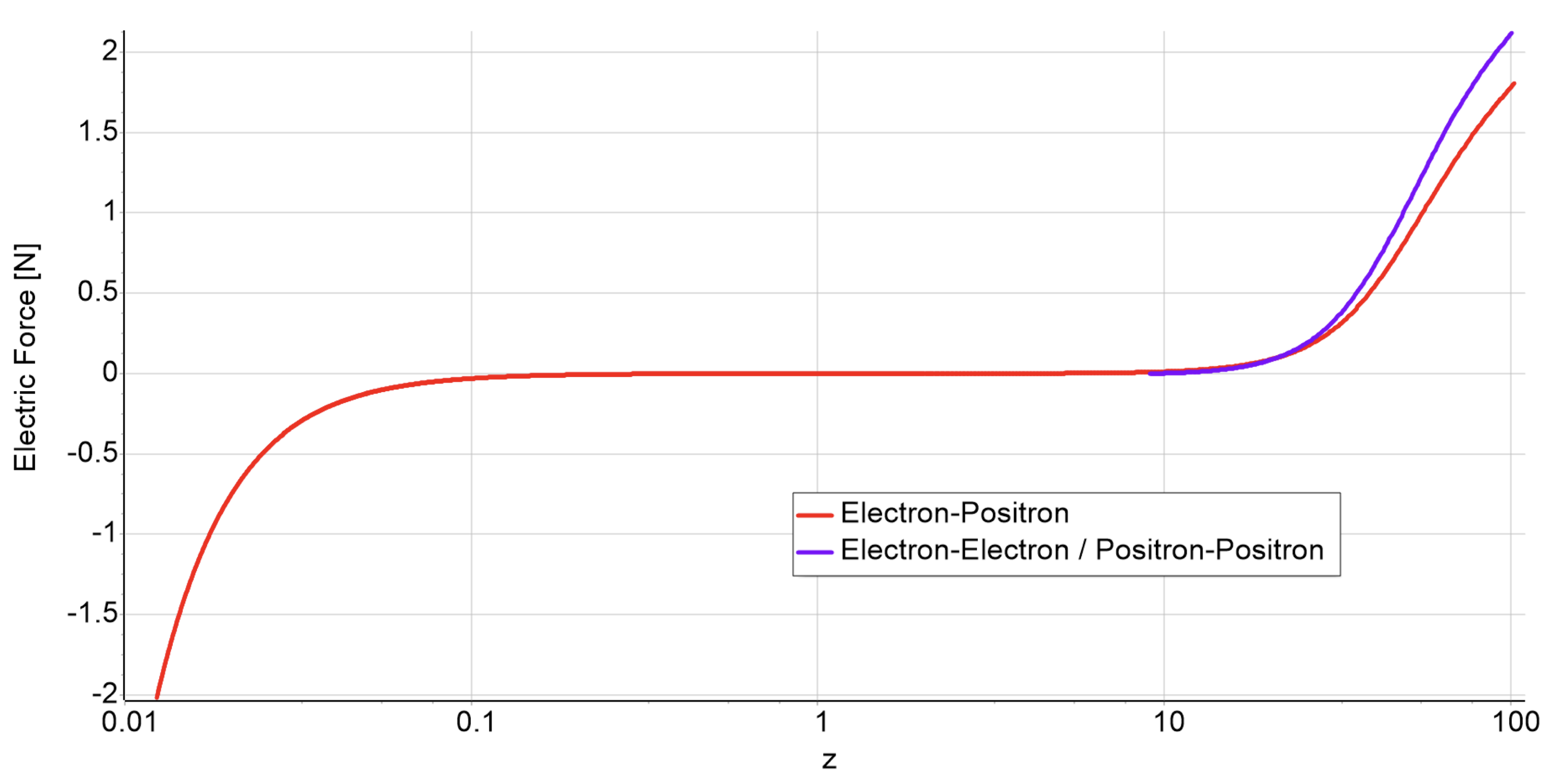}
    \end{center}
    \caption{Within the interval $10^{-2} < z \leq 100$, the electromagnetic interactions involving $e^+e^-$ and combinations of $e^+e^+/e^-e^-$ are examined. At approximately $z = 9$, the $e^+e^+/e^-e^-$ undergo reflection, whereas the $e^+e^-$ pairs experience an onset of negative forces for $z$ values between $1$ and $2$. For $z \leq 10$, these force profiles are independent of the initial invariant mass; however, the corresponding potential energies vary with the invariant mass.}
    \label{fig:FElectronPositron_at10^-15}
\end{figure}

An unforeseen outcome is the repulsion observed between particles with opposite electric charges when they are in close proximity and moving at velocity. Consider an example of a collision $e^+ e^-$ in comparison with a collision of $e^+e^+/e^-e^-$ occurring at relativistic velocity $\beta_{0} = 0.999997$ at $z=10^{16}$, corresponding to an invariant mass of $434\, MeV/c^2$ (Fig \ref{fig:FElectronPositron_at10^6}). 
At about $z=8 \cdot 10^{5}$ deceleration sets in for $e^+e^-$ leading to a behavior very much the same as in the scattering of identical leptons. The further progress of the two force curves depends in a large way on the normalized velocity $\beta$, which depends on the decelerating force. Within this context of lepton-lepton encounters, the resulting force on a lepton is derived from the additive effects of the electric force, both determined based on the lepton's distinct parameters. 
\begin{align}
\label{eq:e+e-Electric_force_full}
\includegraphics[width=1\textwidth]{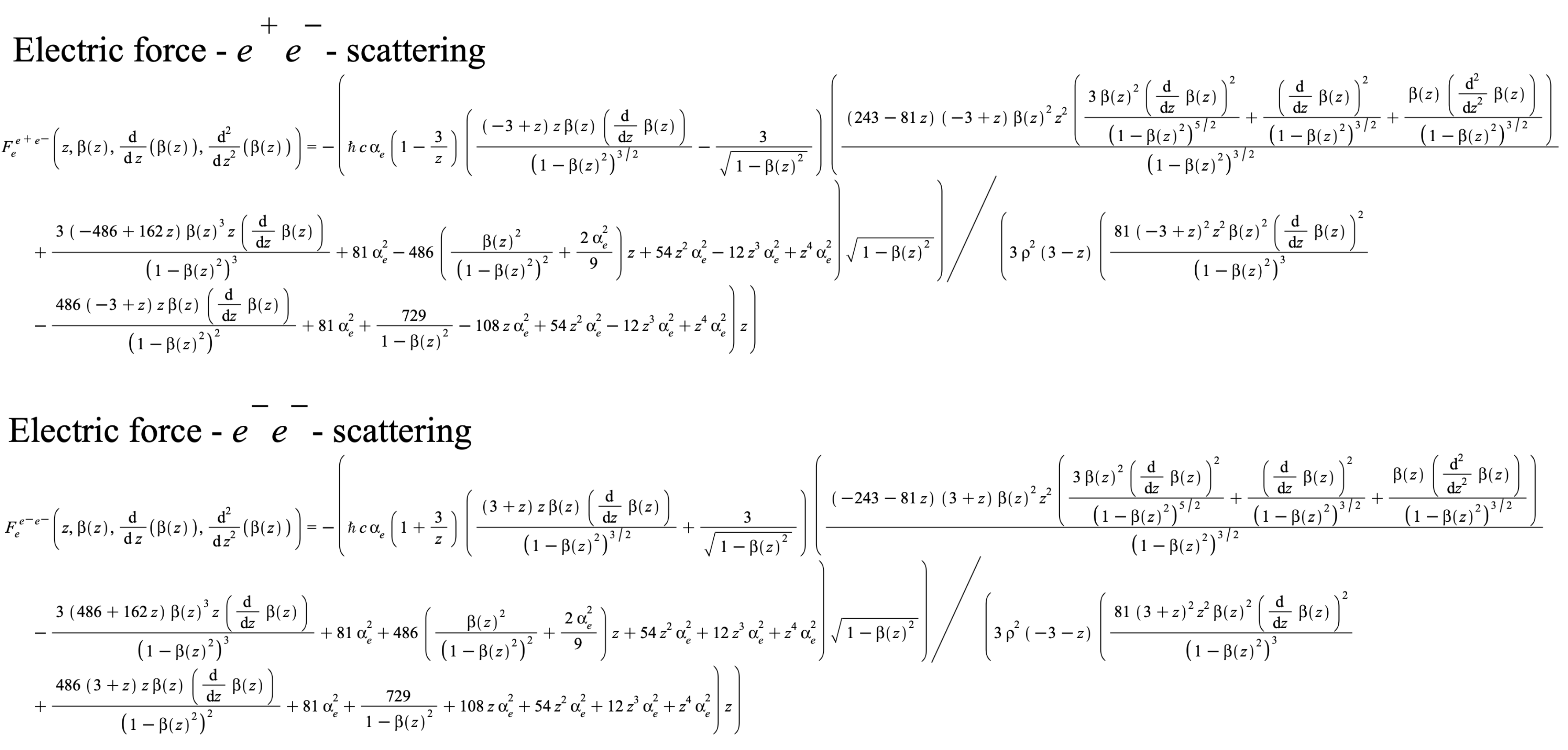}
\end{align}
and the weak force 
\begin{align}
\label{eq:e+e-Weak_force_full}
\includegraphics[width=1\textwidth]{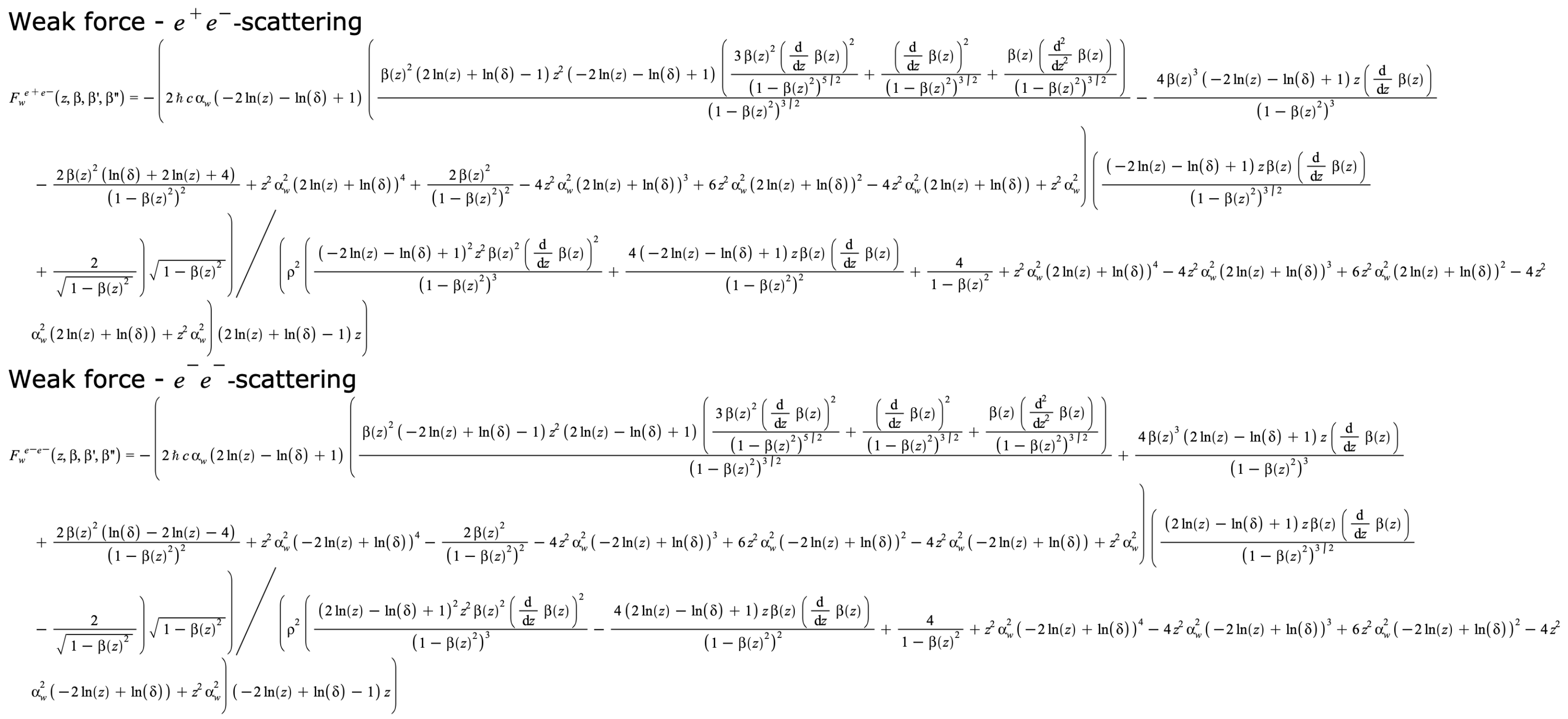}
\end{align}
According to the Standard Model, the weak force is characterized by limited range. In the present framework, the influence of the weak force is also minimal for $z>1$, whereas for $z<1$, its impact can become significant. The full force is

\begin{align}
F \! \left(z , \beta(z) , \frac{d \beta}{d z} \! \left(z \right), \frac{d^{2}\beta}{d z^{2}}\! \left(z \right)\right)  = F_{e} \! \left(z , \beta , \beta', \beta'' \right)  + F_{w} \! \left(z , \beta , \beta', \beta''\right).
\end{align}
This force corresponds to the time derivative of the momentum, yielding a nonlinear second-order differential equation. 
\begin{align}
\label{eq:nonlinearBeta}
F \! \left(z , \beta(z) , \frac{d \beta}{d z} \! \left(z \right), \frac{d^{2}\beta}{d z^{2}}\! \left(z \right)\right)  = \frac{dp}{dt}
 = \frac{d}{dt}\left(m_{e} c \beta  \gamma \right) \\
 = m_{e} c \left(\gamma^{3} \beta^{2}+\gamma \right) \cdot \frac{d \beta}{d t} 
 = m_{e} c \left(\gamma^{3} \beta^{2}+\gamma \right) \cdot \frac{c \beta}{\rho} \cdot \frac{d \beta}{d z} 
 \nonumber
\end{align}
The second-order nonlinear differential equation under consideration lacks both a known analytical solution and a suitable numerical technique to solve it. Consequently, the challenge is to derive an approximation that enables the determination of the correct $\beta(z)$ for certain values of $z$ in the interval ranging from infinity to zero. Our method starts by setting both the first and second derivatives in the force from \eqref{eq:nonlinearBeta} to zero. We then proceed to numerically calculate the reduced form of the equation from $z=10^{16}$ to $z=1$ and from $z=1$ to $z=10^{-9}$ using MATLAB. Thereafter, the computation of the first and second derivatives is done numerically with StatFree. With the established $\beta$ and its derivatives, the full equation is evaluated, keeping in mind that the obtained results \textbf{do not} constitute the solution to the nonlinear differential equation! The results are illustrated in Figs \ref{fig:Figure_FullVersusSimplifiedForces} 
and \ref{fig:Figure_FullVersusSimplifiedForces_z<1}. 
Several insights can be gleaned from these graphs. Firstly, the two equations produce the same results for $z > 10^5$. Secondly, the comprehensive force results in a more substantial reduction in velocity compared to the simplified model. The accurate force, when accounting for the precise $\beta$ and its derivatives, should notably influence the system earlier than the blue trajectory suggests. The full solution induces an earlier deceleration relative to the simplified form, with the area beneath its curve matching that beneath the simplified trajectory. This is because the simplified curve is the minimal possible deceleration, but for the real deceleration setting in earlier, the correct $\beta$ converges with the curve of the minimal possible deceleration at some $z$. As the lepton velocity decreases, the derivatives near zero, leading both the complete and approximated full force to converge with the result of the simplified expression. In these conditions, the simplified equation provides a satisfactory approximation of the nonlinear differential Eq. \eqref{eq:nonlinearBeta} for the range $10^{-1} < z < 10$ and when $z > 10^5$. The scenario differs in the context of $e^+e^+/e^-e^-$ or $e^+e^-$ collisions (Fig \ref{fig:Figure_SimpleForce_ee_pp_ep}). 
Although the curves do not precisely typify the true $\beta$, they offer a suitable approximation over the specified interval. Identical-type leptons (depicted by the blue curve) cease their motion around $z \approx 9$, which hinders them from coming close enough to transform into different particles, thus leading to Møller scattering. In contrast, electron-positron interactions exhibit a distinct dynamic. Their motion diminishes as the smallest $\beta$, valued below $5 \cdot 10^{-4}$, is found within the range $1 < z < 2$, consistently equaling across all curves regardless of the initial $\beta_{0}$. This situation suggests that electron-positron pairs potentially convert at $z \approx 1$ into bosons, leptons, mesons, baryons, or sterile bosons, as dictated by the energy of the initial electron-positron pair. The outcomes of the simplified computation illustrating $\beta$'s behavior are depicted in Fig \ref{fig:Figure_SimpleBeta}. 
The diagram demonstrates $\beta$ for beams with varying initial energies. Regardless of these initial energy differences, the beams achieve an identical velocity around $z \approx 1$. For the interval $10^{-1} < z < 10$, $\beta$ simplified and exact appear nearly identical. However, the precise curves on the right should exhibit a considerably flatter profile, whereas the curve on the left of this interval would present a significantly steeper decline. The forces of the first generation same charge leptons and leptons of opposite charge scattering are depicted in Fig \ref{fig:FElectronPositron_at10^-15}).

\subsection{Particle transformation in collision processes}
\label{collision processes}
Up to this point, the discussion of collision processes with the exception of $e^- e^+$ has been omitted in this article. In the Standard Model of Particle Physics, particle formation is explained via boson exchange mechanisms (involving photons, $Z$, $W$, and Higgs bosons) or through vector-meson exchange. Within this Kaluza-Klein model, transformation of particles occurs when the colliding particles come almost to rest and merge. A commonly recognized process is the conversion of a $e^+e^-$ pair into a $\pi^+\pi^-$ pair, described by:
\begin{align}
e^{+} + e^{-} \rightarrow \gamma^* \rightarrow \pi^{+} + \pi^{-} \nonumber
\end{align}
accompanied by the excitation table:
\begin{align}
\begin{array}{c}
\left(\begin{array}{c}
0 \\
0 \\
0 \\
3 \\
(-2) \\
(0)
\end{array}\right) +
\left(\begin{array}{c}
0 \\
0 \\
0 \\
-3 \\
(2) \\
(0)
\end{array}\right)
\rightarrow
\left(\begin{array}{c}
n-n \\
0 \\
0 \\
3-3 \\
(-2)+(2) \\
(0)+(0)
\end{array}\right) \nonumber \\
\rightarrow
\left(\begin{array}{c}
1 \\
-1 \\
0 \\
2 \\
(-1) \\
(0)
\end{array}\right)
\left(\begin{array}{c}
-1 \\
1 \\
0 \\
1 \\
(-2) \\
(0)
\end{array}\right) +
\left(\begin{array}{c}
1 \\
-1 \\
0 \\
-1 \\
(2) \\
(0)
\end{array}\right)
\left(\begin{array}{c}
-1 \\
1 \\
0 \\
-2 \\
(1) \\
(0)
\end{array}\right)
\end{array}
\end{align}
As demonstrated in Figs \ref{fig:FElectronPositron_at10^6} and \ref{fig:FElectronPositron_at10^-15}, collisions between electrons and positrons exhibit a repulsive interaction, resulting in deceleration and eventual transmutation.

\subsection{Asymmetric forces in electric and weak interactions of \texorpdfstring{\(ud\)}{ud}-hadrons}
\label{random walk}
In $ud$-quark configurations, the electric interaction, along with the weak force, presents an asymmetrical behavior. Consequently, this results in a difference in the force experienced by a $d$-quark compared to that on a $u$-quark (see Fig \ref{fig:Figure_ElectricForceAsymmetrical}). These quarks undergo a form of \textit{Zitterbewegung} that is distinct from the electron’s \textit{Zitterbewegung} \cite{RN222}, and can best be characterized as a form of \textit{random walk} \cite{RN225}.

\subsection{Two possible laboratory experiments and one astronomical measurement to assess the model}

The Kaluza-Klein model offers insights regarding sterile particles and force interactions. To evaluate the model's validity, two laboratory experiments and a cosmological observation are proposed: 
\subsubsection{Creation of sterile bosons}
\label{disc:Creation of dark matter}
Dynamic interactions between particles are expected to generate sterile bosons \eqref{eq:sterile_particle}. In this framework, the rapid incoming particles decelerate to unite and form sterile bosons. The process could yield sterile bosons with contributions principally in any of the lines of the column below. However, sterile bosons are expected to form primarily in a way similar to the creation of second- and third-generation leptons $e^{+} + e^{-} \rightarrow  l^++l^-$. Therefore, there might be a preference for storing energy in the dimensions attributed to the strong force. 
\begin{align}
e^{+} + e^{-} \rightarrow S \nonumber
\end{align}
with the excitation table:
\begin{align}
\begin{array}{c}
\left(\begin{array}{c}
0 \\
0 \\
0 \\
3 \\
(-2) \\
(0)
\end{array}\right) +
\left(\begin{array}{c}
0 \\
0 \\
0 \\
-3 \\
(2) \\
(0)
\end{array}\right)
\rightarrow
\left(\begin{array}{c}
n-n \\
0 \\
0 \\
3-3 \\
(-2)+(2) \\
(0)+(0)
\end{array}\right) \nonumber 
\end{array}
\end{align}
The mass of the sterile boson $S$ is expected to be
\begin{align}
\label{eq:experiment_sterile_particle}
m_{S} &= \frac{\hbar}{c \rho} \left[ \alpha_{s} \cdot 2 \cdot n
+ \frac{\alpha_{e}}{9} \cdot 6 + \alpha_{w} \cdot 4 \right] 
\end{align}
, neglecting possible energy deposition in the electric and weak sectors because of the smallness of the respective coupling constants. To ensure the stability of a sterile boson, the invariant mass of the initial electron-positron pair must match the mass permissible for a sterile boson; otherwise, it would decay into particles that interact through strong and/or electromagnetic and weak forces. Table \ref{tab:invariant_masses_sterile_particles} lists the invariant masses associated with the sterile bosons generated in experiments where the initial mass is above $200 \, MeV/c^2$. 
\begin{table}[h]
    \caption{Masses of sterile bosons}
    \centering
    \begin{tabular}{|c|c|c|}
        \hline
        A& $m [kg]$ & $m[MeV/c^2]$ \\ \hline
        2 = $A_s$& $4.0222 \cdot 10^{-28}$ & $225.62$ \\ \hline
        4& $8.0262 \cdot 10^{-28}$ & $450.24$ \\ \hline
        6& $1.2030 \cdot 10^{-27}$ & $674.84$ \\ \hline
        8& $1.6034 \cdot 10^{-27}$ & $899.45$ \\ \hline
        10& $2.0038 \cdot 10^{-27}$ & $1124.1$ \\ \hline
        12 = $A_c$& $2.4042 \cdot 10^{-27}$ & $1348.7$ \\ \hline
        14& $2.8046 \cdot 10^{-27}$ & $1573.3$ \\ \hline
        16& $3.2050 \cdot 10^{-27}$ & $1797.9$ \\ \hline
        18& $3.6054 \cdot 10^{-27}$ & $2022.5$ \\ \hline
        20& $4.0059 \cdot 10^{-27}$ & $2247.1$ \\ \hline
        & & \\ \hline
        42 = $A_b$& $8.4102 \cdot 10^{-27}$& $4717.8$\\ \hline
    \end{tabular}
    \label{tab:invariant_masses_sterile_particles}
\end{table}
An experiment designed to probe the energy range is expected to reveal an enhancement in the missing mass spectrum close to the resonances at those particular mass values. It remains uncertain whether resonances appear only at masses corresponding to the numerical values of the flavor constants, or whether sterile boson production occurs at every even-valued mass instead.

\subsubsection{Nucleon's random walk}
\label{disc:NucleonsRandomWalk}
Inside nucleons, we observe quark compositions made of one u-quark and one d-quark, primarily bound by the strong force but also subject to electric and weak forces. The strong interaction exerts an equal influence on both quarks, whereas the electric and weak forces exert differing degrees of force. Consequently, nucleons may undergo a sort of erratic motion \cite{RN225}. This dynamic differs from Brownian motion \cite{RN231}, which describes how a fluid exerts thermodynamic effects on a suspended small object. The movement of nucleons occurs on a much smaller scale $s\approx \rho$, with their frequency given by $f = \frac{c \beta}{2\rho} \approx 10^{23} \text{ Hz}$. This renders directly measuring an individual nucleon improbable. Nevertheless, a group of ultra-cold polarized neutrons could exhibit some secondary random walk behavior.

\subsubsection{Decreasing strength of dark energy}
\label{disc:DecreasingDarkEnergy}
The process of hadronization, which involves the formation of protons and neutrons, began approximately $10^{-6}$ seconds following the Big Bang \cite{RN226,RN227,RN228}. If protons and neutrons, as hypothesized, engage in a random walk (see \textit{\nameref{random walk}}), this behavior might contribute to galaxy clusters drifting apart. Although the exact timeline and specifics are not covered in this discussion, if the random walk is essentially synonymous with what is termed dark energy, then it must diminish as matter transitions into sterile bosons (see \textit{\nameref{Sterile elementary Particles}}), essentially becoming dark matter. Consequently, the proportion of matter to dark matter would shift in favor of dark matter. This would lead to a reduction in the rate of galactic acceleration. The initial indication of a decreasing acceleration is offered by DESI \cite{desicollaboration2024desi2024vicosmological}. Additional information is expected to be obtained from the Euclid space telescope, the Vera C. Rubin Observatory, and Nancy Grace Roman Space Telescope when becoming operational.  

\subsection{Testable quantities}
\label{subsec:Testable quantities}
The model predicts quantities (Table \ref{tab:Predicted-quantities}), which might help to review the model. 
\begin{table}[h]
    \caption{Predicted values along with links to the corresponding sections of the paper}
    \centering
    \begin{tabular}{|c|c|c|c|}
        \hline
        No.&Predicted quantity& Numerical values & Link \\ \hline
        1&Neutrino distribution ratio & $\delta=1.0000029100(64)$ & \ref{subsec:Weak_coupling} \\ \hline
        2&z-values& \makecell{$z_{s\_Mesons} = 0.0370376(43)$ \\ $z_{v\_Mesons} = 0.1509(24)$ \\ $z_{\_Baryons} = 0.03880668(59)$} & \ref{subsec:strong_coupling_and_zValues} \\ \hline
        3&Gluon mass & $m_g=449.21473(50) \frac{MeV}{c^2}$ & \ref{subsec:Gluon_mass}\\ \hline
        4&\makecell{Hadron masses \\ 102 calculated \\ 71 measured}& Tables \ref{tab:Scalar_mesons}, \ref{tab:Vector_mesons}, \ref{tab:hbar2_baryons}, \ref{tab:3hbar2_baryons} & \ref{subsec:Hadron_mass} \\ \hline
        5&Sterile bosons& Table \ref{tab:invariant_masses_sterile_particles} & \ref{Sterile elementary Particles}, \ref{disc:Creation of dark matter} \\ \hline
        6&Nucleons' random walk& $f = \frac{c \beta}{2\rho} \approx 10^{23} \text{ Hz}$ & \ref{disc:NucleonsRandomWalk} \\ \hline
        7&$e^-e^-$ stop & $z\approx9$ & \ref{ep-deceleration} \\ \hline
        8&$e^+e^-$ deceleration& $\beta_{min} < 5\cdot10^{-4}$ & \ref{ep-deceleration} \\ \hline
        9&Massive photons& \makecell{$m_{2. gen}=0.0086(1)\frac{eV}{c^2}$ \\ $m_{3. gen}=0.0501(3)\frac{eV}{c^2}$} & \ref{subsubsec:Massive_photons} \\ \hline
    \end{tabular}
    \label{tab:Predicted-quantities}
\end{table}
\begin{enumerate}
    \item In this framework, the electron mass $m_{e^{-}} = \frac{\hbar}{c \rho} \left[ \frac{\alpha_{e}}{3} + 2 \alpha_{w} (1 - \ln (\delta)) \right]$ in Eq. \eqref{eq:electron_mass_no_mag} depends on $\delta = \frac{z_{\nu}}{z_{\bar{\nu}}} = \left(\frac{\rho_{\bar{\nu}}}{\rho_{\nu}}\right)^{\frac{1}{3}}$ as given in Eq. \eqref{eq:delta_neutrino_density}. A uniform background density of neutrinos $\rho _{\nu}$ and antineutrinos $\rho_{\overline{\nu}}$ is assumed to set the numerical value of $\delta$ and, consequently, the resulting electron mass. The relations above are unaffected by neither the neutrino/antineutrino species nor their individual energies. However, while the absolute neutrino and antineutrino densities are not known, their ratio is. If particle densities in the low-energy sea are not several orders of magnitude larger than the excess antineutrino density $\rho_{R\overline{\nu}}$ near a nuclear power reactor, then the electron mass in the vicinity of the reactor could, in principle, be considerably less than the standard value, establishing a new $\delta_R=\left(\frac{\rho_{\overline{\nu}}+\rho_{R\overline{\nu}}}{\rho _{\nu}}\right)^{\frac{1}{3}}$. If $\delta_R$ deviates significantly from $\delta$ the absolute densities of the neutrino and antineutrino seas can be calculated using the following equations:
    \begin{align}
    \label{eq:neutrino_sees}
    \rho _{\nu} = \frac{\rho_{R\overline{\nu}}}{\delta_{R}^{3}-\delta^{3}} &&
    \rho_{\overline{\nu}} = \frac{\delta^{3} \rho_{R\overline{\nu}}}{\delta_{R}^{3}-\delta^{3}}
    \end{align}
    \item The z-values do not correspond directly to the particle radii. Currently, no established relationship between these two quantities is known. Regrettably, in the absence of a more accurate approach tailored to individual hadrons, the z-values may simply represent average values used to approximate the hadron masses.
    \item New measurements, analogous to those carried out at DESY in 1979 and discussed in \ref{subsec:Gluon_mass}, could yield more stringent results for comparison with the prediction of this model.  
    \item The 71 measured hadron masses, compared to the predicted values, show strong agreement. Additional measurements of the 31 remaining hadrons could provide further tests of the model.
    \item If the experiment described in \ref{disc:Creation of dark matter} reveals the missing energy as quantified, it could provide evidence for the existence of sterile bosons. Conversely, if no missing energy is observed, it would pose a challenge to the model.
    \item Asymmetric electric and weak forces might cause random walk behavior of nucleons. If these asymmetric forces cause movements that can be identified with the movements attributed to dark energy, then the dark energy's density must be proportional to the density of ordinary matter. 
    \item Relying solely on the incoming and outgoing momenta together with relativistic kinematics does not allow one to infer a minimum distance. This seems to rule out a direct, non–theory-dependent measurement. Still, for this type of process—Møller scattering—differences in runtime, the time electrons take to travel from some chosen separation before the primary interaction to a separation after it, could help distinguish this model from competing ones. 
    \item The situation mirrors the previous point, except that here a minimum velocity cannot be extracted. Again, runtime differences could potentially be used to discriminate between this model and others. However, analysis is more involved in this case because particle transmutation may occur.
    \item Eq. \eqref{eq:Photon_mass} allows for higher generation massive photons. So far, no massive photons have been detected. Current searches for massive photons are theory driven, which complicates finding an objective result.
\end{enumerate}


\section{Discussion}
\subsection{Model}
The described Kaluza-Klein analogue employs one of the simplest versions of a Calabi-Yau manifold \cite{RN221}, characterized in this context by a 10-dimensional spacetime organized in the following manner: 
\begin{itemize}
    \item \textbf{One dimension for time}  
    \item \textbf{Three standard spatial dimensions}  
    \item \textbf{Six compactified dimensions, each shaped like perpendicular cylinders}  
\end{itemize}    
In this 10D spacetime, there are no strings or additional elements; instead, it acts as a structural framework where disturbances propagate at light speed, altering the topology and breaching Ricci flatness \cite{RN62}. Most disturbances are transient ripples within the 10D spacetime; however, some become stable excitations within the compactified dimensions, manifesting as particles in 4D spacetime yet still traveling at light speed in 10D. A stable excitation within a compactified dimension of radius $r$ can be viewed as a sinusoidal wave with a wavelength $\lambda$, which adheres to condition \eqref{eq:double_phase_space}. By employing the Planck relation, one can describe the energy of a stable excitation, which enables summarization of the total energy of a hadron as comprising:  
\begin{itemize}
    \item \textbf{Excitation energies}  
    \item \textbf{Potential energies}  
    \item \textbf{Energies connected to the flavor constants}
    \item \textbf{Energies from magnetic dipoles within the curled-up dimensions }  
    \item \textbf{Energy from magnetism leaking into the three-dimensional space }
\end{itemize} 

\subsection{Constants}  
During the development of this model, it was both possible and necessary to determine 22 independent physical constants Table \ref{tab:Derived_constants}.  
\begin{table}[tbp]
    \centering
    \caption{Derived constants}
    \includegraphics[width=\linewidth,keepaspectratio]{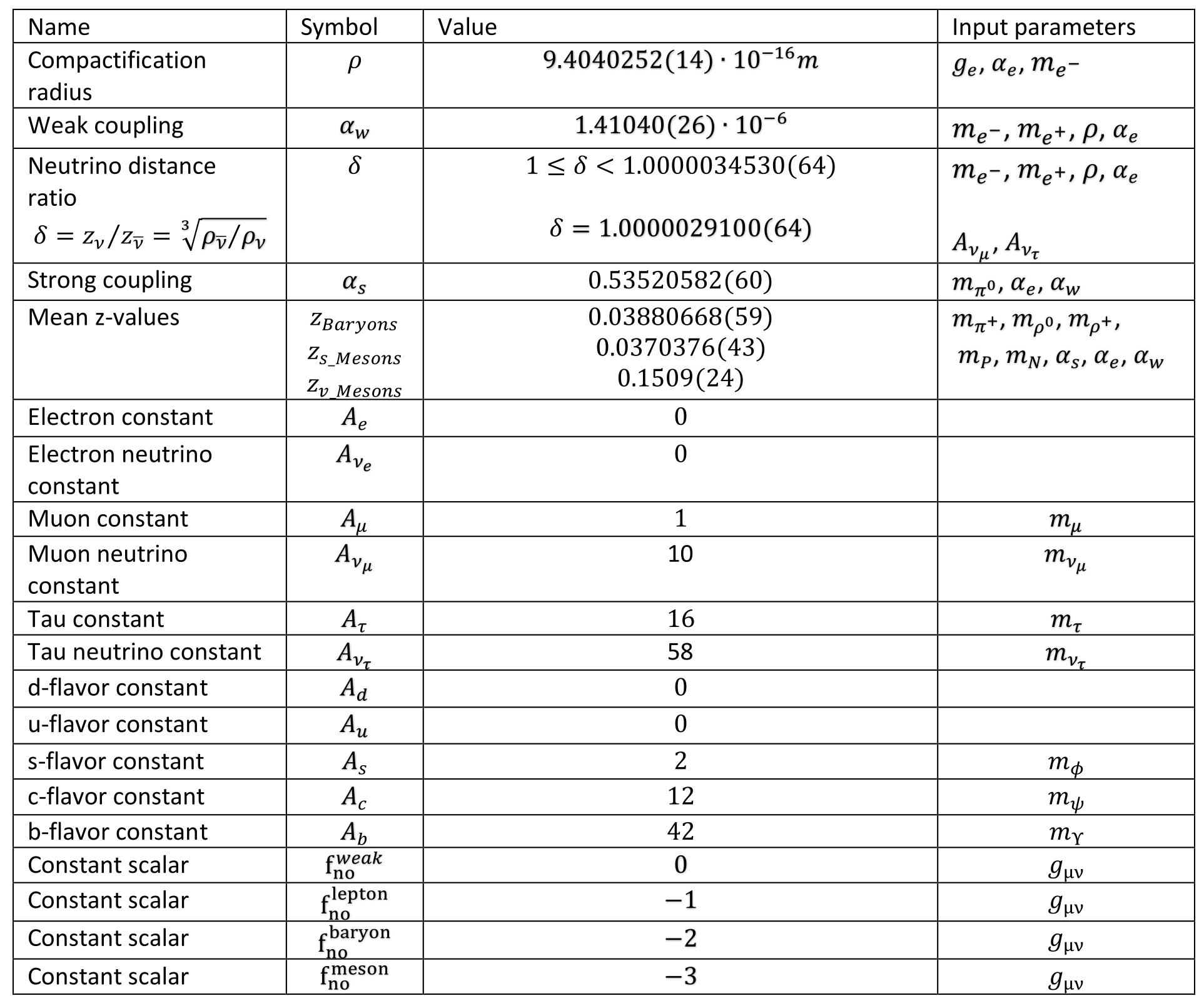}
    \label{tab:Derived_constants}
\end{table}
Fifteen of the constants are integer-valued, twenty-one are dimensionless, and only the compactification radius $\rho$ carries the physical dimension of length.

\subsection{Hadron masses}
Using Einstein's energy-mass equivalence, one can compute the masses of scalar mesons 

(Tables \ref{tab:Scalar_mesons} \& \ref{tab:Vector_mesons}) and baryon (Tables \ref{tab:hbar2_baryons} \& \ref{tab:3hbar2_baryons}) masses. In 70\% of cases, the relative difference between observed and predicted masses is less than 0.05; 89\% have errors below 0.1 and every relative error is less than 0.2. The model is grounded in the framework of the Standard Model of particle physics, encompassing the three foundational interactions—strong, electromagnetic, and weak—while extending the mathematical principles of general relativity to 10 dimensions. Thus, gravity plays an integral role within this framework. The model stipulates that the six compactified dimensions account for the strong (dimensions 4–6), electromagnetic (dimension 7), and weak (dimensions 8 \& 9) interactions. Within this framework, stable waves around compactified cylinders emerge naturally and constitute the primary source of mass. A significant portion of a particle’s mass, particularly for the light hadrons, originates from the magnetic field of the compactified dimension 7. For example, the mass contributions from magnetic fields and dipole interactions are:  
\begin{itemize}
    \item \textbf{Proton ($udu$)}: 28\%  
    \item \textbf{Neutron ($dud$)}: 29\%  
    \item \textbf{Lambda ($sud$)}: 16\%  
    \item \textbf{Bottom Xi ($bus$)}: 1\%  
    \item \textbf{Charged rho ($u \overline{d}$)}: 42\%  
    \item \textbf{Upsilon ($b \overline{b}$)}: 0.1\%  
\end{itemize}
Spin-$\hbar/2$ baryons consisting of three different quark flavors can exist with two slightly different masses. This model successfully accounts for these differences. For example, the $\Lambda^{0}$ and $\Sigma^{0}$ baryons share the same quark content ($d,s,u$) but exhibit different masses—a difference the model is able to calculate. Within this model, the potential to generate sterile bosons \eqref{eq:sterile_particle}, which could be a candidate for dark matter, emerges.

\subsection{Forces}\label{sec:Force calculation}  
In the calculation, the tensor formalism is applied individually to each type of interaction. This distinction is vital as the masses influenced by strong, electric, or weak forces are dictated solely by the corresponding metric tensors. However, quarks experience all three interactions concurrently. Rather than a unified calculation, Eq. \eqref{eq:all_forces} is used to perform three distinct force calculations, each tailored to one interaction. The outputs offer the specific force equations and graphical depictions found in subsection~\textit{\nameref{sec:Force_calculation_and_asymptotic}}. The net force acting on a particle is obtained by summing all contributing forces. However, these contributions are not entirely independent, as each force depends on the separation $z$ between the particle and its interaction partner, as well as the velocity of the particle and the first and second $z$-derivatives of that velocity. In turn, the velocity is affected by all three fundamental force components. In this model, \textbf{all interactions emerge purely from the curvature of the 10-dimensional spacetime}. The asymmetric nature of the electric force results in a continuous movement resembling a random walk. However, matter bonds through multiple forces, particularly the strong force, the overall electric force, and the gravitational force. Consequently, the random walk of hadrons does not lead to disintegration. Yet, a significant cumulative force might cause matter clusters to dislodge from one another, potentially explaining the phenomenon referred to as dark energy. Based on the aforementioned explanation, the density of dark energy ought to be directly proportional to the density of ordinary matter. If other galaxy clusters are in proximity, gravitational interactions between these clusters may counterbalance, leading to a negligible or null effect of dark energy. With growing distance between the clusters dark energy would seem to kick in again.

\subsection{Open questions}  

\subsubsection{Flavor constants}
The quark masses are specified in Eqs. \eqref{eq:reduced_dsb-quark_masses}–\eqref{eq:reduced_uct-anti-quark_masses}. They express a hypothesis regarding the role of the flavor constants, which yields physically reasonable hadron mass predictions. All coupling constants are scaled by the same flavor constant. Nevertheless, the dominant contribution to the resulting masses originates from the strong interaction. Various numerical choices for the three flavor constants are possible without significantly altering the hadron masses. However, no procedure is currently known that can uniquely determine the ratio of the flavor constants associated with the three interactions. 

\subsubsection{Deceleration in particle collisions}
When electrons and positrons collide, they can exhibit deceleration. The complexity of equations provided in Eqs. \eqref{eq:Electric_force_full} and \eqref{eq:Weak_force_full}, due to their reliance on variables $z$ and $\beta$, obstructs straightforward calculations. The numerical approach of subsection \textit{\nameref{ep-deceleration}} describes the mechanisms underlying the collision. Notably, investigating the emergence and role of the weak interaction at elevated beam energies could provide significant insights.

\subsubsection{Handedness}
In this study, handedness is not addressed directly. However, helicity can be derived from Equation $\eqref{eq:velocity}$ in combination with Tables $\ref{tab:fermions_first_generation}$ and \ref{tab:intermediate_bosons}. 

\subsubsection{Mass formula}
The mass of a hadron is expressed as the sum of its excitation energy, potential energy, and the energies arising from the magnetic field and magnetic moments in the electric dimension. The influences from both strong and weak dimensions on magnetism are disregarded because of their contributions canceling. However, there are still observed differences between the calculated and experimental masses that remain unexplained.  

\subsubsection{Massive photons}
\label{subsubsec:Massive_photons}
Eq. \eqref{eq:Photon_mass} points to the possible existence of massive photons: inserting $A_{\nu _{\mu}} = 10$ and $A_{\nu _{\tau}} = 58$ yields masses of $0.0086(1)\,\frac{\mathrm{eV}}{c^2}$ and $0.0501(3)\,\frac{\mathrm{eV}}{c^2}$, respectively. Such massive photons would be governed by a Yukawa-type potential, implying a finite interaction range $\lambda_{\gamma} \approx \frac{\hbar}{m_{\gamma} c}$ on the order of $10^{-5}\,\mathrm{m}$ and $2 \cdot 10^{-6}\,\mathrm{m}$. Consequently, massive photons would be irrelevant for observations involving astronomical photons, consistent with existing experimental constraints \cite{Vasileiou2013LIV,Wei2016FRBdispersion,KosteleckyMewes2013}. Furthermore, no short-range deviations attributable to a nonzero photon mass have been observed so far \cite{Ryutov1997PhotonMass,Luo2003PhotonMass,Williams1971PhotonMassGeomagnetic}, which could indicate the absence of higher-generation (massive) photons. Nevertheless, improved experimental sensitivity may eventually reveal such effects, should they exist.

\subsubsection{Sterile bosons}
The most general expression for sterile bosons is presented in Eq. \eqref{eq:sterile_particle}. The premise underlying the computation leading to the data in Table \ref{tab:invariant_masses_sterile_particles} is that surplus energy is confined within the dimensions of strong interaction. Nonetheless, alternative configurations for energy storage are feasible. Should missing mass resonances manifest in other energy sectors, the approach to energy storage must be adjusted while adhering to \eqref{eq:sterile_particle}.

\subsection{Future perspective}

To date, this model has been applied predominantly to the computation of hadron masses and interaction forces, while phenomena such as particle collisions and production mechanisms have received comparatively limited attention, and decay channels have not been treated at all. It remains an open question whether the framework can be extended to describe particle production and decay dynamics in a quantitatively reliable manner. Clearly, further theoretical refinements and dedicated experimental investigations are required to systematically assess both the capabilities and the limitations of this approach. However, if sterile bosons constitute dark matter and nuclear random walk can be linked to dark energy, the implications are vast. It suggests that an expanding force, resulting from the random walk of nucleons, existed abundantly after the big bang. This likely generated an immense force that separated the masses. Simultaneously, collision processes were frequent, leading, as the model suggests, to the widespread creation of dark matter (sterile bosons), which reduced ordinary matter in favor of dark matter. Therefore, for this model to hold, a variable cosmological parameter is a necessity. Moreover, the future of the universe is also noteworthy because as the proportion of normal matter to dark matter shifts in favor of dark matter, gravity may eventually dominate, leading to a potential final rebound. This could result in a cyclic universe.

\bibliographystyle{Frontiers-Harvard} 
\bibliography{KK-Mass-references}

\end{document}